\documentclass[prd,amsfonts,onecolumn,superscriptaddress,aps,nofootinbib,11pt]{revtex4-1}
\usepackage[top=3cm,bottom=3cm,left=2cm,right=2cm,marginparwidth=1.75cm]{geometry}

\usepackage{ragged2e}
\usepackage{bbold}

\usepackage{amssymb,amsmath}
\usepackage{graphicx}
\usepackage[dvipsnames,table]{xcolor}
\usepackage{hyperref}
\usepackage{listings}
\usepackage{soul}
\usepackage{slashed}
\usepackage{multirow}
\usepackage{subcaption}
\usepackage{xspace}

\usepackage{tikz-feynman}
\tikzfeynmanset{compat=1.1.0}

\usetikzlibrary{shapes.misc}

\tikzset{
  every crossed dot/.style={fill=white},
  every square dot/.style={fill=white}
}

\hypersetup{
	unicode=false,          % non-Latin characters in Acrobat’s bookmarks
	pdftoolbar=true,        % show Acrobat’s toolbar?
	pdfmenubar=true,        % show Acrobat’s menu?
	pdffitwindow=false,     % window fit to page when opened
	pdfauthor={William},     % author
	colorlinks=true,       % false: boxed links; true: colored links
	linkcolor=blue,          % color of internal links
	citecolor=red,        % color of links to bibliography
	urlcolor=blue
}
\graphicspath{{./images/}}

\definecolor{darkpastelpurple}{rgb}{0.59, 0.44, 0.84}
\definecolor{frenchlilac}{rgb}{0.53, 0.38, 0.56}
\definecolor{violet}{rgb}{0.56, 0.0, 1.0}

\definecolor{grey}{cmyk}{0,0,0,0.75}
\definecolor{tangerine}{cmyk}{0,0.5,1,0}
\definecolor{darkgreen}{cmyk}{1,0,1,0.23}
\definecolor{Red}{rgb}{1,0,0}
\definecolor{Blue}{rgb}{0,0,1}
\definecolor{Green}{rgb}{0,1,0}
\definecolor{Grey}{cmyk}{0,0,0,0.75}
\definecolor{Tangerine}{cmyk}{0,0.5,1,0}
\definecolor{Darkgreen}{cmyk}{1,0,1,0.23}
\definecolor{Cyan}{cmyk}{1,0,0,0}
\definecolor{Yellow}{cmyk}{0,0,1,0}
\definecolor{darkblue}{cmyk}{1,0.69,0,0.11}

\newcolumntype{s}{>{\columncolor[HTML]{ffebcd}}c}
\newcolumntype{q}{>{\columncolor[HTML]{fff4e3}}c}
\definecolor{exotic}{HTML}{c41525}
\definecolor{let}{HTML}{764b36}

\usepackage{hyperref}
\hypersetup{
	colorlinks = true,
	linkcolor = violet,
	urlcolor  = blue,
	citecolor = magenta,
	anchorcolor = cyan,
}

\usepackage{amsmath}

\usepackage{comment}
\usepackage{tikz}
\usetikzlibrary{decorations.pathmorphing,decorations.markings}

\tikzset{
	psi/.style={
		decoration={
			markings,
			mark=at position 0.6 with {\arrow{>}}
		},
		postaction={decorate},
		double,
		double distance=1pt
	},
	psiNoArrow/.style={
		decoration={
			markings,
			mark=at position 0.6 with 
		},
		postaction={decorate},
		double,
		double distance=1pt
	},
	nucleon/.style={
		decoration={
			markings,
			mark=at position 0.6 with {\arrow{>}}
		},
		postaction={decorate}
	},
	external/.style={},  
	gluon/.style={
		decorate, draw=black, 
		decoration={coil,amplitude=4pt, segment length=5pt}
	},
	particle/.style={draw=black, postaction={decorate}, decoration={markings,mark=at position .5 with {\arrow[draw=black]{>}}}},
	photon/.style={decorate, decoration={snake,amplitude=2pt, segment length=5pt}, draw=black}
}

\newcommand{\MI}{I\xspace}
\newcommand{\MII}{II\xspace}
\newcommand{\MIII}{III\xspace}
\newcommand{\MIV}{IV\xspace}

\tikzset{
  counterterm/.style={
    draw, thick, circle, fill=white, minimum size=8pt, inner sep=3pt,
    path picture={
      \draw[thick]
        (path picture bounding box.north west) -- (path picture bounding box.south east)
        (path picture bounding box.north east) -- (path picture bounding box.south west);
    }
  },
  big cross/.style={
    circle, minimum size=4mm, inner sep=0pt,
    path picture={
      \fill[white] (path picture bounding box.center) circle (3mm);
      \draw[line width=1.5pt, black]
        (path picture bounding box.south west) -- (path picture bounding box.north east);
      \draw[line width=1.5pt, black]
        (path picture bounding box.north west) -- (path picture bounding box.south east);
    }
  }
}

\newcommand{\fdiag}[2][1]{%
  \vcenter{\hbox{\scalebox{#1}{%
    \begin{tikzpicture}
      \begin{feynman}[every dot={/tikz/fill=black}]
        #2
      \end{feynman}
    \end{tikzpicture}}}}%
}

\newcommand{\ctVZH}{\fdiag[0.7]{
  \vertex (va) at ( 0.0, 0) {};
  \vertex [counterterm] (vb) at ( 2.3, 0) {};
  \vertex (f1) at ( 3.2,  1.0) {\(H\)};
  \vertex (f2) at ( 3.2, -1.0) {\(Z\)};
  \diagram* {
    (va) -- [boson, thick, edge label=\(\gamma/Z\)] (vb),
    (vb) -- [scalar, thick] (f1),
    (vb) -- [boson,  thick] (f2),
  };
}}

\begin{document}
	
	\lstset{frame=tb,
		language=Matlab,
		aboveskip=3mm,
		belowskip=3mm,
		showstringspaces=false,
		columns=flexible,
		basicstyle={\small\ttfamily},
		numbers=none,
		numberstyle=\tiny\color{gray},
		keywordstyle=\color{blue},
		commentstyle=\color{green},
		stringstyle=\color{mauve},
		breaklines=true,
		breakatwhitespace=true
		tabsize=3
	}
	
	\title{Vector-Like Fermions at FCC-ee: NLO Higgs-Strahlung Signatures and Constraints}

	\author{Carlo Marzo}
	\email{carlo.marzo@kbi.ee}
	\affiliation{National Institute of Chemical Physics and Biophysics, R\"{a}vala 10, 10143, Tallinn, Estonia}
	
	\author{Vinicius Padovani}
	\email{vinipadovani@kbfi.ee}
	\affiliation{National Institute of Chemical Physics and Biophysics, R\"{a}vala 10, 10143, Tallinn, Estonia}

	\author{Daniele Rizzo}
	\email{daniele.rizzo@kbfi.ee}
	\affiliation{National Institute of Chemical Physics and Biophysics, R\"{a}vala 10, 10143, Tallinn, Estonia}
	
	\begin{abstract}
		\centerline{\large\bfseries Abstract}
		\vspace{4pt}
		\large We compute the one-loop effects of vector-like fermions (VLFs) on the Higgs-strahlung cross-section, the flagship precision observable of future $e^+e^-$ Higgs factories such as FCC-ee. We consider four benchmark extensions of the Standard Model (SM): two in which a VL quark (VLQ) or VL lepton (VLL) doublet, together with its singlet partners, couples to the Higgs boson through purely internal Yukawa interactions, and two in which a single VL singlet mixes directly with the third-generation charged lepton or neutrino. Working throughout in the on-shell renormalization scheme and imposing perturbativity bounds on the relevant couplings under renormalization-group (RG) running, we compute both the universal self-energy and the non-universal vertex contributions to the Higgs-strahlung amplitude at one loop, together with the oblique parameters $S$ and $T$ and the $H\to\gamma\gamma$ rate as complementary cross-checks. In the two Yukawa-driven scenarios, we find that a sizable fraction of the parameter space still allowed by current LHC searches produces shifts in $\sigma(ZH)$ at or above the per-mille sensitivity envisioned for FCC-ee, offering a radiative probe of these couplings. In the two mixing-driven scenarios, existing electroweak-precision bounds on the relevant mixing angles already preclude an observable effect, although we characterize the underlying loop dynamics in full generality for future reference. These results identify the VL Yukawa couplings, rather than any residual mixing with the SM fermions, as the more promising target for Higgs-strahlung precision measurements at a future $e^+e^-$ collider.
	
	\end{abstract}
	
	\maketitle
	
\section{Introduction}\label{sec:introduction}
The SM of particle physics stands as the most precisely tested theoretical framework to date. Yet, with all its  remarkable success, it is undoubtedly incomplete.
This is true on purely experimental grounds, given that the SM offers no explanation for dark matter, baryonic asymmetry, or the origin of neutrino masses.
When these experimental shortcomings are seen through the eyes of theoretical speculation, the SM predictivity suffers from an overabundance
of free parameters and arbitrarily diverse, disconnected scales which can enjoy a more economical explanation within alternative frameworks, such as Grand Unified Theories or supersymmetry.

Well-motivated avenues that extend the SM while addressing such shortcomings often involve VLFs~\cite{Ellis:2014dza}: new fermions whose left- and right-handed
components transform identically under the SM gauge group. This feature, which distinguishes them from the chiral SM generations, has profound consequences. 
Mass terms for VLFs are gauge-invariant without a Higgs insertion and therefore technically natural, rendering the fermion masses free from fine-tuning even
when they are much heavier than the electroweak (EW) scale. Moreover, VLFs can be added to the SM without generating gauge anomalies, and they are predicted in concrete
ultraviolet (UV) completions such as composite Higgs models, partial compositeness scenarios~\cite{Caracciolo:2012je,Panico:2015jxa}, and extra-dimensional constructions~\cite{Ichinose:2002kg}.
They may also play a role in generating the flavor structure of the SM through mixing with third-generation quarks and leptons~\cite{Adhikary:2024esf,CarcamoHernandez:2023wzf}.
VLF mixing with third-generation leptons has similarly been invoked to timely address possible deviations from the SM as in the Cabibbo angle anomaly~\cite{Crivellin:2020ebi} or the muon anomalous magnetic moment~\cite{deJesus:2023som,CarcamoHernandez:2023wzf}.
Interestingly, attempts to achieve a first-order phase transition within a framework of VLF plus a minimal singlet scalar were met with significant constraints from  double-Higgs production measurements~\cite{Adhikary:2024esf}.

Direct VLF searches at the LHC have so far returned null results, pushing mass limits into the TeV range~\cite{CMS:2024bni,Banerjee:2024zvg}. 
Probing the corresponding loop-induced effects at a precision lepton collider offers therefore a complementary, mass-independent handle on the same states, in line with the broader shift of the field from the LHC's discovery-first program towards precision physics.
A consensus has emerged in the community that an electron-positron Higgs factory represents the most compelling next step~\cite{deBlas:2024bmz,Altmann:2025feg}.
Among the proposals under active study, the circular colliders FCC-ee~\cite{FCC:2018evy,FCC:2025lpp,Agapov:2022bhm} and CEPC~\cite{CEPCStudyGroup:2018ghi}, the linear colliders ILC~\cite{ILC:2013jhg}
and CLIC~\cite{Roloff:2018dqu}, and the LEP-3 concept~\cite{Anastopoulos:2025jyh}, FCC-ee has emerged as the leading candidate for the early phases of this program.
The \emph{Higgs-strahlung} process $e^+e^-\to ZH$ peaks at the next-to-leading order (NLO) cross-section $\sigma_{\rm ZH}^{\rm NLO}\simeq 229$~fb near $\sqrt{s}\simeq 244$~GeV, and the clean,
fully-known initial state of the leptonic collider allows the inclusive $ZH$ rate to be reconstructed through the recoil-mass technique independently of the 
Higgs decay channel~\cite{FCC:2018evy,deBlas:2019rxi,Azzurri:2021nmy,Kahraman:2025myp}. FCC-ee is designed to accumulate $\mathcal{O}(10^6)$ events at this peak run,
with a projected precision of $\sim 0.3\%$ on the inclusive cross-section~\cite{Li:2025kjy}, pushing the ultimate reach of the full dataset approaches 
the one-per-mille level~\cite{Azzurri:2021nmy}. The same facilities also foresee operations at higher center-of-mass energies: FCC-ee adds a stage 
at $\sqrt{s}=365$~GeV near the $t\bar{t}$ threshold, where a second measurement of $\sigma_{ZH}$ is projected at $\sim 0.5\%$ accuracy~\cite{Li:2025kjy},
 while CLIC begins at $380$~GeV and the linear colliders extend further to $500$~GeV and beyond~\cite{ILC:2013jhg,Roloff:2018dqu}. As $\sqrt{s}$ grows the 
 Higgs-strahlung cross-section falls and alternative production channels such as $W$-boson fusion $e^+e^-\to\nu\bar\nu H$ become more competitive, nonetheless 
 the $ZH$ process remains accessible throughout, and the higher-energy measurements provide complementary, more discovery-prone sensitivity to modifications of
the cross-section energy dependence.

This sub-percent precision transforms the Higgs-strahlung cross-section into a precision observable in its own right: a quantity sensitive not only to the overall
coupling strength of the Higgs to the $Z$ boson, but also to radiative corrections induced by any new states that couple to the EW sector.
Heavy VLFs that are kinematically inaccessible at LHC energies, or whose direct-production cross-sections are too small to be distinguished from background, 
can nonetheless leave an imprint in $\sigma(ZH)$ through one-loop virtual contributions. Such loop-level sensitivity to new physics (NP) in $e^+e^-\to ZH$ has recently been demonstrated in the Standard Model Effective Field Theory framework~\cite{Asteriadis:2024xuk}, where NLO weak corrections were shown to shift the cross-section at a level comparable to the projected experimental precision. The VLF-induced loop effects scale as $y_{\rm VL}^2/(16\pi^2)$ relative to the
tree-level cross-section, where $y_{\rm VL}$ is the relevant VLF Yukawa coupling. Within the parameter space allowed by perturbativity, $y_{\rm VL}$ 
can reach $\mathcal{O}(1)$, making the loop-induced correction a potentially observable signal at the per-mille sensitivity of FCC-ee. We will adopt
 $\mathcal{O}(10^{-3})$ as the reference precision against which to test the relevance of the NP effects computed in this work. Such a value may look mildly
 optimistic relative to current analyses, but is a reasonable assumption given the technological advances expected to contribute before the actual
 measurements begin.

In the following, we present our survey of the one-loop VLF contributions to $e^+e^-\to ZH$ for four simplified benchmark submodels that span the qualitatively
distinct behaviors of the general VLF extension of the SM. Model~\MI extends the SM quark sector with a VL doublet and up- and down-type singlets that 
mix only within the NP sector, characterized by Yukawa couplings $y_U$, $y_D$ and a common mass parameter $M_Q$. 
Model~\MII is the leptonic analog: a VL lepton doublet and singlets with couplings $y_E$, $y_N$ and mass $M_L$. 
Model~\MIII adds a single right-handed neutrino singlet that mixes with the $\tau$-neutrino through a Yukawa coupling $\lambda_\nu$, generating a new 
physical mass eigenstate $m_N$ and a neutrino mixing angle $\theta_\nu$. Model~\MIV is the charged-lepton analog, in which a right-handed singlet mixes with
the $\tau$ through $\lambda_e$, yielding a new mass eigenstate $m_E$ and two physical rotation angles $\theta_L$, $\theta_R$. Models~\MI and~\MII contribute to
Higgs-strahlung exclusively through diagrams in which only NP states propagate in the loops. Models~\MIII and~\MIV instead generate diagrams in which SM
and NP fermions circulate simultaneously demanding a particular attention in highlighting the pure NP effects.

Two classes of external constraints bound the parameter space of each submodel. The first consists of direct experimental bounds: lower limits on the VLF masses from pair-production searches at LEP and the LHC for Models~\MI and~\MII, and upper limits on the mixing angles from EW precision fits for Models~\MIII~\cite{Blennow:2023mqx} and~\MIV~\cite{Crivellin:2020ebi}. The second is a theoretical requirement of perturbativity: the Yukawa couplings must remain perturbative under RG evolution up to a
benchmark UV cutoff, implemented numerically using the \texttt{RGBeta} package~\cite{Thomsen:2021ncy}. Together these constraints delimit the region of parameter space in which
the computed cross-section deviations are simultaneously experimentally allowed and theoretically under control. The resulting perturbativity limits are summarized in
Table~\ref{table:perturbativity-merged}.

Beyond the Higgs-strahlung cross-section itself, we evaluate two further observables that probe the same one-loop VLF contributions from complementary angles.
The self-energy corrections to the gauge-boson propagators that enter the universal contribution to $\sigma(ZH)$ also appear in the EW universal fit parametrized by $S$ and $T$ ~\cite{Peskin:1991sw,Cynolter:2008ea}.
Current experimental constraints on $S$ and $T$~\cite{ParticleDataGroup:2024cfk,Freitas:2023xnx} provide a mandatory consistency check. The non-universal vertex corrections sourced from the VLF Yukawa interactions will instead show up in the loop-induced
 $H\to\gamma\gamma$ rate. The latter is constrained by recent ATLAS and CMS measurements~\cite{ATLAS:2022tnm,CMS:2021kom} and provides an independent handle 
 on the impact of the Yukawa couplings.

The paper is organized as follows: Section~\ref{sec:the-model} presents the VLF Lagrangian, the four simplified submodels, and the main phenomenological constraints from  collider searches on each.
Section~\ref{subsec:perturbativity} imposes perturbativity and presents the resulting allowed parameter space.
Section~\ref{sec:cross-section-calculation} details the one-loop computation: the form-factor decomposition and the renormalization procedure,
the universal self-energy corrections and their relation to $S$ and $T$, the non-universal vertex corrections and their impact on the $H\to\gamma\gamma$ calculation.
Section~\ref{sec:numerical-analysis-and-results} presents the numerical results. Section~\ref{sec:conc}
contains our conclusions. Appendix~\ref{sec:1-loop-renormalization} provides details of the on-shell (OS) renormalization scheme we adopted, and Appendix~\ref{sec:RGE} 
collects the RG equations.

\section{SM extension with VLFs}\label{sec:the-model}
In this work we study the SM extended by one full VL family of quarks and leptons in a framework that keeps the scalar sector minimal, with just the SM
Higgs doublet. Closely related realizations of this setup have been analyzed previously in Ref.~\cite{Adhikary:2024esf}, here we simply take the full
renormalizable VL extension as the underlying model and later specify the submodels relevant for the different phenomenological limits.

To support an unambiguous embedding with the SM chiral fields, we will also use left-chiral two-component Weyl spinors for VL states as well.
Therefore, the subscripts $L$ and $R$ are part of the field names and indicate the SM gauge quantum
numbers associated with the usual left- and right-handed SM fermions, rather than Lorentz chirality. The VL partners transforming in the conjugate gauge
representations, and therefore completing the left-handed counterpart in a 4-component Dirac state, are denoted by a tilde.

The SM fermions are collectively denoted by
\begin{equation}
	\psi^i = \left\{Q_L^i,\,u_R^i,\,d_R^i,\,L_L^i,\,e_R^i \right\}, \qquad i=1,2,3,
    \label{eq:VLF_doublet}
\end{equation}
where $i$ is a generation index. The VL sector is taken to contain one quark doublet, one lepton doublet, and the corresponding singlets,
\begin{equation}
	\Psi_4 = \left\{Q_L^d,\,U_R^s,\,D_R^s,\,L_L^d,\,N_R^s,\,E_R^s \right\},
    \label{eq:VLF_singlet}
\end{equation}
together with their conjugate partners
\begin{equation}
	\widetilde \Psi_4 = \left\{\widetilde Q_R^d,\,\widetilde U_L^s,\,\widetilde D_L^s,\,\widetilde L_R^d,\,\widetilde N_L^s,\,\widetilde E_L^s \right\}.
    \label{eq:VLF_singlet_conjugate}
\end{equation}
The gauge quantum numbers of the new fields are summarized in Table~\ref{table:Fermions-of-the-general-model}.
With these conventions, the most general renormalizable fermionic Lagrangian can be decomposed as
\begin{equation}
	\mathcal{L}_{\rm fermion} \sim Y_{ij}\, \psi^i \psi^j \Phi + M_4\, \Psi_4 \widetilde \Psi_4 + y_4\, \Psi_4 \widetilde \Psi_4 \Phi + \lambda_i\, \psi^i \Psi_4 \Phi + \widetilde{\lambda}_i\, \psi^i \widetilde \Psi_4 \Phi + \text{h.c.},
\end{equation}
or, equivalently, expanding the collective fields using Eqs.~(\ref{eq:VLF_doublet}-\ref{eq:VLF_singlet_conjugate})
\begin{equation}
	\mathcal{L}_{\rm fermion} = \mathcal{L}_{\rm SM}^{Y} + \mathcal{L}_{\rm VL}^{M} + \mathcal{L}_{\rm VL}^{Y} + \mathcal{L}_{\rm mix}^{Y} + \text{h.c.},
    \label{eq:generic_lagrangian}
\end{equation}
where the various contractions are understood to be chosen according to the gauge quantum numbers of the corresponding fields. The fully expanded renormalizable Lagrangian then reads
\begin{equation}
	\begin{split}
		\mathcal{L}_{\rm SM}^{Y} ={}& (Y_u)_{ij}\, Q_L^i \cdot \widetilde{\Phi}\, u_R^j + (Y_d)_{ij}\, Q_L^i \cdot \Phi\, d_R^j + (Y_e)_{ij}\, L_L^i \cdot \Phi\, e_R^j ,
	\end{split}
\end{equation}
\begin{equation}
	\begin{split}
		\mathcal{L}_{\rm VL}^{M} ={}& M_Q\, Q_L^d \widetilde Q_R^d + M_U\, \widetilde U_L^s U_R^s + M_D\, \widetilde D_L^s D_R^s \\
		&+ M_L\, L_L^d \widetilde L_R^d + M_N\, \widetilde N_L^s N_R^s + M_E\, \widetilde E_L^s E_R^s ,
	\end{split}
\end{equation}
\begin{equation}
	\label{general-fermion-lagrangian}
	\begin{split}
		\mathcal{L}_{\rm VL}^{Y} ={}& y_U\, Q_L^d \cdot \widetilde{\Phi}\, U_R^s + \widetilde y_U\, \widetilde Q_R^d \cdot \Phi\, \widetilde U_L^s
		+ y_D\, Q_L^d \cdot \Phi\, D_R^s + \widetilde y_D\, \widetilde Q_R^d \cdot \widetilde{\Phi}\, \widetilde D_L^s \\
		&+ y_N\, L_L^d \cdot \widetilde{\Phi}\, N_R^s + \widetilde y_N\, \widetilde L_R^d \cdot \Phi\, \widetilde N_L^s
		+ y_E\, L_L^d \cdot \Phi\, E_R^s + \widetilde y_E\, \widetilde L_R^d \cdot \widetilde{\Phi}\, \widetilde E_L^s ,
	\end{split}
\end{equation}
and
\begin{equation}
	\begin{split}
		\mathcal{L}_{\rm mix}^{Y} ={}& (\lambda_u)_i\, Q_L^i \cdot \widetilde{\Phi}\, U_R^s + (\lambda_d)_i\, Q_L^i \cdot \Phi\, D_R^s + (\lambda_e)_i\, L_L^i \cdot \Phi\, E_R^s + (\lambda_\nu)_i\, L_L^i \cdot \widetilde{\Phi}\, N_R^s \\
		&+ (\kappa_u)_i\, Q_L^d \cdot \widetilde{\Phi}\, u_R^i + (\kappa_d)_i\, Q_L^d \cdot \Phi\, d_R^i + (\kappa_e)_i\, L_L^d \cdot \Phi\, e_R^i .
	\end{split}
\end{equation}

The term $\mathcal{L}_{\rm SM}^{Y}$ contains the ordinary SM Yukawa interactions, $\mathcal{L}_{\rm VL}^{M}$ collects the VL mass terms,
 $\mathcal{L}_{\rm VL}^{Y}$ describes Higgs-induced interactions entirely within the VL sector, while $\mathcal{L}_{\rm mix}^{Y}$ contains 
 the Yukawa interactions that mix the VL states with the SM fermions. 
 Since no right-handed neutrino is introduced in the SM sector, there is no term of the form $L_L^d \cdot \widetilde{\Phi}\, \nu_R^i$. 
 Notice that we are not addressing massive neutrino phenomenology with our VLFs models, which must therefore be thought of as first-approximation massless all 
 along. Consequently, bounds coming from neutrino mixing cannot be used to rule out our parameter space, since this choice is dictated by a theoretical
 assumption: the mechanism responsible for neutrino masses is not affecting in any sizable way the Higgs-strahlung process, which is the main target of this work.

Since here we do not consider explicit CP violation, and as a simplifying assumption, we take the Yukawa couplings to be real. Moreover, following \cite{Adhikary:2024esf}, we reduce the number of free 
Yukawa parameters by requiring that in each sector they are equal:
\begin{equation}
	\widetilde y_S = y_S, \qquad S = U,D,N,E,
\end{equation}
so that each VL Yukawa sector can be parameterized in terms of just one real coupling.

\begin{table}[t]
\centering
\setlength{\tabcolsep}{6pt}
\begin{tabular}{c c c c c c c c c c c c c}
\hline
Field & $Q^d_L$ & $U^s_R$ & $D^s_R$ & $L^d_L$ & $N^s_R$ & $E^s_R$ & $\widetilde Q^d_R$ & $\widetilde U^s_L$ & $\widetilde D^s_L$ & $\widetilde L^d_R$ & $\widetilde N^s_L$ & $\widetilde E^s_L$ \\
\hline
$SU(3)_C$ & $\mathbf{3}$ & $\overline{\mathbf{3}}$ & $\overline{\mathbf{3}}$ & $\mathbf{1}$ & $\mathbf{1}$ & $\mathbf{1}$ & $\overline{\mathbf{3}}$ & $\mathbf{3}$ & $\mathbf{3}$ & $\mathbf{1}$ & $\mathbf{1}$ & $\mathbf{1}$ \\
$SU(2)_L$ & $\mathbf{2}$ & $\mathbf{1}$ & $\mathbf{1}$ & $\mathbf{2}$ & $\mathbf{1}$ & $\mathbf{1}$ & $\mathbf{2}$ & $\mathbf{1}$ & $\mathbf{1}$ & $\mathbf{2}$ & $\mathbf{1}$ & $\mathbf{1}$ \\
$U(1)_Y$ & $+\frac{1}{6}$ & $-\frac{2}{3}$ & $+\frac{1}{3}$ & $-\frac{1}{2}$ & $0$ & $+1$ & $-\frac{1}{6}$ & $+\frac{2}{3}$ & $-\frac{1}{3}$ & $+\frac{1}{2}$ & $0$ & $-1$ \\
\hline
\end{tabular}
\caption{VLF multiplets of the model studied in this work. The superscripts $s$ and $d$ denote singlets and doublets under $SU(2)_L$, respectively.}
\label{table:Fermions-of-the-general-model}
\end{table}

The various submodels considered later are then obtained from the complete Lagrangian above by selecting a given subset among the VLF family presented. 
As a further simplifying assumption, we consider the mixing couplings negligible or close to zero in the case of Model~\MI and Model~\MII. 
Such mixing could be small simply because the corresponding couplings happen to be tiny, or because a symmetry forbids $\mathcal{L}_{\rm mix}^{Y}$ altogether, e.g.\ an additional gauge symmetry under which the VL multiplets are charged and the SM fermions are not, broken at some high scale.
We do not commit to a specific completion here, but the distinction matters for how strictly the zero-mixing submodels introduced below should be interpreted, as discussed below for Models~\MI and~\MII. In the following subsections we introduce the models and describe their main features.

\subsection{Model \MI}
Model~\MI extends the SM with just the VLQ sector of Eq.~(\ref{eq:generic_lagrangian}), and the SM fermions are not mixed with the VLQs, so that the only non-zero NP Lagrangian parameters are the VL doublet mass $M_Q$, the VL singlet masses $M_U$, $M_D$, and the Yukawa couplings $y_U$, $y_D$. Since there is no mixing between the SM and the VLQ sector, the up- and down-type quark mass matrices are block diagonal. The SM blocks take the usual form, while the VLQ blocks take the form
\begin{equation}
	M_{\rm S} = \begin{pmatrix}
		M_{Q} & \frac{1}{\sqrt{2}} \, v \, y_{S} \\
		\frac{1}{\sqrt{2}} \, v \, y_{S}^*  & M_{S}
	\end{pmatrix} \, , \qquad S = U,D \, .
    \label{eq:model_1_quark_mass_matrices}
\end{equation}
Physical masses are obtained by diagonalizing the mass matrix, with the result
\begin{equation}
    m_{S^{\pm}} = \frac{1}{2}\left(M_{Q} + M_{S} \pm \sqrt{(M_{Q} - M_{S})^2 + 2 \, v^2 \, |y_S|^2} \right) \, , \qquad S = U,D \, .
\end{equation}
Treating $M_Q$ as a free input, these four equations can be inverted to express the two singlet masses and the two Yukawa couplings as functions of the four physical masses $m_{U^\pm}$, $m_{D^\pm}$ and $M_Q$:
\begin{equation}
    M_S = m_{S^+} + m_{S^-} - M_Q \,, \qquad
    y_S = \frac{\sqrt{2}}{v}\sqrt{(m_{S^+} - M_Q)(M_Q - m_{S^-})} \,, \qquad S = U, D \,.
    \label{eq:lagrangian_masses_in_terms_of_physical_ones}
\end{equation}
Real solutions for $y_S$ require $m_{S^-} \leq M_Q \leq m_{S^+}$ in each sector.

The rotation angle that diagonalizes the mass matrix is easily computed, and reads
\begin{equation}
	\label{eq:rotation_angle_mass_matrices_model_1}
	\tan 2 \, \theta_{S} = \frac{\sqrt{2}\; v \; y_S}{M_{S} -M_{Q}} = \frac{2\sqrt{(m_{S^+}-M_Q)(M_Q-m_{S^-})}}{m_{S^+}+m_{S^-}-2M_Q} \, , \qquad S = U,D \, ,
\end{equation}
where in the last step we have substituted Eq.~\eqref{eq:lagrangian_masses_in_terms_of_physical_ones} to express the mixing angle directly in terms of the physical masses and $M_Q$.

Model~\MI is defined with exactly vanishing SM-mixing Yukawas, so it carries no direct mixing-driven exclusion. As a reference lower bound we nonetheless adopt the pair-production limit that applies once any small residual mixing is present, since any realistic completion is expected to retain one. Recent ATLAS and CMS pair-production searches exclude VLQs up to $1300$--$1500$~GeV~\cite{Benbrik:2024fku}, essentially independently of the (small) mixing, so we take this as the lower bound on the lighter physical masses $m_{U^-}$, $m_{D^-}$, while the derived Yukawa couplings $y_U$ and $y_D$ are bounded from above by perturbativity (Sec.~\ref{subsec:perturbativity}).

\subsection{Model \MII}
Model~\MII is the leptonic counterpart of Model~\MI: the SM is extended with the VLL sector of Eq.~(\ref{eq:generic_lagrangian}) alone, and, as in Model~\MI, the SM fermions are not mixed with the new states. The only non-zero NP Lagrangian parameters are therefore the VL doublet mass $M_L$, the VL singlet masses $M_E$, $M_N$, and the Yukawa couplings $y_E$, $y_N$. The mass matrix takes the form
\begin{equation}
	M_{\rm S} = \begin{pmatrix}
		M_{L} & \frac{1}{\sqrt{2}} \, v \, y_{S} \\
		\frac{1}{\sqrt{2}} \, v \, y_{S}^*  & M_{S}
	\end{pmatrix} \, , \qquad S = E,N \, .
    \label{eq:model_2_lepton_mass_matrices}
\end{equation}
Diagonalization proceeds exactly as in Model~\MI, with $M_Q\to M_L$, giving the physical masses
\begin{equation}
    m_{S^{\pm}} = \frac{1}{2}\left(M_{L} + M_{S} \pm \sqrt{(M_{L} - M_{S})^2 + 2 \, v^2 \, |y_S|^2} \right) \, , \qquad S = E,N \, ,
\end{equation}
Treating $M_L$ as a free input, the inversion gives the singlet masses $M_E$, $M_N$ and the Yukawa couplings $y_E$, $y_N$ as functions of the four physical masses $m_{E^\pm}$, $m_{N^\pm}$ and $M_L$:
\begin{equation}
    M_S = m_{S^+} + m_{S^-} - M_L \,, \qquad
    y_S = \frac{\sqrt{2}}{v}\sqrt{(m_{S^+} - M_L)(M_L - m_{S^-})} \,, \qquad S = E, N \,,
    \label{eq:lagrangian_masses_in_terms_of_physical_ones_model2}
\end{equation}
with real solutions requiring $m_{S^-} \leq M_L \leq m_{S^+}$ in each sector. The mixing angle reads
\begin{equation}
	\label{eq:rotation_angle_mass_matrices_model_2}
	\tan 2 \, \theta_{S} = \frac{\sqrt{2}\; v \; y_S}{M_{S} -M_{L}} = \frac{2\sqrt{(m_{S^+}-M_L)(M_L-m_{S^-})}}{m_{S^+}+m_{S^-}-2M_L} \, , \qquad S = E,N \, .
\end{equation}

Like Model~\MI, Model~\MII carries no direct mixing-driven exclusion. The appropriate model-independent reference is the branching-ratio scan of Ref.~\cite{Dermisek:2014qca}, which recasts LHC multilepton searches over the full space of VLL decay branching ratios and yields $M_L \gtrsim 300$--$500$~GeV once any non-zero mixing is assumed (dedicated third-generation VLL searches reach $130$--$900$~GeV~\cite{ATLAS:2023sbu,CMS:2019hsm} but assume a single dominant decay channel), so we take this as the lower bound on the lighter physical masses $m_{E^-}$, $m_{N^-}$, while the derived Yukawa couplings $y_E$ and $y_N$ are bounded from above by perturbativity (Sec.~\ref{subsec:perturbativity}).

\subsection{Model \MIII}
While Models~\MI and~\MII probe the cross-section's sensitivity to the VL Yukawa couplings alone, in this and the following model we focus instead on its sensitivity to the mixing between VLFs and the SM fermions. Model~\MIII introduces only a neutral leptonic singlet, $N$, coupled to the $\tau$ generation of SM neutrinos. The only non-zero NP Lagrangian parameter is therefore $(\lambda_\nu)_3$, which from now on we simply denote $\lambda_\nu$. The Majorana-type mass matrix in the subspace spanned by the $\tau$-neutrino and the VL neutrino states $N_L^s$ and $N_R^s$ takes the form
\begin{equation}
	M_{\rm \nu} = 
    \begin{pmatrix}
		0 & 0 & \frac{1}{\sqrt{2}} \, v \, \lambda_\nu\\
        0 & 0 & M_{N} \\
		\frac{1}{\sqrt{2}} \, v \, \lambda_\nu^* & M_{N} & 0
	\end{pmatrix}  \, .
    \label{eq:model_3_neutrino_mass_matrices}
\end{equation}
Evidently, one of the physical states remains massless, and we can associate it with the $\tau$-neutrino, while the VL neutrino gets a Dirac mass 
\begin{equation}
    m_{N} =  \sqrt{M_{N} ^2 + \frac{1}{2} \, v^2 \, |\lambda_\nu|^2} \, .
\end{equation}
The mixing angle that diagonalizes Eq.~\eqref{eq:model_3_neutrino_mass_matrices} is also easily found
\begin{equation}
	\label{eq:rotation_angle_mass_matrices_model_3}
	\tan \theta_{\nu} =\frac{v \, \lambda_\nu}{\sqrt{2} \, M_{N}} \, .
\end{equation}
The independent inputs of our scan are the physical VL neutrino mass $m_N$ and the mixing Yukawa $\lambda_\nu$. The Lagrangian mass $M_N$ is then determined by inverting the $m_N$ formula above:
\begin{equation}
    M_N = \sqrt{m_N^2 - \frac{v^2\lambda_\nu^2}{2}}\,,
    \label{eq:lagrangian_mass_model3}
\end{equation}
and the mixing angle $\theta_\nu$ follows from Eq.~\eqref{eq:rotation_angle_mass_matrices_model_3}. Real solutions require
\begin{equation}
    |\lambda_\nu| \leq \frac{\sqrt{2}\,m_N}{v} \, .
    \label{eq:model3_lambda_bound}
\end{equation}
Model~\MIII mixes with SM leptons by construction, so that the strongest experimental constraint is an upper bound on the mixing angle rather than a lower bound on the mass: a global EW fit to non-unitarity of the effective neutrino mixing matrix~\cite{Blennow:2023mqx} gives $\eta_{\tau\tau} \leq 8.9\times10^{-4}$ (95\% confidence level, general-seesaw scenario), translating to $\theta_\nu^{\max} = \sqrt{2\eta_{\tau\tau}} \simeq 0.042$, essentially mass-independent above the EW scale (the dedicated CMS search for tau-coupled heavy neutral leptons~\cite{CMS:2024xdq} is at least one order of magnitude weaker than this floor and does not improve the bound). Together with the kinematic bound of Eq.~\eqref{eq:model3_lambda_bound}, $\lambda_\nu$ is therefore bounded from above by this mixing-angle constraint and by perturbativity (Sec.~\ref{subsec:perturbativity}).

\subsection{Model \MIV}
Model~\MIV is the analog of Model~\MIII for the charged-lepton sector: instead of extending the SM with a neutral VLL, we introduce a charged state, $E$ mixed with the $\tau$ lepton. The only non-zero NP Lagrangian parameter is therefore $(\lambda_e)_3$, which from now on we simply denote $\lambda_e$. Similarly, the $33$ element of the SM Yukawa matrix, $(Y_e)_{33}$, is denoted $Y_e$. The mass matrix takes the form
\begin{equation}
	M_{\rm e} = 
    \begin{pmatrix}
		 \frac{1}{\sqrt{2}} \, v \, Y_e & \frac{1}{\sqrt{2}} \, v \, \lambda_e \\
		 0 & M_E
	\end{pmatrix}  \, ,
    \label{eq:model_4_charged_lepton_mass_matrix}
\end{equation}
and is diagonalized by the bi-unitary transformation
\begin{equation}
U_L^\dagger M_{\rm e} U_R =
\begin{pmatrix}
m_\tau & 0\\
0 & m_E
\end{pmatrix},
\end{equation}
where (we assume real parameters) 
\begin{equation}
U_L =
\begin{pmatrix}
\cos\theta_L & \sin\theta_L\\
-\sin\theta_L & \cos\theta_L
\end{pmatrix},
\qquad
U_R =
\begin{pmatrix}
\cos\theta_R & \sin\theta_R\\
-\sin\theta_R & \cos\theta_R
\end{pmatrix}.
\label{eq:model4_UL_UR}
\end{equation}

After standard calculations, one obtains
\begin{equation}
\tan 2\theta_L =
\frac{\sqrt{2}v\lambda_e M_E}
{M_E^2-\frac{v^2}{2}\left( \lambda_e^2 + Y_e^2 \right)} \, ,
\label{eq:model4_tan2thetaL}
\end{equation}
and
\begin{equation}
\tan 2\theta_R
= \frac{v^2 Y_e \lambda_e}{M_E^2+\frac{v^2}{2}\left(\lambda_e^2-Y_e^2\right)} \, .
\end{equation}
The physical squared masses are the eigenvalues of either
$M_{\rm e}M_{\rm e}^\dagger$ or $M_{\rm e}^\dagger M_{\rm e}$:
\begin{equation}
m_{\tau,E}^2 =
\frac{1}{2}\left[M_E^2+\frac{v^2}{2}\left(Y_e^2+\lambda_e^2\right) \mp \sqrt{ \left[ M_E^2+\frac{v^2}{2}\left( Y_e^2 + \lambda_e^2 \right) \right]^2
-2v^2Y_e^2M_E^2} \right].
\label{eq:model4_tau_E_masses}
\end{equation}
Here the upper sign corresponds to the mostly-SM charged lepton, while the lower sign corresponds to the mostly-VL charged lepton.
Since $M_E$ is not directly observable, Eq.~\eqref{eq:model4_tau_E_masses} can instead be inverted for $M_E^2$ at fixed $m_\tau,m_E$. Requiring the result to be real and positive bounds $|\lambda_e|$. From the sum and product of the two eigenvalues,
\begin{equation}
2M_E^2 + v^2\left(\lambda_e^2+Y_e^2\right) = 2\left(m_\tau^2+m_E^2\right)\,, \qquad
M_E^2\,v^2 Y_e^2 = 2\,m_\tau^2 m_E^2\,,
\label{eq:model4_sum_product}
\end{equation}
eliminating $Y_e$ gives a quadratic equation for $M_E^2$,
\begin{equation}
2\left(M_E^2\right)^2 + M_E^2\left[v^2\lambda_e^2-2\left(m_\tau^2+m_E^2\right)\right] + 2\,m_\tau^2 m_E^2 = 0\,,
\label{eq:model4_ME_quadratic}
\end{equation}
whose roots are both real and positive precisely when
\begin{equation}
v^2\lambda_e^2 - 2\left(m_\tau^2+m_E^2\right) \;\leq\; -4\,m_\tau m_E\,,
\end{equation}
or
\begin{equation}
|\lambda_e| \;\leq\; \frac{\sqrt{2}\,\left(m_E-m_\tau\right)}{v}\,.
\label{eq:model4_lambda_bound}
\end{equation}
This bound is independent of the perturbativity ceiling of Sec.~\ref{subsec:perturbativity}.

As in Model~\MIII, the operative experimental constraint on Model~\MIV is a mixing-angle upper bound rather than a direct mass limit. A combined fit to $Z\ell\ell$ and $W\ell\nu$ couplings~\cite{Crivellin:2020ebi} (exploiting LEP precision data~\cite{ALEPH:2005ab}) constrains the left-handed mixing angle to $\theta_L^{\max} \simeq \sqrt{2\times10^{-3}} \simeq 0.045$, again in a mass-independent way. The right-handed angle $\theta_R \simeq (m_\tau/M_E)\,\theta_L$ is parametrically suppressed and unconstrained by these data. Together with the kinematic bound of Eq.~\eqref{eq:model4_lambda_bound}, $\lambda_e$ is therefore bounded from above by this mixing-angle constraint and by perturbativity (Sec.~\ref{subsec:perturbativity}).

\section{Perturbativity}\label{subsec:perturbativity}
Sec.~\ref{sec:the-model} presented the main phenomenological bounds on each submodel. We now turn to a different constraint we imposed upon our parameter space: perturbativity, up to a benchmark UV cutoff, of the VL Yukawa couplings under RG evolution. 
While lack of perturbativity is not fatal per se, any extension of the SM which significantly narrows the range of applicability of perturbative methods inevitably poses additional problems, demanding a further search for a broader UV completion. 
Given that we aim at improving the SM predictability, we must require our VL extension to establish a valid effective field theory for any foreseeable future experimental application. 

Because the four submodels are benchmark constructions rather than realistic, full-fledged UV completions, we compute the resulting limits also for representative masses somewhat below the discussed collider bounds, mapping how the bound itself varies across the VL mass scale.

No chiral symmetry protects the VL Yukawa couplings, which can in principle grow large under RG evolution. We bound them using the one-loop $\beta$-functions of Appendix~\ref{sec:RGE}, obtained with the \texttt{RGBeta} package~\cite{Thomsen:2021ncy}. These evolve the coupled system formed by the three SM gauge couplings $\{g_1,g_2,g_3\}$, the top Yukawa $y_t$, and the NP Yukawa coupling(s) of the submodel under consideration, up to a benchmark cutoff $\Lambda_i$. The running of the remaining SM Yukawa couplings is neglected, as they are too small to affect the cross-section. The Higgs quartic $\lambda_H$ is likewise left out of the running: assessing the (meta)stability of the EW vacuum, for which $\lambda_H$ is the relevant coupling, lies beyond the scope of this work, and we refer the interested reader to Ref.~\cite{Adhikary:2024esf} for a detailed study of vacuum stability in closely related VLF models.

We adopt a decoupling scheme: below the VL mass threshold $M_{\rm VL}$ the heavy states are integrated out and the running follows the pure-SM $\beta$-functions, initialized at $\mu=m_t$ to their $\overline{\rm MS}$ values, above $M_{\rm VL}$ we switch to the full SM+NP system.

We define the \emph{perturbativity ceiling} on the VL Yukawa as the largest value it can take at $M_{\rm VL}$ such that no coupling in the running system exceeds $|y|<\sqrt{4\pi}$ before $\Lambda_i$ is reached. In practice the VL Yukawa is always the first coupling to saturate this bound while the SM gauge couplings and $y_t$ remain comfortably sub-critical throughout. We evaluate the ceiling at three benchmark cutoffs, $\Lambda_1 = 10^5$~GeV, $\Lambda_2 = 10^{10}$~GeV, and $\Lambda_3 = 10^{15}$~GeV and collect the results in  Table~\ref{table:perturbativity-merged}.

\begin{table}[h]
\begingroup
\centering
\setlength{\tabcolsep}{14pt}
\renewcommand{\arraystretch}{0.85}
\begin{tabular}{| c | c | c c | c c | c c |}
\hline
 & & \multicolumn{2}{c|}{$\Lambda_1 = 10^5 \, \text{GeV}$} & \multicolumn{2}{c|}{$\Lambda_2 = 10^{10} \, \text{GeV}$} & \multicolumn{2}{c|}{$\Lambda_3 = 10^{15} \, \text{GeV}$} \\
\hline\hline
 & $M_Q=M_U=M_D$ [GeV] & $y_U^{\max}$ & $y_D^{\max}$ & $y_U^{\max}$ & $y_D^{\max}$ & $y_U^{\max}$ & $y_D^{\max}$ \\
\hline
\multirow{4}{*}{Model~\MI} & 1000 & 1.236 & 1.232 & 0.794 & 0.785 & 0.687 & 0.672 \\
 & 1500 & 1.277 & 1.274 & 0.796 & 0.786 & 0.684 & 0.669 \\
 & 2000 & 1.310 & 1.307 & 0.797 & 0.788 & 0.682 & 0.667 \\
 & 3000 & 1.364 & 1.360 & 0.799 & 0.790 & 0.679 & 0.665 \\
\hline\hline
 & $M_L=M_E=M_N$ [GeV] & $y_E^{\max}$ & $y_N^{\max}$ & $y_E^{\max}$ & $y_N^{\max}$ & $y_E^{\max}$ & $y_N^{\max}$ \\
\hline
\multirow{4}{*}{Model~\MII} & 200  & 1.520 & 1.503 & 0.926 & 0.897 & 0.727 & 0.687 \\
 & 500  & 1.635 & 1.619 & 0.959 & 0.931 & 0.746 & 0.706 \\
 & 1000 & 1.736 & 1.721 & 0.985 & 0.957 & 0.761 & 0.721 \\
 & 1500 & 1.803 & 1.789 & 1.001 & 0.973 & 0.769 & 0.729 \\
\hline\hline
 & $M_N$ [GeV] & \multicolumn{2}{c|}{$\lambda_\nu^{\max}$} & \multicolumn{2}{c|}{$\lambda_\nu^{\max}$} & \multicolumn{2}{c|}{$\lambda_\nu^{\max}$} \\
\hline
\multirow{4}{*}{Model~\MIII} & 200  & \multicolumn{2}{c|}{1.833} & \multicolumn{2}{c|}{1.168} & \multicolumn{2}{c|}{0.921} \\
 & 500  & \multicolumn{2}{c|}{1.952} & \multicolumn{2}{c|}{1.205} & \multicolumn{2}{c|}{0.943} \\
 & 1000 & \multicolumn{2}{c|}{2.056} & \multicolumn{2}{c|}{1.234} & \multicolumn{2}{c|}{0.959} \\
 & 1500 & \multicolumn{2}{c|}{2.122} & \multicolumn{2}{c|}{1.252} & \multicolumn{2}{c|}{0.969} \\
\hline\hline
 & $M_E$ [GeV] & \multicolumn{2}{c|}{$\lambda_e^{\max}$} & \multicolumn{2}{c|}{$\lambda_e^{\max}$} & \multicolumn{2}{c|}{$\lambda_e^{\max}$} \\
\hline
\multirow{4}{*}{Model~\MIV} & 200  & \multicolumn{2}{c|}{1.852} & \multicolumn{2}{c|}{1.200} & \multicolumn{2}{c|}{0.963} \\
 & 500  & \multicolumn{2}{c|}{1.970} & \multicolumn{2}{c|}{1.237} & \multicolumn{2}{c|}{0.985} \\
 & 1000 & \multicolumn{2}{c|}{2.072} & \multicolumn{2}{c|}{1.265} & \multicolumn{2}{c|}{1.001} \\
 & 1500 & \multicolumn{2}{c|}{2.138} & \multicolumn{2}{c|}{1.283} & \multicolumn{2}{c|}{1.011} \\
\hline
\end{tabular}
\endgroup
\caption{Maximal new-physics Yukawa couplings $y_F^{\max}$ compatible with perturbativity up to the benchmark cutoffs $\Lambda_1<\Lambda_2<\Lambda_3$, for representative VL masses in each submodel ($M_Q=M_U=M_D$ for Model~\MI, $M_L=M_E=M_N$ for Model~\MII, $M_N$ for Model~\MIII, $M_E$ for Model~\MIV. Models~\MIII and~\MIV have only one coupling, spanning both sub-columns).}
\label{table:perturbativity-merged}
\end{table}

The four submodels fall into two structurally distinct classes. In Models~\MI and~\MII the two new Yukawa couplings run as a \emph{mutually coupled} pair, each enters the other's $\beta$-function, so both grow faster than either would alone, and the ceiling is reached sooner. 
Model~\MI (cf.~\eqref{eq:beta-yU-MI} and~\eqref{eq:beta-yD-MI}) is further enhanced by the $SU(3)_C$ color factor of the VLQ multiplet, making it the most tightly constrained submodel in the table. Model~\MII (cf.~\eqref{eq:beta-yE-MII} and~\eqref{eq:beta-yN-MII}) shares the coupled structure but not the color enhancement, and follows close behind.
Models~\MIII and~\MIV instead carry a single new Yukawa each ($\lambda_\nu$ and $\lambda_e$, respectively), which runs in isolation (cf.~\eqref{eq:beta-lambdanu-MIII} and~\eqref{eq:beta-lambdae-MIV}): its $\beta$-function is driven only by the SM gauge couplings and $y_t$, with a slower growth. This is why Models~\MIII and~\MIV consistently show the weaker constraints of Table~\ref{table:perturbativity-merged}.

These structural differences also show up in the table's numerical trends. At fixed cutoff, the ceiling loosens as $M_{\rm VL}$ grows, the coupling starts closer to $\Lambda_i$ and has correspondingly less room to run before the cutoff is reached, a pattern that is clearest at $\Lambda_1$ and holds throughout Models~\MII--\MIV. Model~\MI is the exception (cf.~\eqref{eq:beta-yU-MI} and~\eqref{eq:beta-yD-MI}): its color-enhanced, coupled running flattens this trend at $\Lambda_2$ and mildly reverses it at $\Lambda_3$, where $y_U^{\max}$ and $y_D^{\max}$ instead decrease slightly with $M_{\rm VL}$. At fixed $M_{\rm VL}$, conversely, the ceiling tightens as the cutoff is raised, since the coupling must then remain perturbative over a wider energy range. As a result, the heaviest states tabulated retain a sizable perturbative window even at $\Lambda_3$ in Models~\MII--\MIV, while the lightest states are the most constrained overall.

\begin{figure}[h]
\centering
\includegraphics[width=0.4\textwidth]{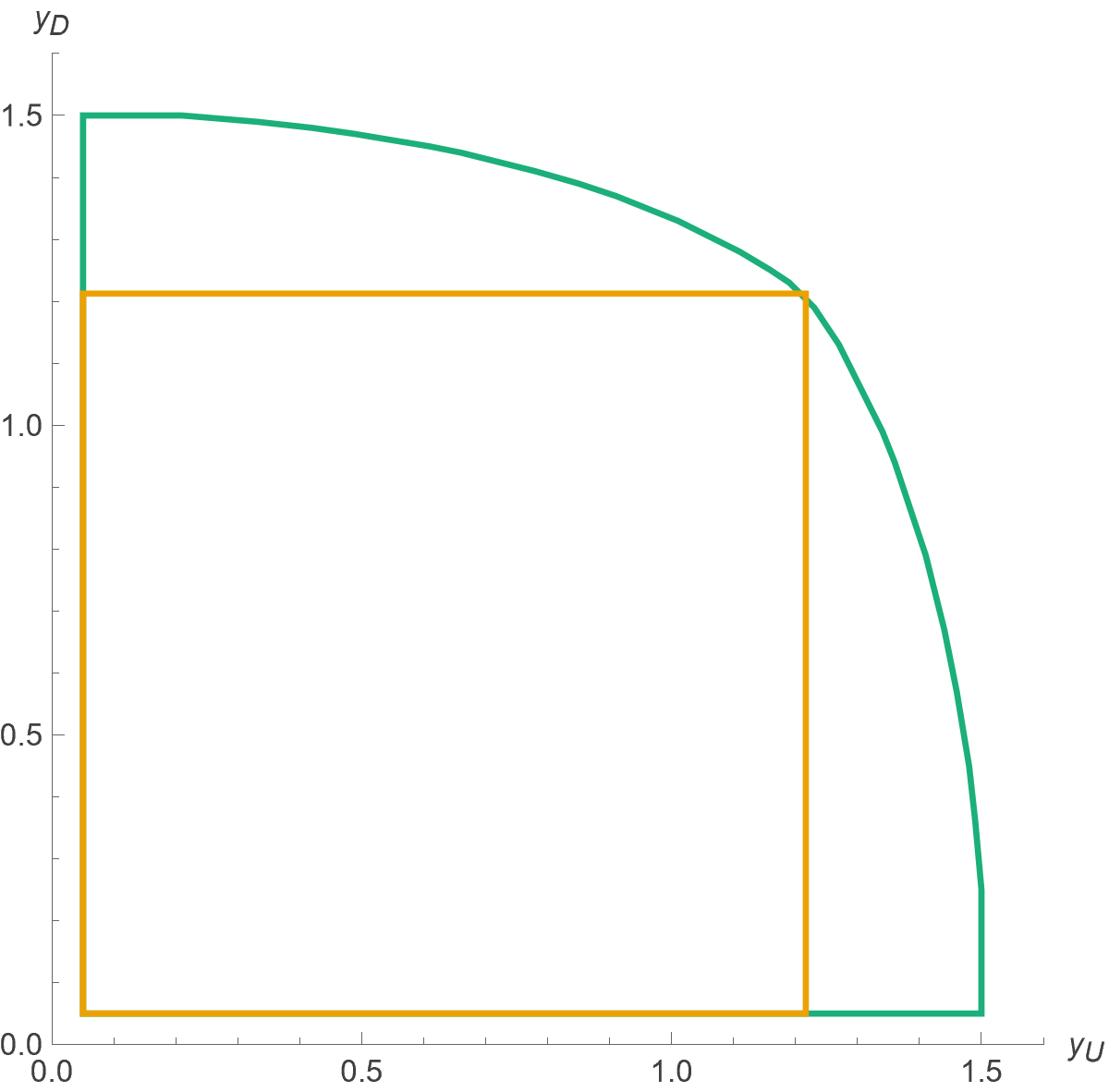}
\caption{The $(y_U,y_D)$ plane for Model~\MI at $M_Q=M_U=M_D=1000$~GeV and $\Lambda_1=10^5$~GeV: the rectangular cut of Table~\ref{table:perturbativity-merged} (orange) against the true jointly-perturbative region allowed by the coupled RG running (green).}
\label{fig:validpoints}
\end{figure}

The perturbativity ceiling can be compared directly against the other constraints of Sec.~\ref{sec:the-model}. For Models~\MI and~\MII, perturbativity and the collider bounds on the mass scale act on different axes, one bounds the coupling, the other the mass, so within the collider-allowed mass window the ceiling remains a comparatively loose bound, of order unity throughout. For Models~\MIII and~\MIV the comparison is more direct, since the bound on the mixing angle also constrains $\lambda_\nu$ and $\lambda_e$ algebraically, through Eqs.~\eqref{eq:rotation_angle_mass_matrices_model_3} and~\eqref{eq:model4_tan2thetaL}. Evaluated over the mass range of Table~\ref{table:perturbativity-merged}, this translates into $\lambda_\nu,\lambda_e \lesssim 0.05$--$0.4$ , roughly an order of magnitude below both the perturbativity ceiling and the purely kinematic reality bound (Eq.~\eqref{eq:model3_lambda_bound} for Model~\MIII and Eq.~\eqref{eq:model4_lambda_bound} for Model~\MIV). The bound on the mixing is therefore by far the strictest of the three for these two submodels, and perturbativity only becomes the operative constraint once one moves beyond the collider-allowed mixing angle.

We use the rectangular cut of Table~\ref{table:perturbativity-merged}, rather than a more precise jointly-perturbative one, for simplicity: bounding each Yukawa coupling independently understates the true perturbative region for Models~\MI and~\MII. Fig.~\ref{fig:validpoints} illustrates this for Model~\MI at $M_Q=M_U=M_D=1000$~GeV and $\Lambda_1$: the red square marks the rectangular cut $y_U\leq y_U^{\max}$, $y_D\leq y_D^{\max}$ that we impose throughout the numerical scan, while the green region shows the full set of $(y_U,y_D)$ points for which the coupled system remains perturbative up to $\Lambda_1$. Because $y_U$ and $y_D$ enter each other's $\beta$-function (cf.~\eqref{eq:beta-yU-MI} and~\eqref{eq:beta-yD-MI}), a smaller value of one coupling allows the other to run further before hitting $\sqrt{4\pi}$, curving the true boundary outward relative to the naive rectangle. The green area is therefore not actually excluded by perturbativity, only by our simplified, coupling-by-coupling criterion, and the effect is mild enough not to affect our conclusions. This is why, in the scan plots of Sec.~\ref{sec:numerical-analysis-and-results}, a handful of points sit slightly beyond the nominal ceiling of Table~\ref{table:perturbativity-merged} without violating the underlying perturbativity bound itself.

\section{Vector-like states @NLO}\label{sec:cross-section-calculation}

\subsection{Higgs-strahlung}\label{sec:higgs-strahlung}
As mentioned, different proposals of future lepton colliders are expected to produce large numbers of Z-bosons via the Higgs-strahlung process $e^+ e^- \rightarrow ZH$. The corresponding precise measurement would then present an appealing theoretical opportunity only if matched with a comparably precise profiling of the model-dependent signature. By complementing the discovery-first approach of hadronic machines such as LHC, and the consequent establishing of the SM as the leading phenomenological contribution, leptonic colliders are expected to explore a class of models manifesting in sub-leading radiative signatures. In other words, the $e^+ e^- \rightarrow ZH$ process is not expected to be modified at tree-level due to some light resonance coupled to electrons, which has accidentally escaped direct past probing. Instead, we are required to study the impact that new states have on the SM's leading order (LO) prediction at NLO. 

Although more cumbersome compared to the straightforward tree-level computation, the procedure for determining the one-loop contribution is well-known and enjoys powerful consistency checks in the request of UV finiteness and decoupling of heavy masses. While it is possible to adopt equivalent alternatives \cite{Sirlin:1980nh,Passarino:1989ta,Novikov:1993ic,Marciano:1980be,Degrassi:1990tu}, we followed, as also done in \cite{Marzo:2022nrw}, an OS renormalization prescription supported by gauge-invariant dimensional regularization. Details of the application of this scheme to Higgs-strahlung can be found in 
\cite{Denner:1991kt,Denner:1992bc,Abouabid:2020eik,Freitas:2023xnx,Marzo:2022nrw}. Here we present the main features of the computation and address its analytical dependence on the reduced parameter space of the four simplified scenarios discussed in Sec.~\ref{sec:the-model}. 
Further details on the counterterms structure and definition, in connection with the provided ancillary files,  can be found in the dedicated Appendix~\ref{sec:1-loop-renormalization}. 

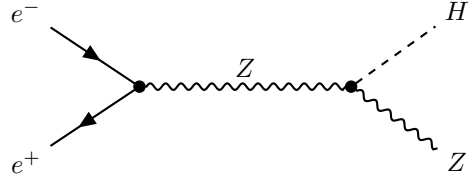
\begin{figure}[t]
\centering
\begin{tikzpicture}
  \begin{feynman}[every dot={/tikz/fill=black}]
 
    \vertex (i1) at (-2.6,  1.0) {\(e^-\)};
    \vertex (i2) at (-2.6, -1.0) {\(e^+\)};
    \vertex (f1) at ( 3.1,  1.0) {\(H\)};
    \vertex (f2) at ( 3.1, -1.0) {\(Z\)};
 
    \vertex [dot] (va) at (-1.1, 0) {};   % ee Z* vertex  (was -1.8)
    \vertex [dot] (vb) at ( 1.7, 0) {};   % ZZH vertex     (was  4.1)
 
    \diagram* {
      %--- Incoming leptons
      (i1) -- [fermion, thick]                             (va),
      (i2) -- [anti fermion, thick]                        (va),
 
      %--- Off-shell Z from ee vertex to self-energy
      (va) -- [boson, thick, edge label=\(Z\)]           (vb),

      %--- Outgoing Higgs (dashed) and Z (wavy)
      (vb) -- [scalar, thick]            (f1),
      (vb) -- [boson, thick]            (f2),
    };

  \end{feynman}
\end{tikzpicture}
\caption{LO Higgs-strahlung in the SM via Z-boson exchange. }
\label{fig:eeZHLO}
\end{figure}

The Higgs-strahlung process is largely dominated by the tree-level exchange of a Z-boson Fig~(\ref{fig:eeZHLO}), the scalar mediation of the Higgs particle is also present, but highly suppressed by the tiny Yukawa couplings.
The kinematic of the four-point scattering process is captured by the Lorentz invariant Mandelstam variables $s,t$ and $u$ as well as the spin $s_{\pm}$ of the incoming fermions and the polarization $\lambda$ of the $Z$. Of the three Mandelstam variables, only two are independent, given that momentum conservation forces $s+t+u=2m^2_e + m^2_Z+m^2_H$. It is convenient to trade the two independent invariants for the beam energy and the cosine of the angle $\theta$ between the incoming positron and the $Z$ as measured in the center-of-mass frame
\begin{align}
& E = \sqrt{s}/2\, , \quad (s\geq m^2_Z+m^2_H) \, , \nonumber \\
& 2\, s \,\beta_{s}\cos\theta = t-u \, , \nonumber
\end{align}
with $\beta_{s}$ the two-particle phase space function (in the $m_e = 0$ approximation) 
\begin{align}
&\beta_{s} = \frac{\sqrt{(m_H^2-m_Z^2)^2 - 2 (m_H^2 + m_Z^2) s + s^2}}{s}. \nonumber
\end{align}
Lorentz invariance can also be exploited at the amplitude-level through a decomposition in a basis of matrix elements $ \mathcal M_j\left(\lambda; s_+, s_-\right)$ living in the space of spin and polarizations
\begin{align} \label{SPdec}
\mathcal M\left(\lambda; s_+, s_-\right) &= \sum_j F_j \mathcal M_j\left(\lambda; s_+, s_-\right) \,,
\end{align}
with the scalar form factors $F_j$ parametrizing the model-dependent signature \cite{Fleischer:1982af,Denner:1991kt}. Although the vector nature of the extra NP fermions can exempt us from studying polarization-induced effects, we will refer to the form factor of the decomposition Eq.~(\ref{SPdec}) to efficiently present the reader with the analytical formulas. 
For instance, the dominant tree-level contribution of Fig~(\ref{fig:eeZHLO}) involves the following helicity\footnote{The electron can safely be treated as massless.} elements 
\begin{align}
\label{eq:HelBasis}
& \mathcal M_L\left(\lambda; s_+, s_-\right) = \bar v(s_+)  \slashed{\epsilon}^*(\lambda) P_L u(s_-)\, , \\
& \mathcal M_R\left(\lambda; s_+, s_-\right) = \bar v(s_+) \slashed{\epsilon}^*(\lambda) P_R u(s_-) \, , \quad (P_{L,R} = \left( 1 \mp \gamma_5\right)/2) \,,
\end{align}
and defines the form factors   
\begin{align}
\label{eq:CLRLO}
&F^{\rm LO}_{L,R} = \frac{4 \alpha \pi\, m_Z\,g^{Ze}_{L,R}}{c_Ws_W(s-m^2_Z)}, \quad
 \bigg(\text{with} \,\, g^{Ze}_L=\frac{s^2_W-\frac{1}{2}}{c_Ws_W} \, ,   \quad \text{and} \quad g^{Ze}_R=\frac{s_W}{c_W}\bigg)
\end{align}
where $\alpha = \tfrac{e^2}{4\pi}$ is the fine structure constant and $s_W$, $c_W$ are the sine and cosine of the Weinberg angle\footnote{In the OS scheme we adopted, this quantity is not running and is defined, at all orders in terms of the physical masses of the $Z$ and $W$ boson.}. Using these expressions for the integrated unpolarized cross-section,
\begin{align} 
\label{eq:sigmaZhLO}
 &\sigma^{\rm LO}_{ZH} (s) = \frac{\beta_{s}}{32\pi} \left(1 + \frac{s}{12} \,\frac{\beta^2_{s}}{m_Z^2}\right) \bigg[ |F^{\rm LO}_L|^2+|F^{\rm LO}_R|^2\bigg] .
\end{align}
one recovers the well-known maximum cross-section of $\sim 254$ fb at a center-of-mass energy of $E \sim 244$ GeV, thus motivating future Higgs factory colliders operating near this energy regime.

\begin{figure}[t]
\centering
\includegraphics[scale=0.6]{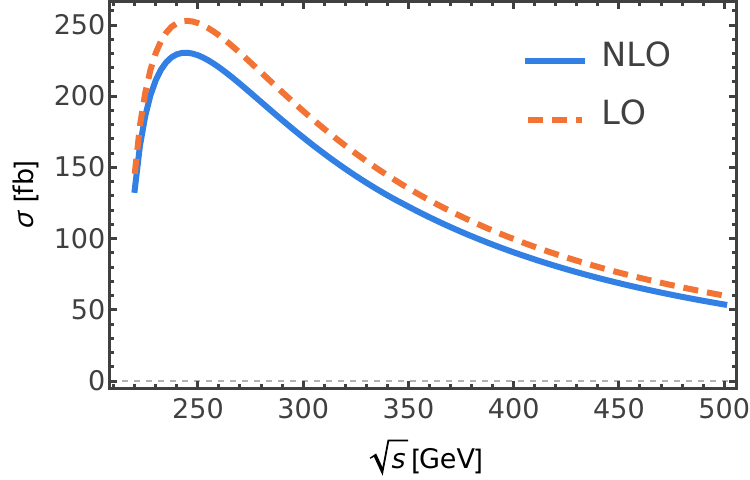} \qquad
\includegraphics[scale=0.6]{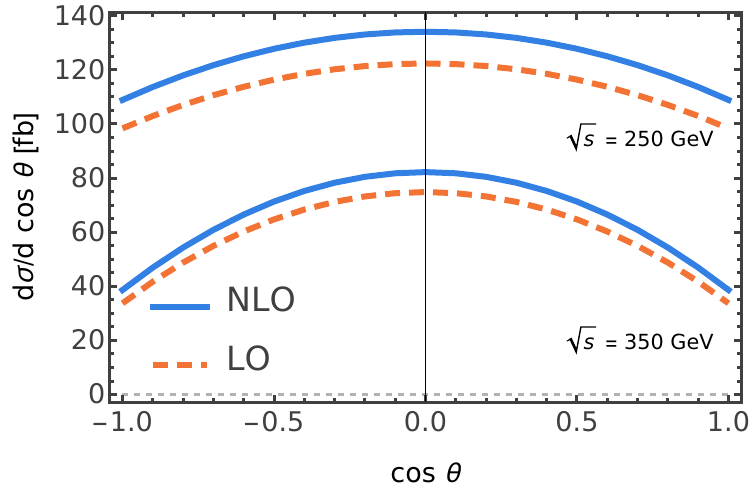}
\caption{\small Integrated and differential cross-section, LO (dashed) vs NLO (continue). }
\label{fig:SMNLO}
\end{figure}

Moving up to one-loop order, the relevant form of $|\mathcal{M}|^2$ in terms of tree-level $F^{\rm LO}$ and loop-level $\delta F_j$ form factors can be written as 
\begin{align}
|\mathcal{M}^{\rm NLO}|^2 \simeq
\sum_{ij} (F^{\rm LO}_i)^* (F^{\rm LO}_j +2 \delta F_j){\rm Re}[\mathcal{M}^*_i \mathcal{M}_j]\, . 
\end{align}

The SM radiative corrections to $e^+e^- \rightarrow ZH$ have been thoroughly studied in the literature \cite{Denner:1991kt,Fleischer:1982af,Denner:1992bc} to which we refer for a  more detailed presentation. We just recall here that they can be qualitatively framed into two distinct contributions. The first is of infrared nature, arising from virtual loops of charged particles and requiring the inclusion of soft-photon radiation to yield a finite cross-section \cite{Denner:1991kt,Denner:1992bc}. Being a universal quantum electrodynamics (QED) effect, it is unaffected by any NP we intend to study. The second is of EW origin, generating UV dependence on regulators that can be handled within the same OS renormalization framework we adopt for the NP contributions. Altogether, the full NLO SM prediction, shown in Fig.~\ref{fig:SMNLO}, amounts to a reduction of $\sim 25$ fb near the production peak at $\sqrt{s} \sim 244$ GeV with respect to the LO cross-section of $\sim 254$ fb. Notice how the impact of radiative corrections is also maximized around the peak, where it accounts for $\sim 10\%$ of the LO value.  

\subsection{Universal vs non-universal effects of VL states @NLO}\label{sec:non_universal_VL_at_NLO}
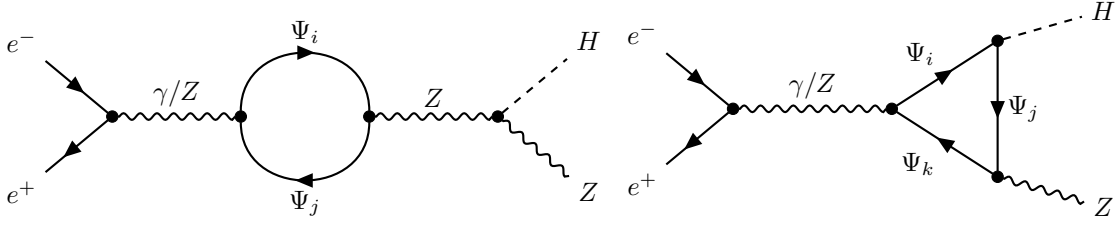
\begin{figure}[t]
\centering
\begin{tikzpicture}
  \begin{feynman}[every dot={/tikz/fill=black}]
 
    \vertex (i1) at (-2.6,  1.0) {\(e^-\)};
    \vertex (i2) at (-2.6, -1.0) {\(e^+\)};
    \vertex (f1) at ( 4.9,  1.0) {\(H\)};
    \vertex (f2) at ( 4.9, -1.0) {\(Z\)};
 
    \vertex [dot] (va) at (-1.4, 0) {};   % ee Z* vertex  (was -1.8)
    \vertex [dot] (v1) at ( 0.3, 0) {};   % left  self-energy vertex (unchanged)
    \vertex [dot] (v2) at ( 2., 0) {};   % right self-energy vertex (unchanged)
    \vertex [dot] (vb) at ( 3.7, 0) {};   % ZZH vertex     (was  4.1)
 
    \diagram* {
      %--- Incoming leptons
      (i1) -- [fermion, thick]                             (va),
      (i2) -- [anti fermion, thick]                        (va),
 
      %--- Off-shell Z from ee vertex to self-energy
      (va) -- [boson, thick, edge label=\(\gamma / Z\)]           (v1),
 
      %--- Z self-energy: fermion (f, fbar) loop
      (v1) -- [fermion, thick, half left, looseness=1.55,
               edge label=\(\Psi_i\)]                           (v2),
      (v2) -- [fermion, thick, half left, looseness=1.55,
               edge label=\(\Psi_j\)]                     (v1),
 
      %--- Off-shell Z from self-energy to production vertex
      (v2) -- [boson, thick, edge label=\(Z\)]           (vb),
 
      %--- Outgoing Higgs (dashed) and Z (wavy)
      (vb) -- [scalar, thick]            (f1),
      (vb) -- [boson, thick]            (f2),
    };

  \end{feynman}
\end{tikzpicture}
\begin{tikzpicture}
  \begin{feynman}[every dot={/tikz/fill=black}]

    \vertex (i1) at (-2.6,  1.0)  {\(e^-\)};
    \vertex (i2) at (-2.6, -1.0)  {\(e^+\)};
    \vertex (f1) at ( 3.5,  1.3)  {\(H\)};
    \vertex (f2) at ( 3.5, -1.3)  {\(Z\)};
 
    \vertex [dot] (va)  at (-1.4,  0.00) {};
    \vertex [dot] (vt1) at ( 0.7,  0.00) {};  % Z* = 2.4 units
    \vertex [dot] (vt2) at ( 2.1,  0.90) {};
    \vertex [dot] (vt3) at ( 2.1, -0.90) {};

    \diagram* {
      (i1) -- [fermion, thick]                       (va),
      (i2) -- [anti fermion, thick ]                  (va),
      (va) -- [boson, thick, edge label=\(\gamma / Z\)]     (vt1),
      (vt1) -- [fermion, thick, edge label=\(\Psi_i\)]   (vt2),
      (vt2) -- [fermion, thick, edge label=\(\Psi_j\)]                      (vt3),
      (vt3) -- [fermion, thick, edge label=\(\Psi_k\)]                      (vt1),
      (vt2) -- [scalar, thick]     (f1),
      (vt3) -- [boson, thick]    (f2),
    };
  \end{feynman}
\end{tikzpicture}
\caption{\small The universal self-energy (sunset) and the flavor-dependent (penguin) corrections
 to $e^+ e^- \rightarrow ZH$, generated by the models of Sec.~\ref{sec:the-model}. $\Psi$ can be 
 any of the VLFs. }
\label{fig:VLNLO}
\end{figure}
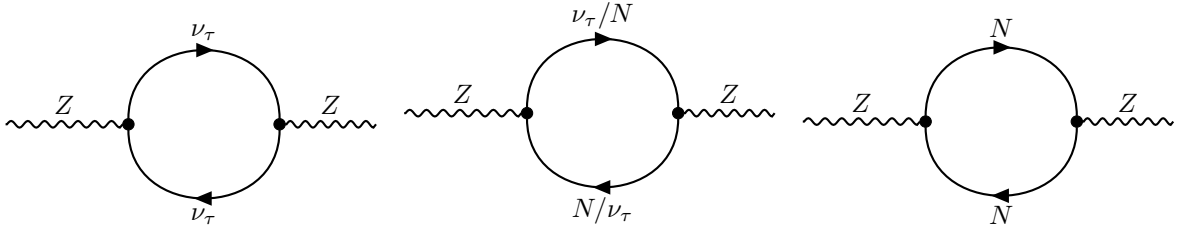
\begin{figure}[t]
\centering
\begin{tikzpicture}
  \begin{feynman}[every dot={/tikz/fill=black}]
 
      \vertex [dot] (v1) at ( 0.3, 0) {};   % left  self-energy vertex (unchanged)
    \vertex [dot] (v2) at ( 2.3, 0) {};   % right self-energy vertex (unchanged)
    \vertex (vb) at ( 3.7, 0) {};   % ZZH vertex     (was  4.1)
 
    \diagram* {
      %--- Incoming leptons
 
      %--- Off-shell Z from ee vertex to self-energy
      (va) -- [boson, thick, edge label=\(Z\)]           (v1),
 
      %--- Z self-energy: fermion (f, fbar) loop
      (v1) -- [fermion, thick, half left, looseness=1.55,
               edge label=\(\nu_{\tau}\)]                           (v2),
      (v2) -- [fermion, thick, half left, looseness=1.55,
               edge label=\(\nu_{\tau}\)]                     (v1),
 
      %--- Off-shell Z from self-energy to production vertex
      (v2) -- [boson, thick, edge label=\(Z\)]           (vb),
 
    };
  \end{feynman}
\end{tikzpicture}
\begin{tikzpicture}
  \begin{feynman}[every dot={/tikz/fill=black}]

    \vertex [dot] (v1) at ( 0.3, 0) {};   % left  self-energy vertex (unchanged)
    \vertex [dot] (v2) at ( 2.3, 0) {};   % right self-energy vertex (unchanged)
    \vertex  (vb) at ( 3.7, 0) {};   % ZZH vertex     (was  4.1)
 
    \diagram* {
      %--- Incoming leptons
 
      %--- Off-shell Z from ee vertex to self-energy
      (va) -- [boson, thick, edge label=\(Z\)]           (v1),
 
      %--- Z self-energy: fermion (f, fbar) loop
      (v1) -- [fermion, thick, half left, looseness=1.55,
               edge label=\(\nu_{\tau} / N\)]                           (v2),
      (v2) -- [fermion, thick, half left, looseness=1.55,
               edge label=\(N / \nu_{\tau}\)]                     (v1),
 
      %--- Off-shell Z from self-energy to production vertex
      (v2) -- [boson, thick, edge label=\(Z\)]           (vb),
 
    };

  \end{feynman}
\end{tikzpicture}
\begin{tikzpicture}
  \begin{feynman}[every dot={/tikz/fill=black}]
 
    \vertex [dot] (v1) at ( 0.3, 0) {};   % left  self-energy vertex (unchanged)
    \vertex [dot] (v2) at ( 2.3, 0) {};   % right self-energy vertex (unchanged)
    \vertex (vb) at ( 3.7, 0) {};   % ZZH vertex     (was  4.1)
 
    \diagram* {
      %--- Incoming leptons
 
      %--- Off-shell Z from ee vertex to self-energy
      (va) -- [boson, thick, edge label=\(Z\)]           (v1),
 
      %--- Z self-energy: fermion (f, fbar) loop
      (v1) -- [fermion, thick, half left, looseness=1.55,
               edge label=\(N\)]                           (v2),
      (v2) -- [fermion, thick, half left, looseness=1.55,
               edge label=\(N\)]                     (v1),
 
      %--- Off-shell Z from self-energy to production vertex
      (v2) -- [boson, thick, edge label=\(Z\)]           (vb),
 
    };

  \end{feynman}
\end{tikzpicture}
\caption{\small Z-boson self-energy corrections as generated in Model~\MIII. The second diagram is representative of two different alternative contributions, with both SM neutrino and the VL state. }
\label{fig:Mod3ZZ}
\end{figure}

All models profiled in this work exhibit mixing among eigenstates, arising either purely within the NP sector or through a third-generation SM state. Consequently, a concise presentation of the NP effects over the Higgs-strahlung's form factors is complicated by the intricate interdependence among the many
 contributing terms\footnote{Related Mathematica notebooks can be obtained from the authors upon request.}. Nevertheless, some general observations can be drawn to reveal the main differences, as well as the common features, of the four VL models.
First, all models contribute to corrections of the vector boson propagation, including the singlet extension of Model~\MIII, due to mixing. As a result, each model generates diagrams of the type shown in the left panel of Fig.~(\ref{fig:VLNLO}), correcting both the Z-boson self-energy and generating 
$\gamma-Z$ mixed propagation. This contribution is directly linked to a parallel EW-scale signature probed at LEP and parameterized by deviations from the best-fit values of the EW precision observables S, T and U \cite{Peskin:1991sw,Maksymyk:1993zm}. Neglecting the subdominant U, the precision observables employed in this work are obtained from UV-finite combinations of the same corrections that enter the \emph{universal}\footnote{Given the mixing patterns involved, the term `universal' may appear as a stretch, most evidently in the Model~\MIII scenario, where different Yukawa values determine the impact of the extra sterile states. We retain the standard terminology for simplicity.} contribution to Higgs-strahlung:
\begin{align} \label{STU}
& \frac{\alpha}{4 (c_W s_W)^2} \,S = \frac{\Pi^{VL}_{ZZ}(m_Z^2) - \Pi^{VL}_{ZZ}(0)}{m_Z^2} - \frac{c_W^2 - s_W^2}{c_W s_W} \frac{\Pi^{VL}_{Z\gamma}(m_Z^2)}{m_Z^2} - \frac{\Pi^{VL}_{\gamma \gamma }(m_Z^2)}{m_Z^2} \,, \notag \\
& \alpha \,T =  \,\,\frac{\Pi^{VL}_{WW}(0)}{m_W^2} - \frac{\Pi^{VL}_{ZZ}(0)}{m_Z^2} ,
\end{align}
where $\Pi_{\mu \nu}(p^2)$ stands for the transverse component of the two-point function
\begin{align}
    \Pi_{\mu \nu}(p^2) = \bigg(\eta_{\mu \nu} - \frac{p_{\mu}p_{\nu}}{p^2}\bigg) \Pi(p^2) + \frac{p_{\mu}p_{\nu}}{p^2} \Pi'(p^2)\, \, .
\end{align}
For this work, we have carefully checked our parameter space against the 1-sigma range of the latest fit \cite{ParticleDataGroup:2024cfk}. We found the collider constraints (barring some fine-tuned avoidance of the NP collider signal, which we do not consider in order to favor a broader analysis) to provide the strongest bounds and, accordingly, to select the viable parameter
space probed in this work. For illustrative purposes, it is instructive to compare the qualitatively distinct impact of the different VL models. At one extreme, Model~\MI and Model~\MII generate corrections exclusively through NP states propagating in the loops, with no direct mixing with SM fermions. Taking Model~\MII as representative, we find\footnote{This and all the analytical loop expressions in this paper are obtained using the \texttt{FeynArts} and \texttt{FormCalc} packages \cite{Hahn:2000jm,Hahn:2000kx}, with the loop integrals evaluated numerically through \texttt{LoopTools} \cite{Hahn:1998yk}. The definitions and the conventions for the scalar loop functions
are, consequently, the ones of said packages.} an already complicated structure displaying the mixing patterns and multiple scales typical of doublet extensions: 
{\small{
\begin{equation}
\begin{aligned} \label{Mod2ZZ}
\big(\tfrac{\alpha_{EM}}{4 \pi c_W^2 s_W^2}\big)^{-1} \Pi^2_{ZZ}(p^2) = &g_{E^-}^2\, A_0\!\left(m_{E^-}^2\right)
+ \Big(4 c_{\theta_E}^4 s_W^4 + 2 c_{\theta_E}^2 \big(c_W^4 + 3 s_W^4\big) s_{\theta_E}^2 + \big(c_W^2 - s_W^2\big)^2 s_{\theta_E}^4\Big) A_0\!\left(m_{E^+}^2\right) \\
&+ c_{\theta_N}^4\, A_0\!\left(m_{N^-}^2\right) + s_{\theta_N}^2\big(2 c_{\theta_N}^2 + s_{\theta_N}^2\big) A_0\!\left(m_{N^+}^2\right) \\
&+ 2 c_{\theta_E}^2 s_{\theta_E}^2\, m_{E^-}\big(m_{E^-} - m_{E^+}\big) B_0\!\left(p^2, m_{E^-}^2, m_{E^+}^2\right) \\
&+ 2 c_{\theta_N}^2 s_{\theta_N}^2\, m_{N^-}\big(m_{N^-} - m_{N^+}\big) B_0\!\left(p^2, m_{N^-}^2, m_{N^+}^2\right) \\
&- 2 g_{E^-}^2\, B_{00}\!\left(p^2, m_{E^-}^2, m_{E^-}^2\right) - 4 c_{\theta_E}^2 s_{\theta_E}^2\, B_{00}\!\left(p^2, m_{E^-}^2, m_{E^+}^2\right) \\
&- 2 g_{E^+}^2\, B_{00}\!\left(p^2, m_{E^+}^2, m_{E^+}^2\right) - 2 c_{\theta_N}^4\, B_{00}\!\left(p^2, m_{N^-}^2, m_{N^-}^2\right) \\
&- 4 c_{\theta_N}^2 s_{\theta_N}^2\, B_{00}\!\left(p^2, m_{N^-}^2, m_{N^+}^2\right) - 2 s_{\theta_N}^4\, B_{00}\!\left(p^2, m_{N^+}^2, m_{N^+}^2\right) \\
&+ p^2 g_{E^-}^2\, B_1\!\left(p^2, m_{E^-}^2, m_{E^-}^2\right) + 2 p^2 c_{\theta_E}^2 s_{\theta_E}^2\, B_1\!\left(p^2, m_{E^-}^2, m_{E^+}^2\right) \\
&+ p^2 g_{E^+}^2\, B_1\!\left(p^2, m_{E^+}^2, m_{E^+}^2\right) + p^2 c_{\theta_N}^4\, B_1\!\left(p^2, m_{N^-}^2, m_{N^-}^2\right) \\
&+ 2 p^2 c_{\theta_N}^2 s_{\theta_N}^2\, B_1\!\left(p^2, m_{N^-}^2, m_{N^+}^2\right) + p^2 s_{\theta_N}^4\, B_1\!\left(p^2, m_{N^+}^2, m_{N^+}^2\right)
\end{aligned}
\end{equation}
}}
where we used the shorthands 
\begin{align}
& g_{E^-} \equiv c_W^2 c_{\theta_E}^2 - s_W^2\big(c_{\theta_E}^2 + 2 s_{\theta_E}^2\big), \quad
 g_{E^+} \equiv 2 c_{\theta_E}^2 s_W^2 + \big(s_W^2 - c_W^2\big) s_{\theta_E}^2 \, . \\ \notag
\end{align}  
A simpler expression is instead found for Model~\MIII, which only involves the VL singlet through its mixing with SM states, so that the magnitude of the correction is entirely governed by the degree of that mixing. 
It should be noted that the mixing with SM states in Models~\MIII and~\MIV requires a more careful definition of what constitutes the NP contribution to the universal EW fit.
In Model~\MIII, for instance, the diagrams carrying NP parameters are those of Fig.~(\ref{fig:Mod3ZZ}), all of which depend on the mixing angle and, consequently, on the NP Yukawa coupling. 
The transverse self-energy contribution is promptly found in the concise form

\begin{equation} \label{model3ZZ}
\begin{aligned}  \big(\tfrac{\alpha_{EM}}{4 \pi c_W^2 s_W^2}\big)^{-1} \Pi^3_{ZZ}(p^2) = &\tfrac{1}{2} s_{\theta_{\nu}}^2\big(2 c_{\theta_{\nu}}^2 + s_{\theta_{\nu}}^2\big) A_0\!\left(m_{N}^2\right)
+ \tfrac{1}{2} m_{N}^2 s_{\theta_{\nu}}^4\, B_0\!\left(p^2, m_{N}^2, m_{N}^2\right) \\
&- c_{\theta_{\nu}}^4\, B_{00}\!\left(p^2, 0, 0\right)
- 2 c_{\theta_{\nu}}^2 s_{\theta_{\nu}}^2\, B_{00}\!\left(p^2, 0, m_{N}^2\right)
- s_{\theta_{\nu}}^4\, B_{00}\!\left(p^2, m_{N}^2, m_{N}^2\right) \\
&+ \tfrac{1}{2} c_{\theta_{\nu}}^4\, p^2\, B_1\!\left(p^2, 0, 0\right)
+ c_{\theta_N}^2 s_{\theta_{\nu}}^2\, p^2\, B_1\!\left(p^2, 0, m_{N}^2\right)
+ \tfrac{1}{2} p^2 s_{\theta_{\nu}}^4\, B_1\!\left(p^2, m_{N}^2, m_{N}^2\right)
\end{aligned}
\end{equation}  
However, care must be taken not to include this expression directly in the combination defining the EW precision observables, Eq.~(\ref{STU}), as doing so would result in 
double-counting: at zero mixing, the first diagram of Fig.~(\ref{fig:Mod3ZZ}) reduces to a pure SM contribution already accounted for. For the purposes of the EW precision
 observables, the genuine NP contribution is therefore obtained by subtracting this  overlap, or equivalently by retaining only those terms in Eq.~(\ref{model3ZZ}) that
vanish at zero mixing. 
Alongside the self-energy corrections, which source a possible tension between a universal enhancement of the Higgs-strahlung signal and the precision EW fit, the Yukawa interactions give rise to a qualitatively distinct effect, captured by the \emph{penguin}-type diagrams at the right panel
of Fig.~(\ref{fig:VLNLO}). These arise from vertex corrections to the gauge-Higgs coupling, as well as from a radiatively generated, finite contribution to the $\gamma HZ$ vertex of Fig.~(\ref{fig:VLNLOvertex}). Notably, the models considered here have been selected precisely to highlight contrasting behaviors:
these vertices can be mediated either by purely NP fermions or through a mixed interplay involving third-generation SM fermion states. To facilitate independent testability as well as provide a comparison among the different VL realizations within the space constraints of this paper,
we focus on the purely radiative $\gamma ZH$ vertex for Models~\MII and \MIV. 
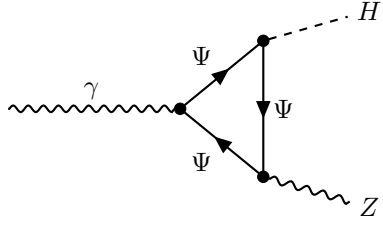
\begin{figure}[t]
\centering
\begin{tikzpicture}
  \begin{feynman}[every dot={/tikz/fill=black}]
 
    \vertex (f1) at ( 3.5,  1.3)  {\(H\)};
    \vertex (f2) at ( 3.5, -1.3)  {\(Z\)};
 
    \vertex (va)  at (-1.4,  0.00) {};
    \vertex [dot] (vt1) at ( 1.0,  0.00) {};  % Z* = 2.4 units
    \vertex [dot] (vt2) at ( 2.1,  0.90) {};
    \vertex [dot] (vt3) at ( 2.1, -0.90) {};

    \diagram* {
      (va) -- [boson, thick, edge label=\(\gamma\)]     (vt1),
      (vt1) -- [fermion, thick, edge label=\(\Psi\)]   (vt2),
      (vt2) -- [fermion, thick, edge label=\(\Psi\)]                      (vt3),
      (vt3) -- [fermion, thick, edge label=\(\Psi\)]                      (vt1),
      (vt2) -- [scalar, thick]     (f1),
      (vt3) -- [boson, thick]    (f2),
    };
  \end{feynman}
\end{tikzpicture}
\caption{\small Generic VL corrections to the vertex $\gamma H Z $. The loop can be mediated by any of the VL \emph{charged} states.}
\label{fig:VLNLOvertex}
\end{figure}
In particular, we consider the \emph{OS} form factor, as the full lengthy off-shell vertex, which is indeed what enters the Higgs-strahlung amplitude, would simply obscure the simpler message.
Picking the form factor $F_{\gamma Z H }$ of the corresponding operator  $\gamma_{\mu} Z^{\mu} H $ which emerges in the broken Higgs phase, Model~\MII contributes with the following structure:
{\small{
 \begin{equation}
\begin{aligned} \label{Mod2Vertex}
F^2_{\gamma Z H} & = \frac{\alpha_{EM} \, y_{E}\, t_{\theta_E} }{\sqrt{2}\,\pi c_W s_W \big(m_H^2 - m_Z^2\big)\big(1 + t_{\theta_E}^2\big)^2}\Bigg[\; \\
&-\big(m_{E^-}-m_{E^+}\big)\big(m_H^2-m_Z^2\big)\big(c_W^2-3 s_W^2\big)\big(1+t_{\theta_E}^2\big) \\
&+ 2 m_{E^-} m_Z^2\big(-c_W^2+s_W^2+2 s_W^2 t_{\theta_E}^2\big)\, B_0\!\left(m_H^2, m_{E^-}^2, m_{E^-}^2\right) \\
&- \big(m_{E^-}+m_{E^+}\big) m_Z^2\big(-1+t_{\theta_E}^2\big)\, B_0\!\left(m_H^2, m_{E^-}^2, m_{E^+}^2\right) \\
&+ 2 m_{E^+} m_Z^2\big(c_W^2 t_{\theta_E}^2 - s_W^2(2+t_{\theta_E}^2)\big)\, B_0\!\left(m_H^2, m_{E^+}^2, m_{E^+}^2\right) \\
&+ 2 m_{E^-} m_Z^2\big(c_W^2 - s_W^2(1+2 t_{\theta_E}^2)\big)\, B_0\!\left(m_Z^2, m_{E^-}^2, m_{E^-}^2\right) \\
&+ \big(m_{E^-}+m_{E^+}\big) m_Z^2\big(-1+t_{\theta_E}^2\big)\, B_0\!\left(m_Z^2, m_{E^-}^2, m_{E^+}^2\right) \\
&+ 2 m_{E^+} m_Z^2\big(2 s_W^2 + (-c_W^2+s_W^2) t_{\theta_E}^2\big)\, B_0\!\left(m_Z^2, m_{E^+}^2, m_{E^+}^2\right) \\
&- \tfrac{1}{2} m_{E^-}\big(m_H^2-m_Z^2\big)\big(-m_H^2+2 m_{E^-}(m_{E^-}+m_{E^+})+m_Z^2\big)\big(-1+t_{\theta_E}^2\big) \\
&\qquad\times C_0\!\left(0, m_Z^2, m_H^2, m_{E^-}^2, m_{E^-}^2, m_{E^+}^2\right) \\
&- \tfrac{1}{2} m_{E^+}\big(m_H^2-m_Z^2\big)\big(-m_H^2+2 m_{E^+}(m_{E^-}+m_{E^+})+m_Z^2\big)\big(-1+t_{\theta_E}^2\big) \\
&\qquad\times C_0\!\left(m_H^2, 0, m_Z^2, m_{E^-}^2, m_{E^+}^2, m_{E^+}^2\right) \\
&+ m_{E^-}\big(m_H^2-m_Z^2\big)\big(-m_H^2+4 m_{E^-}^2+m_Z^2\big)\big(-c_W^2+s_W^2+2 s_W^2 t_{\theta_E}^2\big) \\
&\qquad\times C_0\!\left(m_H^2, m_Z^2, 0, m_{E^-}^2, m_{E^-}^2, m_{E^-}^2\right) \\
&+ m_{E^+}\big(m_H^2-m_Z^2\big)\big(-m_H^2+4 m_{E^+}^2+m_Z^2\big)\big(c_W^2 t_{\theta_E}^2 - s_W^2(2+t_{\theta_E}^2)\big) \\
&\qquad\times C_0\!\left(m_H^2, m_Z^2, 0, m_{E^+}^2, m_{E^+}^2, m_{E^+}^2\right)
\;\Bigg]
\end{aligned}
\end{equation}}}
where $t_{\theta_E} = \tan\theta_{E}$.
This contribution is entirely mediated by VL states (in particular, as expected, notice that $N$ states do not contribute to $F_{\gamma Z H}$), as the proportionality with the NP Yukawa $y_E$ coupling makes it obvious, and should be contrasted with the one of, for instance, Model~\MIV,
which has instead a more intricate structure involving SM and NP Yukawa couplings: 
{\small{
\begin{equation}
\begin{aligned} \label{Mod4Vertex}
& \bigg(\frac{\alpha_{EM}}{2 \sqrt{2} \pi c_W s_W (1+t_{\theta_L}^2)^{3/2}\sqrt{1+t_{\theta_R}^2}} \bigg)^{-1} F^4_{\gamma Z H} = \\
   Y_{e}\bigg\{\;
&-m_{\tau}\big(\kappa + t_{\theta_L}^2\big) - m_E\, t_{\theta_L} t_{\theta_R} \\
&+ \frac{m_Z^2}{m_H^2 - m_Z^2}\bigg[
-m_{\tau}\kappa\Big(B_0\!\left(m_H^2, m_{\tau}^2, m_{\tau}^2\right) - B_0\!\left(m_Z^2, m_{\tau}^2, m_{\tau}^2\right)\Big) \\
&\qquad\qquad\quad - t_{\theta_L}\big(m_{\tau} t_{\theta_L} + m_E t_{\theta_R}\big)\Big(B_0\!\left(m_H^2, m_{\tau}^2, m_E^2\right) - B_0\!\left(m_Z^2, m_{\tau}^2, m_E^2\right)\Big)\bigg] \\
&+ \tfrac{1}{2} t_{\theta_L}\big(m_{\tau}(m_H^2 - 2 m_{\tau}^2 - m_Z^2) t_{\theta_L} - 2 m_{\tau}^2 m_E\, t_{\theta_R}\big)\, C_0\!\left(0, m_Z^2, m_H^2, m_{\tau}^2, m_{\tau}^2, m_E^2\right) \\
&- \tfrac{1}{2} m_E\, t_{\theta_L}\big(2 m_{\tau} m_E\, t_{\theta_L} + (-m_H^2 + 2 m_E^2 + m_Z^2) t_{\theta_R}\big)\, C_0\!\left(m_H^2, 0, m_Z^2, m_{\tau}^2, m_E^2, m_E^2\right) \\
&+ \tfrac{1}{2} m_{\tau}(m_H^2 - 4 m_{\tau}^2 - m_Z^2)\,\kappa\; C_0\!\left(m_H^2, m_Z^2, 0, m_{\tau}^2, m_{\tau}^2, m_{\tau}^2\right)
\;\bigg\} \\[6pt]
+\; \lambda_{e}\bigg\{\;
&m_E\, t_{\theta_L} - m_{\tau}\big(\kappa + t_{\theta_L}^2\big) t_{\theta_R} \\
&+ \frac{m_Z^2}{m_H^2 - m_Z^2}\bigg[
-m_{\tau}\kappa\, t_{\theta_R}\Big(B_0\!\left(m_H^2, m_{\tau}^2, m_{\tau}^2\right) - B_0\!\left(m_Z^2, m_{\tau}^2, m_{\tau}^2\right)\Big) \\
&\qquad\qquad\quad + t_{\theta_L}\big(m_E - m_{\tau} t_{\theta_L} t_{\theta_R}\big)\Big(B_0\!\left(m_H^2, m_{\tau}^2, m_E^2\right) - B_0\!\left(m_Z^2, m_{\tau}^2, m_E^2\right)\Big)\bigg] \\
&+ \tfrac{1}{2} t_{\theta_L}\big(2 m_{\tau}^2 m_E + m_{\tau}(m_H^2 - 2 m_{\tau}^2 - m_Z^2) t_{\theta_L} t_{\theta_R}\big)\, C_0\!\left(0, m_Z^2, m_H^2, m_{\tau}^2, m_{\tau}^2, m_E^2\right) \\
&+ \tfrac{1}{2} m_E\, t_{\theta_L}\big(-m_H^2 + m_Z^2 + 2 m_E(m_E - m_{\tau} t_{\theta_L} t_{\theta_R})\big)\, C_0\!\left(m_H^2, 0, m_Z^2, m_{\tau}^2, m_E^2, m_E^2\right) \\
&+ \tfrac{1}{2} m_{\tau}(m_H^2 - 4 m_{\tau}^2 - m_Z^2)\,\kappa\, t_{\theta_R}\; C_0\!\left(m_H^2, m_Z^2, 0, m_{\tau}^2, m_{\tau}^2, m_{\tau}^2\right)
\;\bigg\} \, . 
\end{aligned}
\end{equation}
}}
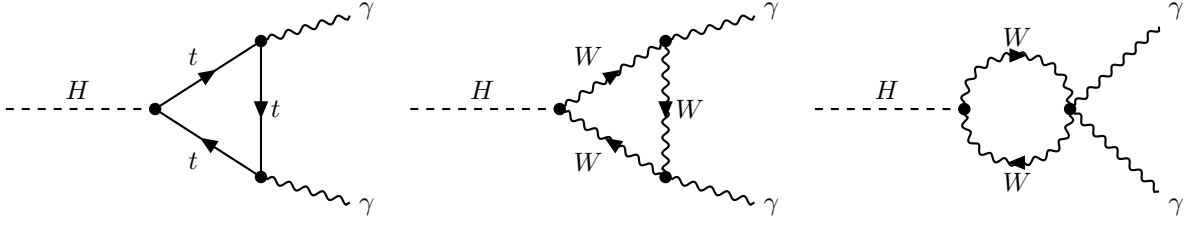
\begin{figure}[t]
\centering
\begin{tikzpicture}
  \begin{feynman}[every dot={/tikz/fill=black}]
 
    \vertex (f1) at ( 3.5,  1.3)  {\(\gamma\)};
    \vertex (f2) at ( 3.5, -1.3)  {\(\gamma\)};
 
    \vertex (va)  at (-1.4,  0.00) {};
    \vertex [dot] (vt1) at ( 0.7,  0.00) {};  % Z* = 2.4 units
    \vertex [dot] (vt2) at ( 2.1,  0.90) {};
    \vertex [dot] (vt3) at ( 2.1, -0.90) {};

    \diagram* {
      (va) -- [scalar, thick, edge label=\(H\)]     (vt1),
      (vt1) -- [fermion, thick, edge label=\(t\)]   (vt2),
      (vt2) -- [fermion, thick, edge label=\(t\)]                      (vt3),
      (vt3) -- [fermion, thick, edge label=\(t\)]                      (vt1),
      (vt2) -- [boson, thick]     (f1),
      (vt3) -- [boson, thick]    (f2),
    };
  \end{feynman}
\end{tikzpicture}
\begin{tikzpicture}
  \begin{feynman}[every dot={/tikz/fill=black}]
 
    \vertex (f1) at ( 3.5,  1.3)  {\(\gamma\)};
    \vertex (f2) at ( 3.5, -1.3)  {\(\gamma\)};
 
    \vertex (va)  at (-1.4,  0.00) {};
    \vertex [dot] (vt1) at ( 0.7,  0.00) {};  % Z* = 2.4 units
    \vertex [dot] (vt2) at ( 2.1,  0.90) {};
    \vertex [dot] (vt3) at ( 2.1, -0.90) {};

    \diagram* {
      (va) -- [scalar, thick, edge label=\(H\)]     (vt1),
      (vt1) -- [charged boson, thick, edge label=\(W\)]   (vt2),
      (vt2) -- [charged boson, thick, edge label=\(W\)]                      (vt3),
      (vt3) -- [charged boson, thick, edge label=\(W\)]                      (vt1),
      (vt2) -- [boson, thick]     (f1),
      (vt3) -- [boson, thick]    (f2),
    };
  \end{feynman}
\end{tikzpicture}
\begin{tikzpicture}
  \begin{feynman}[every dot={/tikz/fill=black}]
 
    \vertex (f1) at ( 3.5,  1.3)  {\(\gamma\)};
    \vertex (f2) at ( 3.5, -1.3)  {\(\gamma\)};
 
    \vertex (va)  at (-1.4,  0.00) {};
    \vertex [dot] (vt1) at ( 0.7,  0.00) {};  % Z* = 2.4 units
    \vertex [dot] (vt2) at ( 2.1,  0.00) {};

    \diagram* {
      (va) -- [scalar, thick, edge label=\(H\)]   (vt1),

	  (vt1) -- [charged boson, thick, half left, looseness=1.55,
               edge label=\(W\)]                           (vt2),
      (vt2) -- [charged boson, thick, half left, looseness=1.55,
               edge label=\(W\)]                     (vt1),

      (vt2) -- [boson, thick]     (f1),
      (vt2) -- [boson, thick]    (f2),
    };
  \end{feynman}
\end{tikzpicture}
\caption{\small Main contributions (swapped triangles included) to $H \rightarrow \gamma \gamma $ in the SM.}
\label{fig:SMHgg}
\end{figure}
\begin{figure}[t]
\centering
\begin{tikzpicture}
  \begin{feynman}[every dot={/tikz/fill=black}]
 
    \vertex (f1) at ( 3.5,  1.3)  {\(\gamma\)};
    \vertex (f2) at ( 3.5, -1.3)  {\(\gamma\)};
 
    \vertex (va)  at (-1.4,  0.00) {};
    \vertex [dot] (vt1) at ( 0.7,  0.00) {};  % Z* = 2.4 units
    \vertex [dot] (vt2) at ( 2.1,  0.90) {};
    \vertex [dot] (vt3) at ( 2.1, -0.90) {};

    \diagram* {
      (va) -- [scalar, thick, edge label=\(H\)]     (vt1),
      (vt1) -- [fermion, thick, edge label=\(\Psi\)]   (vt2),
      (vt2) -- [fermion, thick, edge label=\(\Psi\)]                      (vt3),
      (vt3) -- [fermion, thick, edge label=\(\Psi\)]                      (vt1),
      (vt2) -- [boson, thick]    (f1),
      (vt3) -- [boson, thick]    (f2),
    };
  \end{feynman}
\end{tikzpicture}
\caption{\small Representative of the NP diagonal contribution to $H \rightarrow \gamma \gamma$.}
\label{fig:HHgammaNP}
\end{figure}
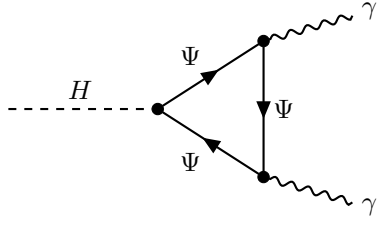
%
%%%%%%%%%%%%%%%
Finally, unmixed purely non-universal contributions can be tested via radiative Higgs decays into photons $H \rightarrow \gamma \gamma$,  whose deviations from the pure SM prediction are conveniently assessed through the ratio of the corresponding Higgs width: 
\begin{align}
R_{\gamma} = \frac{\Gamma(H \rightarrow \gamma \gamma)}{\Gamma(H \rightarrow \gamma \gamma)_{SM}}\, . 
\end{align}
Such values are bound by the 1-sigma range of $ 0.86 < R_{\gamma}^{ATLAS} < 1.24$  and  $ 0.94 < R_{\gamma}^{CMS} < 1.30$ as determined by recent ATLAS \cite{ATLAS:2022tnm} and  CMS \cite{CMS:2021kom} surveys. 
Under the natural assumption of a subdominant NP contribution compared to the SM, we can study deviations in VL models with the following formula
%, 
\begin{align}
\Delta R = R_{H \rightarrow \gamma \gamma} - 1 = \frac{|\mathcal M_{SM} + \mathcal M_{NP}|^2}{|\mathcal M_{SM}|^2} - 1  \sim  2 \frac{Re(\mathcal M^{\dagger}_{SM} \mathcal M_{NP})}{|\mathcal M_{SM}|^2} \, , 
\end{align}
depending on the square of the amplitudes $\mathcal M$, but independent of any common prefactor linked to kinematic or polarization sums.  
Model~\MII, which as in the rest of our survey we use to represent also the behavior of Model~\MI, modifies the SM expectation via 
\begin{align}
\Delta R_2 &=   \frac{\sqrt{2}}{3 \pi} \frac{m_Z}{m_W} \frac{e \,  \alpha^2 \, c_{\theta_E} s_{\theta_E}}{\sqrt{m_Z^2 - m_W^2 }}  \frac{r_{SM}^{\dagger}}{|\mathcal M_{SM}|^2} y_{E}  \bigg( 2\, (m_{E^+} - m_{E^-}) \nonumber \\
&+ m_{E^-} (m_H^2 - 4 m_{E^-}^2) C_0(0,0,m_H^2, m_{E^-}^2, m_{E^-}^2, m_{E^-}^2) \nonumber \\
&- m_{E^+} (m_H^2 - 4 m_{E^+}^2) C_0(0,0,m_H^2, m_{E^+}^2, m_{E^+}^2, m_{E^+}^2)  \bigg) \, .   \nonumber \\
\end{align}
where we used 
\begin{align}
 r_{SM} = & \quad \big(3\, m_H^2 - 16\, m_t^2 + 18 m_W^2\big) + 8 (m_H^2 - 4 m_t^2) \, m_t^2  \,C_0(0,0,m_H^2, m_t^2, m_t^2, m_t^2) \nonumber \\ 
&  - 18 \,(m_H^2 - 2 m_W^2) \,m_W^2  \,C_0(0, 0, m_H^2, m_W^2, m_W^2, m_W^2)  \, , 
\end{align}
and the OS definitions from Eq.(~\ref{eq:OSparameters}) are implicitly used. That we should not mistakenly assume a linear dependence on the NP Yukawa can be made evident in studying the corresponding expression for Model~\MIV, where the subtraction of the SM contribution due to the tau lepton is necessary to isolate the NP contribution.
We found the best route to be the substitution of the explicit formulas for the mixing angles and a series expansion in small NP Yukawa. This reveals, at leading order, the quadratic dependence : 
\begin{align}
\Delta R_4 &= \frac{\alpha^2 m_E^2 \, \lambda_E^2}{(m_E^2-m_{\tau}^2)^2\, \pi^2} \frac{r_{SM}^{\dagger}}{|\mathcal M_{SM}|^2} \bigg( 2 \big(m_{\tau}^2 - m_{E}^2\big) - m_{\tau}^2 (m_H^2 - 4 m_{\tau}^2) C_0(0,0,m_H^2, m_{\tau}^2, m_{\tau}^2, m_{\tau}^2) \nonumber \\
& + m^2_{E} (m_H^2 - 4 m_{E}^2) C_0(0, 0, m_H^2, m_{E}^2, m_{E}^2, m_{E}^2)  \bigg)  \, . 
\end{align}
Again, we found that the collider bounds adopted supersede also the constraining power of the Higgs decay into photons, which is therefore not included in our numerical analysis.

\section{Numerical results}\label{sec:numerical-analysis-and-results}
We now turn to a numerical evaluation of the NP contributions derived analytically in Sec.~\ref{sec:cross-section-calculation}. The primary observable is the fractional shift in the Higgs-strahlung integrated cross-section induced by the VL sector,
\begin{equation}
    \delta_{\rm NP}
    \;\equiv\;
    100\times\frac{
        \sigma^{\rm NLO}_{e^+e^-\to ZH}\big|_{\rm SM+VL}
        \;-\;
        \sigma^{\rm NLO}_{e^+e^-\to ZH}\big|_{\rm SM}
    }{
        \sigma^{\rm LO}_{e^+e^-\to ZH}\big|_{\rm SM}
    }\;\%\,,
    \label{eq:delta-NP}
\end{equation}
quoted as a percentage and evaluated at the candidate ZH-run center-of-mass energy $\sqrt{s}$. 

In each submodel the scan is performed over the parameter ranges detailed in the respective subsections below, the perturbativity bound from Table~\ref{table:perturbativity-merged} is imposed on all derived Yukawa couplings, and the collider and precision-EW bounds of Sec.~\ref{sec:the-model} provide the main selection criterion. The scan is not restricted to the collider-allowed region, however: masses and couplings that fall below the bounds of Sec.~\ref{sec:the-model} are deliberately retained and flagged throughout, so long as they remain within the perturbativity ceiling and satisfy the reality conditions, in order to map the full loop sensitivity of the observables.

Although $S$, $T$, and $R_\gamma$ are evaluated at every scanned point, none of the three turns out to be the operative constraint once the bounds of Sec.~\ref{sec:the-model} and the perturbativity ceiling of Sec.~\ref{subsec:perturbativity} are already imposed: $R_\gamma$ remains within its experimental range across the resulting parameter space, and the EW-precision fit to $S$ and $T$ excludes only a handful of points. Since these observables add essentially no further constraining power, the discussion in the following subsections focuses on the Higgs-strahlung cross-section shift $\delta_{\rm NP}$.

\subsection{Model~\MI: VLQ doublet-singlet}\label{subsec:results-MI}

The five independent scan inputs are the four physical VLQ masses $m_{U^\pm}$, $m_{D^\pm}$ and the doublet Lagrangian mass $M_Q$. The singlet masses $M_U$, $M_D$ and the Yukawa couplings $y_U$, $y_D$ are then obtained from Eq.~\eqref{eq:lagrangian_masses_in_terms_of_physical_ones}. Real solutions require $m_{S^-}\leq M_Q\leq m_{S^+}$ for $S=U,D$ simultaneously. We scan
\begin{equation}
    m_{S^-} \;\in\; [1300,\;3000]~\text{GeV}\,,\qquad
    m_{S^+} \;\in\; [m_{S^-}+100,\;6000]~\text{GeV}\,.
    \label{eq:scan-MI}
\end{equation}
The lower edge of $1300$~GeV for the lighter eigenstates reflects the LHC pair-production exclusion of Sec.~\ref{sec:the-model}, points below this threshold are included despite being excluded at the collider.

\begin{figure}[t]
\centering
\begin{subfigure}[b]{0.45\textwidth}
    \centering
    \includegraphics[width=\textwidth]{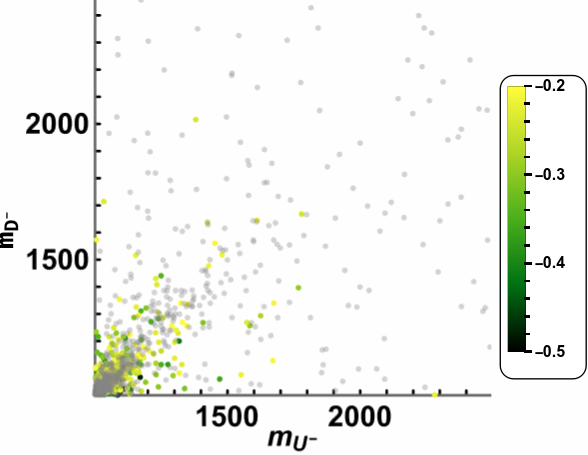}
    \label{fig:results-MI-mu2-md2}
\end{subfigure}
\hfill
\begin{subfigure}[b]{0.45\textwidth}
    \centering
    \includegraphics[width=\textwidth]{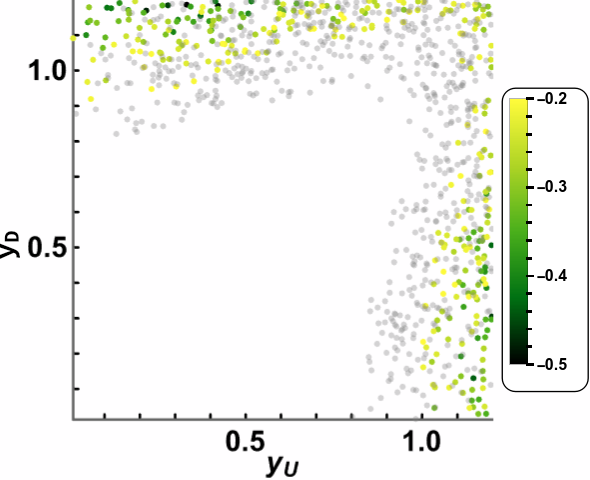}
    \label{fig:results-MI-yu-yd}
\end{subfigure}
\\[6pt]
\begin{subfigure}[b]{0.45\textwidth}
    \centering
    \includegraphics[width=\textwidth]{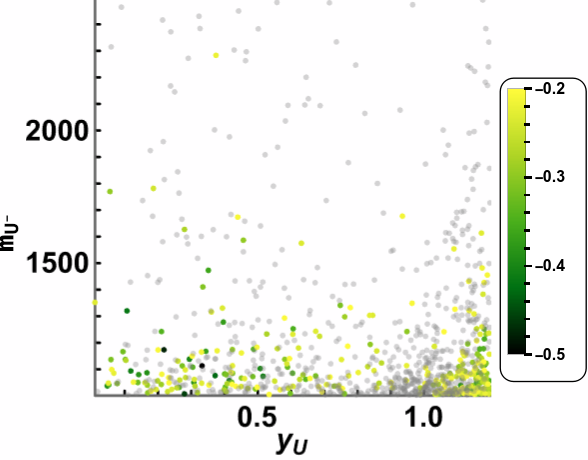}
    \label{fig:results-MI-yu-mu2}
\end{subfigure}
\hfill
\begin{subfigure}[b]{0.45\textwidth}
    \centering
    \includegraphics[width=\textwidth]{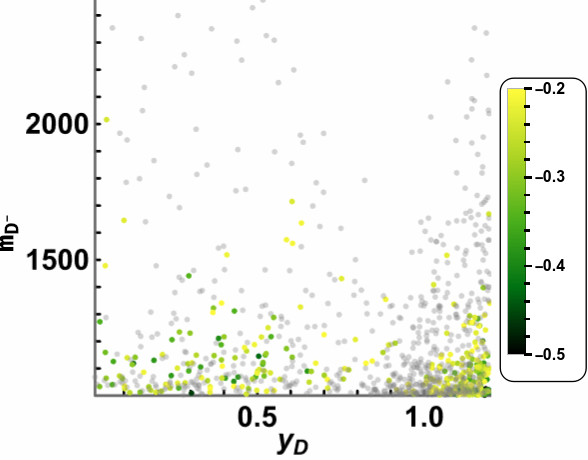}
    \label{fig:results-MI-yd-md2}
\end{subfigure}
\caption{Model~\MI scan projected onto four parameter planes: $m_{U^-}$ vs.\ $m_{D^-}$ (top left), $y_U$ vs.\ $y_D$ (top right), $y_U$ vs.\ $m_{U^-}$ (bottom left), $y_D$ vs.\ $m_{D^-}$ (bottom right). Points with $|\delta_{\rm NP}|<0.1\%$ are not shown, points with $0.1\%\leq|\delta_{\rm NP}|<0.2\%$ are gray and points with $|\delta_{\rm NP}|\geq0.2\%$ are colored by the signed value of $\delta_{\rm NP}$.}
\label{fig:results-MI-scan}
\end{figure}

In Model~\MI the NP contribution to Higgs-strahlung arises entirely from VLQ loops, with no direct coupling to SM fermions. Both the universal correction to the $Z$-boson propagator (the sunset diagram, left panel of Fig.~\ref{fig:VLNLO}) and the non-universal vertex correction (triangle diagram, right panel of Fig.~\ref{fig:VLNLO}) contribute, with their relative importance depending on the up-down mass splitting and the size of the Yukawa couplings, as detailed below. The electrically charged $U^\pm$ and $D^\pm$ states also shift the $H\to\gamma\gamma$ amplitude. Because the up- and down-type sectors share the same doublet mass $M_Q$ but generically differ in their singlet masses, an isospin splitting is produced between the two sectors, generating a positive contribution to the $T$ parameter. The color factor $N_c=3$ amplifies both $\delta_{\rm NP}$ and the oblique corrections relative to the leptonic counterpart of Model~\MII.

Figure~\ref{fig:results-MI-scan} shows the full Model~\MI scan projected onto four parameter planes.

The $\geq0.2\%$ population revealed by the color scale presents an interesting concentrated pattern across the scan. In the $(m_{U^-},m_{D^-})$ plane it sits almost entirely where both masses are near the lower edge of the scan range, with only a sparse tail of colored points at larger masses,
 an expected behavior due to decoupling of NP mass scale. In the $(y_U,y_D)$ plane the colored points instead populate two bands along the top and right edges of the plane, showing that a large correction requires at least one of the two Yukawa couplings to be sizable (roughly $\gtrsim0.5$, most densely near the perturbativity limits) rather than both simultaneously\footnote{We have checked this with a quadratic fit, $\delta_{\rm NP}\propto A\,y_U^2 + B\,y_D^2 + C\,y_U y_D$, to the scanned cross-section shift: both $A$ and $B$ come out large, while $A+B+C$ is comparatively small. Along the diagonal $y_U\to y_D\equiv y$ the shift therefore reduces to $(A+B+C)\,y^2$ and stays small, whereas along either axis, where the other coupling vanishes, it grows as the uncanceled $A\,y_U^2$ or $B\,y_D^2$ term.}. Most of the interior falls below the $0.1\%$ threshold and is not plotted at all, with only a thin bunch of gray points just below the $0.2\%$  cut. The $(y_U,m_{U^-})$ and $(y_D,m_{D^-})$ planes show the same qualitative pattern, with the colored population concentrated at low mass across, essentially, the full coupling range. This illustrates that the mass dependence, not the coupling alone, dominates which points reach the $0.2\%$ threshold.

Table~\ref{table:results-MI-benchmarks} defines four benchmark points (BPs) spanning qualitatively different corners of the Model~\MI parameter space. Their full $\sqrt{s}$-dependence, decomposed into the sunset and triangle contributions, is shown in Fig.~\ref{fig:results-MI-sunset-triangle} below.

BP1 sits at moderate masses and moderate Yukawa couplings, representative of a generic point away from any boundary of the allowed region. BP2 combines the heaviest mass scale considered here with the largest isospin splitting between the up- and down-type singlet sectors and a Yukawa coupling close to the perturbativity limits of Table~\ref{table:perturbativity-merged}, chosen to probe the largest correction attainable in this submodel. BP3 is close to the opposite limit: an almost degenerate up-type sector paired with a large down-type splitting and a strongly asymmetric coupling pattern, isolating the down-sector contribution. BP4 has comparable, sizable splittings in both sectors together with large and similar couplings, representing the most symmetric high-coupling benchmark of the four.

\begin{table}[t]
\centering
\begingroup
\small
\setlength{\tabcolsep}{8pt}
\renewcommand{\arraystretch}{1.2}
\begin{tabular}{|c|c c c c c|c c|c c c|}
\hline
BP & $m_{U^-}$ & $m_{U^+}$ & $m_{D^-}$ & $m_{D^+}$ & $M_Q$ & $y_U$ & $y_D$ & $\delta_{\rm NP} ^{\rm sunset}$ & $\delta_{\rm NP} ^{\rm triangle}$ & $\delta_{\rm NP}$ \\
\hline
BP1 & 1000 & 1500 & 1010 & 1520 & 1015 & 0.50 & 0.30 & $-0.010\%$ & $0.002\%$ & $-0.008\%$ \\
BP2 & 2000 & 3000 & 2950 & 2965 & 2960 & 1.14 & 0.04 & $-0.001\%$ & $-0.099\%$ & $-0.100\%$ \\
BP3 & 2978 & 2981 & 1816 & 2989 & 2980 & 0.01 & 0.63 & $-0.001\%$ & $-0.007\%$ & $-0.008\%$ \\
BP4 & 1115 & 1944 & 1114 & 1907 & 1145 & 0.90 & 0.89 & $-0.008\%$ & $-0.057\%$ & $-0.065\%$ \\
\hline
\end{tabular}
\endgroup
\caption{Benchmark points for Model~\MI used in Fig.~\ref{fig:results-MI-sunset-triangle}. Mass parameters in GeV. Cross-section columns to 3 decimal places. Sunset, triangle, and total cross-section shifts evaluated at $\sqrt{s}=240$~GeV.}
\label{table:results-MI-benchmarks}
\end{table}

\begin{figure}[h]
\centering
\begin{subfigure}[b]{0.48\textwidth}
    \centering
    \includegraphics[width=\textwidth]{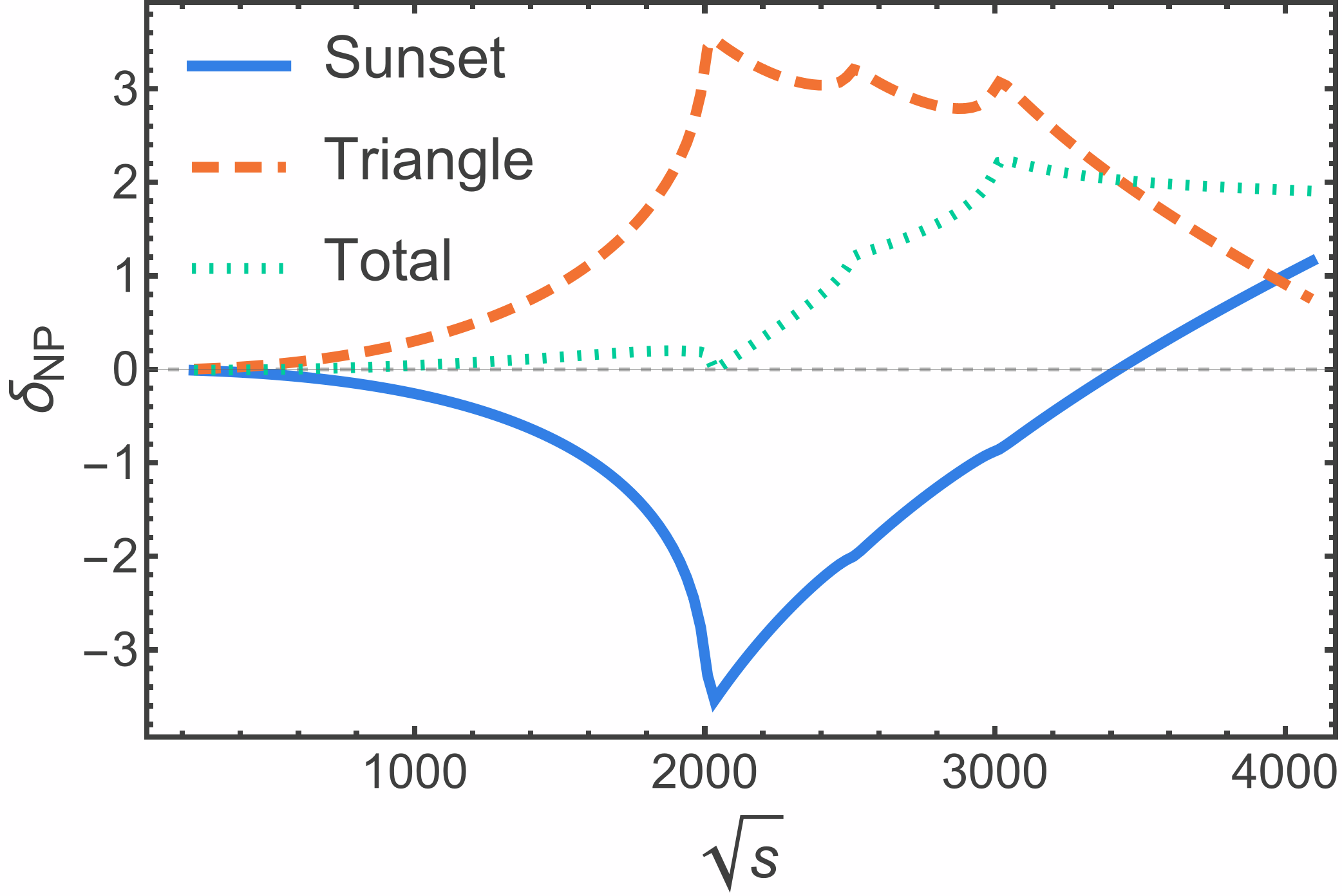}
    \label{fig:results-MI-sunset-triangle-BP1}
\end{subfigure}
\hfill
\begin{subfigure}[b]{0.48\textwidth}
    \centering
    \includegraphics[width=\textwidth]{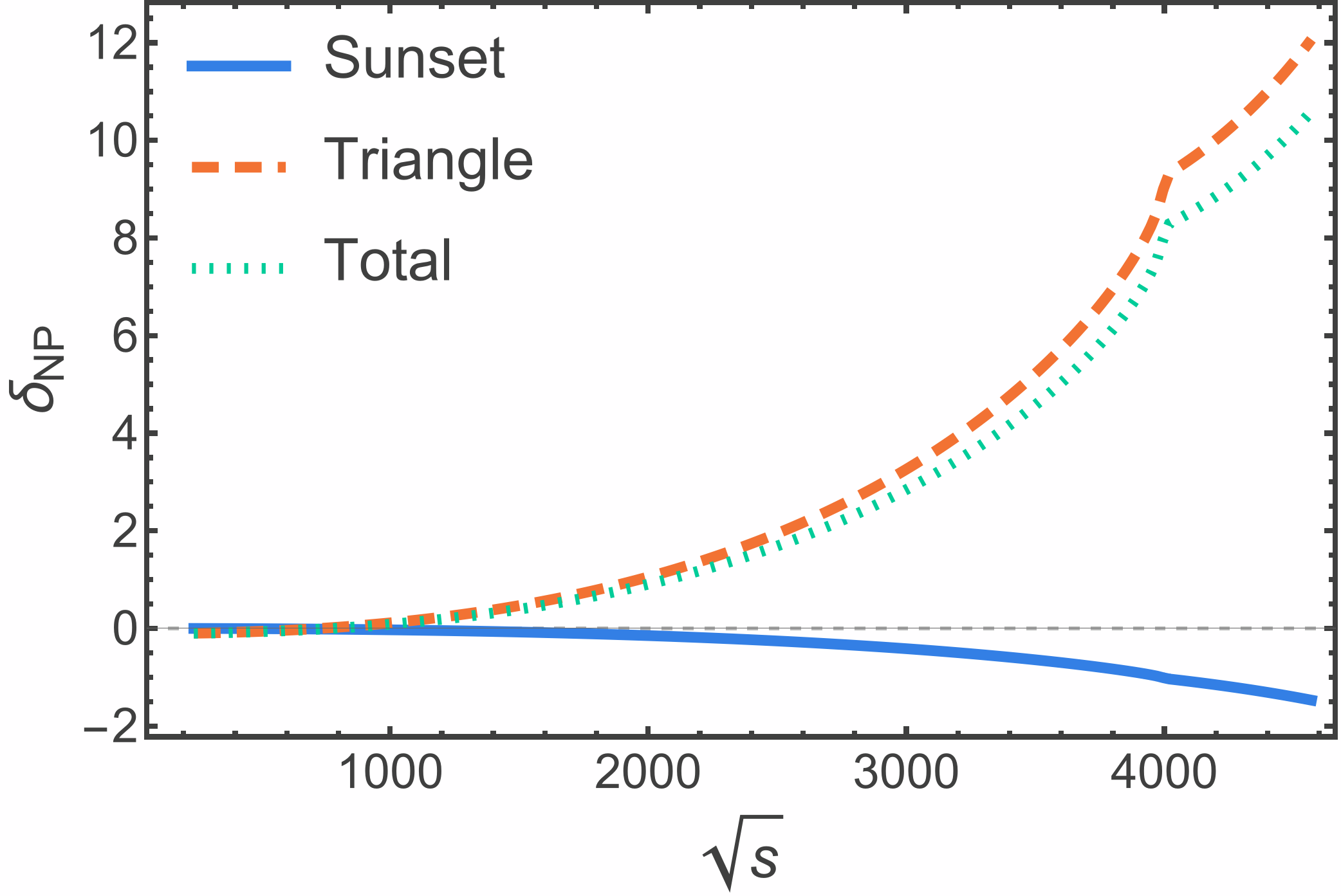}
    \label{fig:results-MI-sunset-triangle-BP2}
\end{subfigure}
\\[6pt]
\begin{subfigure}[b]{0.48\textwidth}
    \centering
    \includegraphics[width=\textwidth]{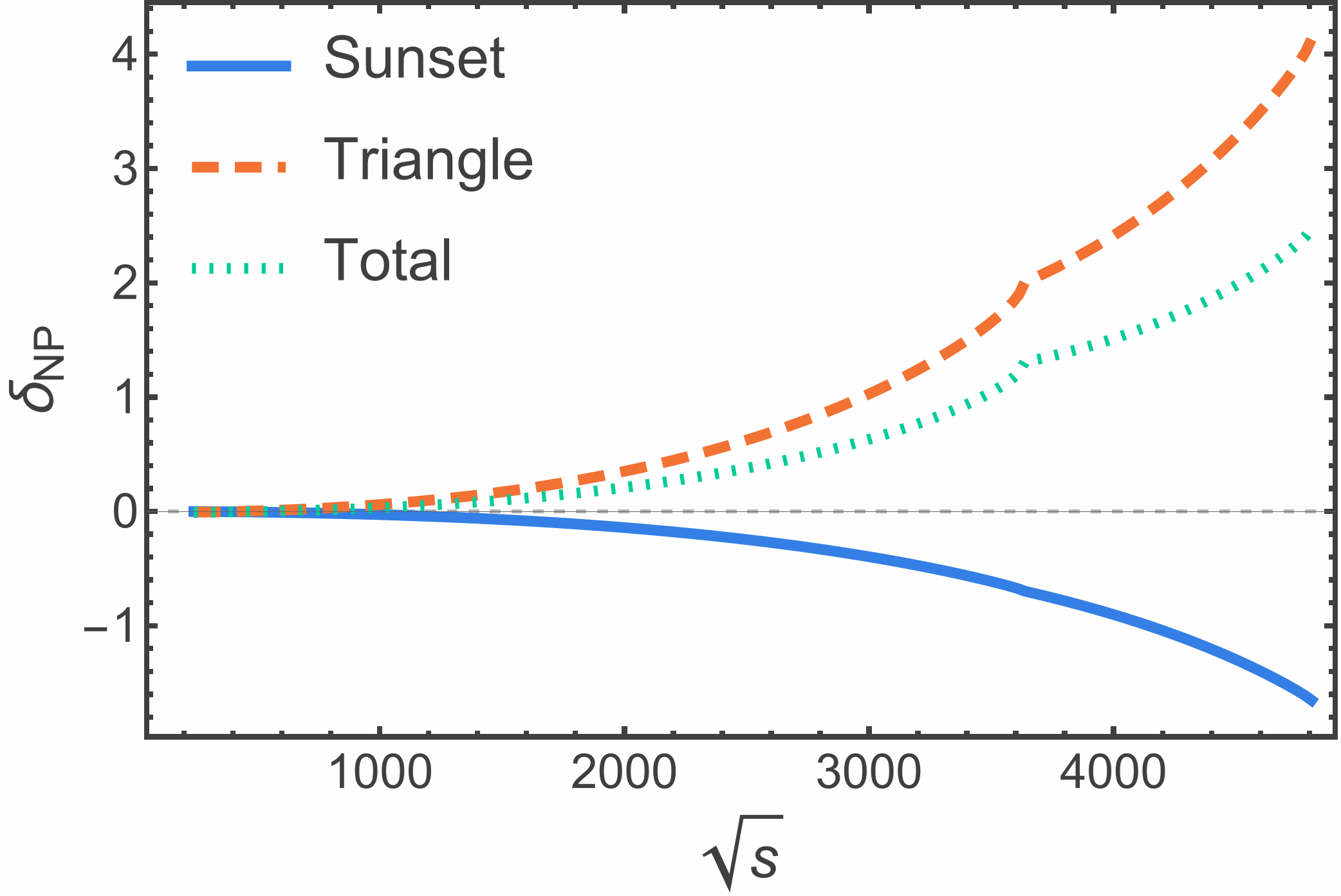}
    \label{fig:results-MI-sunset-triangle-BP3}
\end{subfigure}
\hfill
\begin{subfigure}[b]{0.48\textwidth}
    \centering
    \includegraphics[width=\textwidth]{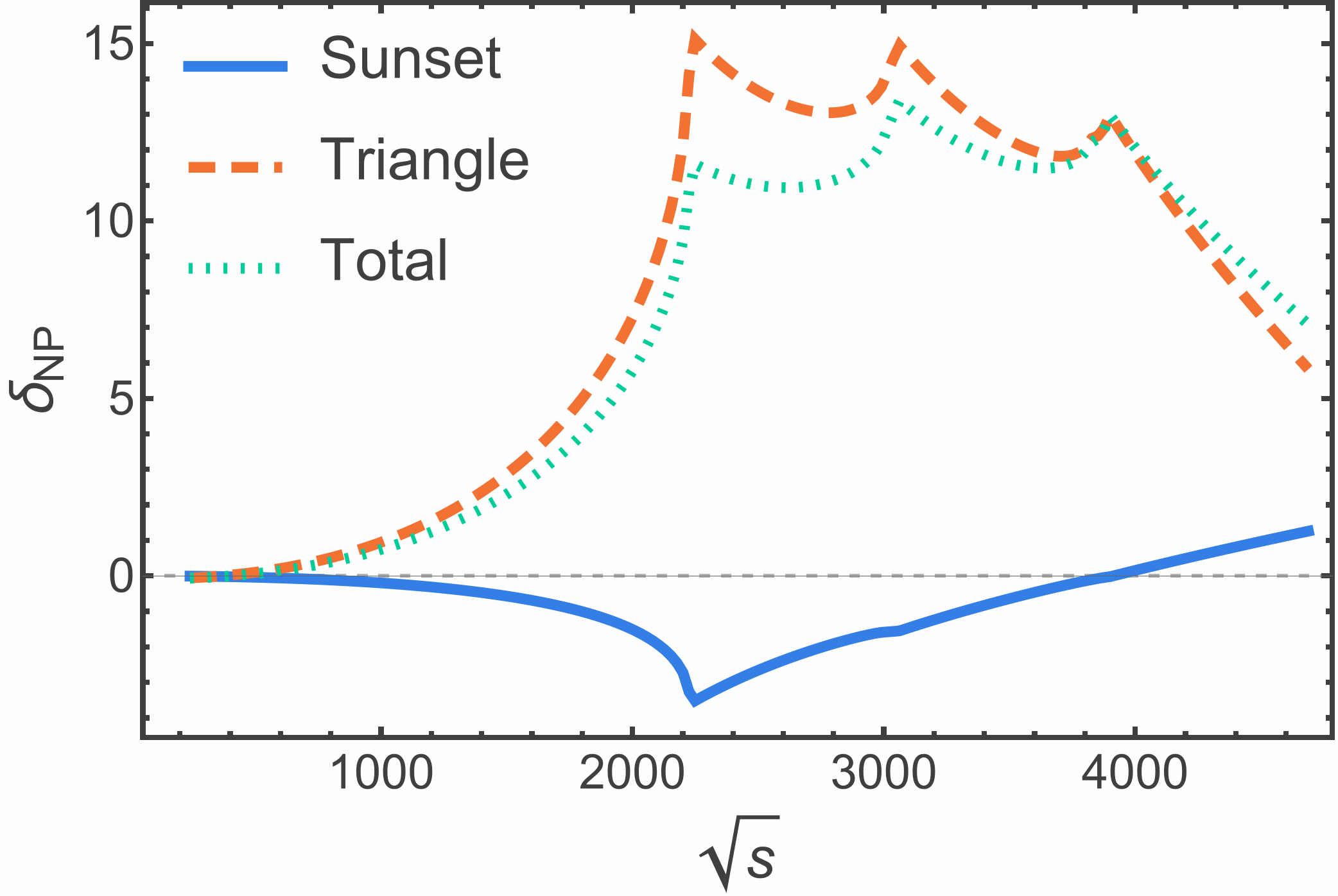}
    \label{fig:results-MI-sunset-triangle-BP4}
\end{subfigure}
\caption{Sunset and triangle diagram contributions to $\delta_{\rm NP}$, and their sum, for the benchmark points detailed in Table~\ref{table:results-MI-benchmarks}. The contribution to $\delta_{\rm NP}$ coming from sunset diagram is in solid blue, the contribution from the triangle diagram is in dashed orange, and the total $\delta_{\rm NP}$ is in dotted green.}
\label{fig:results-MI-sunset-triangle}
\end{figure}

The sunset/triangle breakdown of Fig.~\ref{fig:results-MI-sunset-triangle} shows that the two diagram topologies do not always add constructively. At $\sqrt{s}=240$~GeV, for BP1 the self-energy contribution dominates and is negative while the vertex contribution is smaller and of \emph{opposite} sign, partially canceling to give the smallest total deviation among the four points. For BP2, BP3, and BP4, sunset and triangle are instead both negative and interfere constructively, with the triangle contribution dominating throughout. This is consistent with the sunset (universal, isospin-splitting-driven) contribution being comparatively small whenever the up-down mass splitting alone is modest, leaving the non-universal vertex correction as the dominant channel once the Yukawa couplings are sizable. Beyond $\sqrt{s}=240$~GeV, Fig.~\ref{fig:results-MI-sunset-triangle} shows that the four BPs reach very different overall scales: BP2 and BP4 grow to roughly an order of magnitude larger than BP1 and BP3 by the upper end of the plotted range, and not in the same ranking as at $\sqrt{s}=240$~GeV, where BP2 and BP4 were merely the two largest rather than qualitatively different in scale. All four BPs display resonance-like spikes, matching the on-shell pair-production thresholds set by their own mass spectrum (the light-light, light-heavy, and heavy-heavy combinations of $m_{U^\pm}$ and $m_{D^\pm}$), though with very different strength. For BP1 and BP4 the triangle and total curves develop genuine peaks that partially decline before the next threshold, with BP4's peaks notably sharper than BP1's, consistent with BP4's larger, more symmetric Yukawa couplings (Table~\ref{table:results-MI-benchmarks}) driving a stronger resonant enhancement than BP1's smaller, more asymmetric pair. BP2 and BP3 instead show only a single, much milder kink near the edge of their respective plotted ranges, since their heavier and more widely split spectra push most thresholds beyond the range shown.

\subsection{Model~\MII: VLL doublet-singlet}\label{subsec:results-MII}

The five independent inputs are the four physical VLL masses $m_{E^\pm}$, $m_{N^\pm}$ and the doublet Lagrangian mass $M_L$; the derived parameters $M_E$, $M_N$, $y_E$, $y_N$ follow from Eq.~\eqref{eq:lagrangian_masses_in_terms_of_physical_ones_model2}. Real solutions require $m_{S^-}\leq M_L\leq m_{S^+}$ for $S=E,N$ simultaneously. We scan
\begin{equation}
    m_{S^-} \;\in\; [300,\;2000]~\text{GeV}\,,\qquad
    m_{S^+} \;\in\; [m_{S^-}+100,\;4000]~\text{GeV}\,,
    \label{eq:scan-MII}
\end{equation}
with $M_L$ drawn from $\bigl[\max(m_{E^-},m_{N^-}),\,\min(m_{E^+},m_{N^+})\bigr]$. The lower edge reflects the multilepton VLL bound of Sec.~\ref{sec:the-model}.

In Model~\MII the new-physics contribution to Higgs-strahlung arises entirely from VLL loops, with no direct coupling to SM fermions, mirroring Model~\MI's algebraic structure but with the $SU(3)_C$ color factor removed.  Losing the color factor weakens both the self-energy corrections and the perturbativity ceilings relative to Model~\MI (see the two blocks of Table~\ref{table:perturbativity-merged}), though the correspondingly weaker LHC bounds on VLLs let the scan probe lighter spectra, somewhat compensating the absent color enhancement.

\begin{figure}[h]
\centering
\begin{subfigure}[b]{0.48\textwidth}
    \centering
    \includegraphics[width=\textwidth]{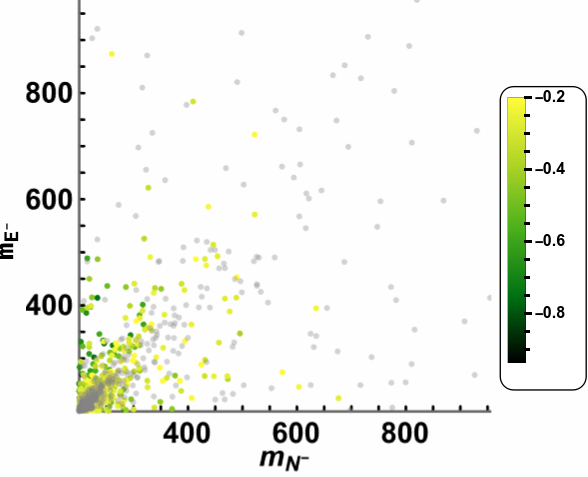}
    \label{fig:results-MII-mn2-me2}
\end{subfigure}
\hfill
\begin{subfigure}[b]{0.48\textwidth}
    \centering
    \includegraphics[width=\textwidth]{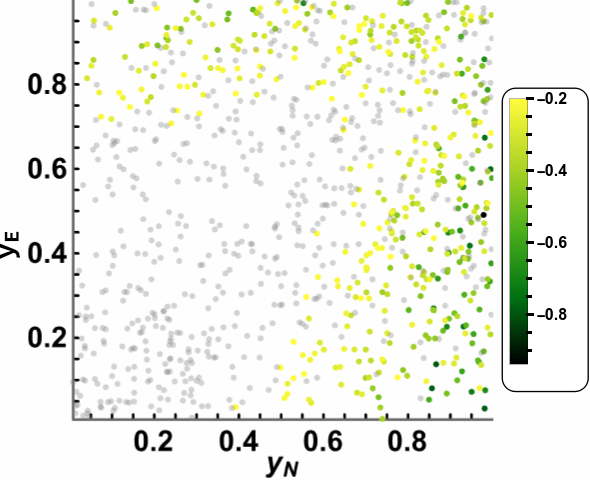}
    \label{fig:results-MII-yn-ye}
\end{subfigure}
\\[6pt]
\begin{subfigure}[b]{0.48\textwidth}
    \centering
    \includegraphics[width=\textwidth]{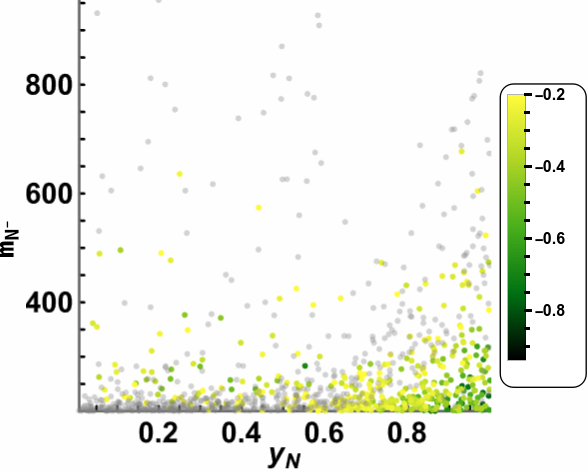}
    \label{fig:results-MII-yn-mn2}
\end{subfigure}
\hfill
\begin{subfigure}[b]{0.48\textwidth}
    \centering
    \includegraphics[width=\textwidth]{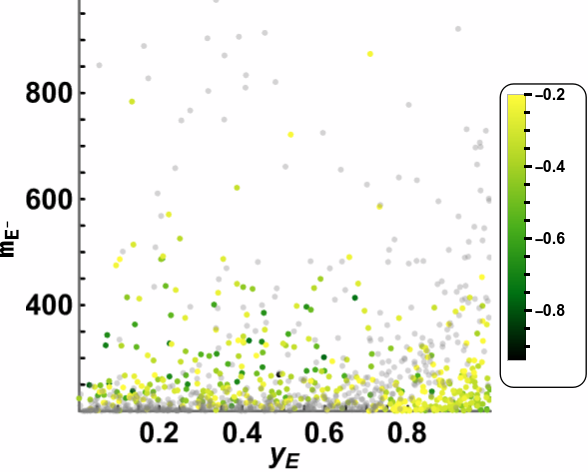}
    \label{fig:results-MII-ye-me2}
\end{subfigure}
\caption{Model~\MII scan projected onto four parameter planes: $m_{N^-}$ vs.\ $m_{E^-}$ (top left), $y_N$ vs.\ $y_E$ (top right), $y_N$ vs.\ $m_{N^-}$ (bottom left), $y_E$ vs.\ $m_{E^-}$ (bottom right). Points with $|\delta_{\rm NP}|<0.1\%$ are not shown, points with $0.1\%\leq|\delta_{\rm NP}|<0.2\%$ are gray and points with $|\delta_{\rm NP}|\geq0.2\%$ are colored by the signed value of $\delta_{\rm NP}$.}
\label{fig:results-MII-scan}
\end{figure}

Figure~\ref{fig:results-MII-scan} shows the analog of Fig.~\ref{fig:results-MI-scan} for Model~\MII.

The pattern is similar to Model~\MI: colored points concentrate at low mass in the $(m_{N^-},m_{E^-})$ plane. The $(y_N,y_E)$ plane departs from Model~\MI's pattern, however: the colored region is clearly asymmetrical, dominated by large $y_N$ rather than balanced between the two couplings. The fact that both sectors share the same floor from the multilepton bound hints that the asymmetry should be blamed on how $y_N$ and $y_E$ enter the loop functions. The $(y_N,m_{N^-})$ and $(y_E,m_{E^-})$ planes again show the colored population concentrated at low mass across essentially the full coupling range, as in Model~\MI.

Table~\ref{table:results-MII-benchmarks} defines four benchmark points (BPs) spanning qualitatively different corners of the Model~\MII parameter space. Their full $\sqrt{s}$-dependence, decomposed into the sunset and triangle contributions, is shown in Fig.~\ref{fig:results-MII-sunset-triangle} below. 
BP1 sits at light masses and moderate, comparable Yukawa couplings, representative of a generic point away from any boundary of the allowed region. BP2 combines the heaviest mass scale considered here with a strongly asymmetric coupling pattern dominated by $y_E$, isolating the charged-lepton sector contribution. BP3 is close to the opposite limit: light-to-moderate masses with the coupling asymmetry reversed, dominated instead by $y_N$ and isolating the neutral-lepton sector. BP4 has the lightest masses together with the largest splittings in both sectors and large, similar couplings, representing the most symmetric high-coupling benchmark of the four.

\begin{table}[t]
\centering
\begingroup
\small
\setlength{\tabcolsep}{8pt}
\renewcommand{\arraystretch}{1.2}
\begin{tabular}{|c|c c c c c|c c|c c c|}
\hline
BP & $m_{E^-}$ & $m_{E^+}$ & $m_{N^-}$ & $m_{N^+}$ & $M_L$ & $y_E$ & $y_N$ & $\delta_{\rm NP} ^{\rm sunset}$ & $\delta_{\rm NP} ^{\rm triangle}$ & $\delta_{\rm NP}$ \\
\hline
BP1 & 200 & 600 & 201 & 830 & 200 & 0.23 & 0.25 & $-0.110\%$ & $-0.036\%$ & $-0.146\%$ \\
BP2 & 543 & 1187 & 1145 & 1219 & 1150 & 0.89 & 0.08 & $-0.003\%$ & $-0.109\%$ & $-0.112\%$ \\
BP3 & 341 & 715 & 215 & 525 & 340 & 0.11 & 0.89 & $-0.035\%$ & $-0.511\%$ & $-0.546\%$ \\
BP4 & 232 & 900 & 246 & 1312 & 270 & 0.89 & 0.90 & $-0.069\%$ & $-0.265\%$ & $-0.333\%$ \\
\hline
\end{tabular}
\endgroup
\caption{Benchmark points for Model~\MII used in Fig.~\ref{fig:results-MII-sunset-triangle}. Mass parameters in GeV. Cross-section columns to 3 decimal places. Sunset, triangle, and total cross-section shifts evaluated at $\sqrt{s}=240$~GeV.}
\label{table:results-MII-benchmarks}
\end{table}

\begin{figure}[h]
\centering
\begin{subfigure}[b]{0.48\textwidth}
    \centering
    \includegraphics[width=\textwidth]{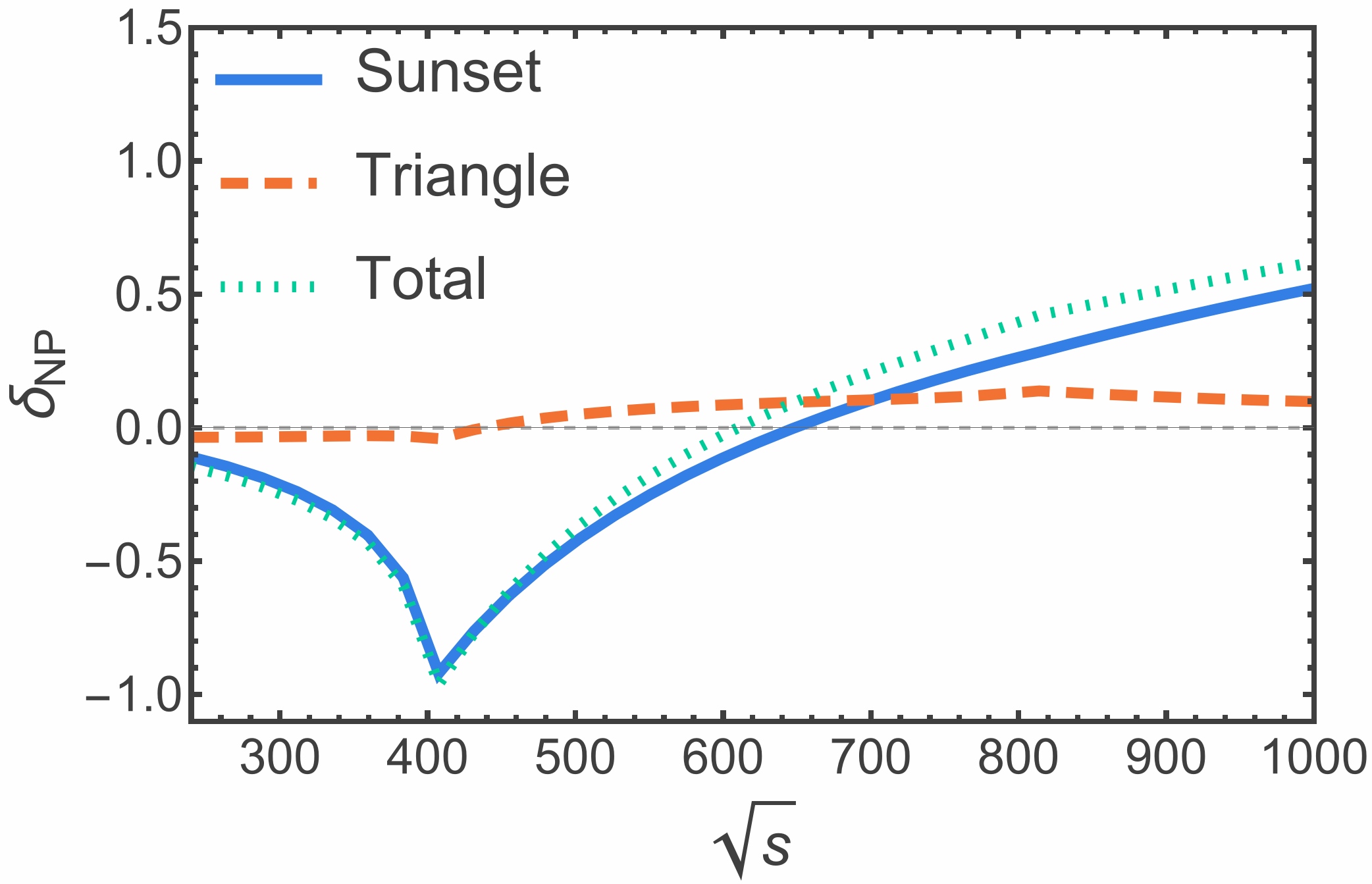}
    \label{fig:results-MII-sunset-triangle-BP1}
\end{subfigure}
\hfill
\begin{subfigure}[b]{0.48\textwidth}
    \centering
    \includegraphics[width=\textwidth]{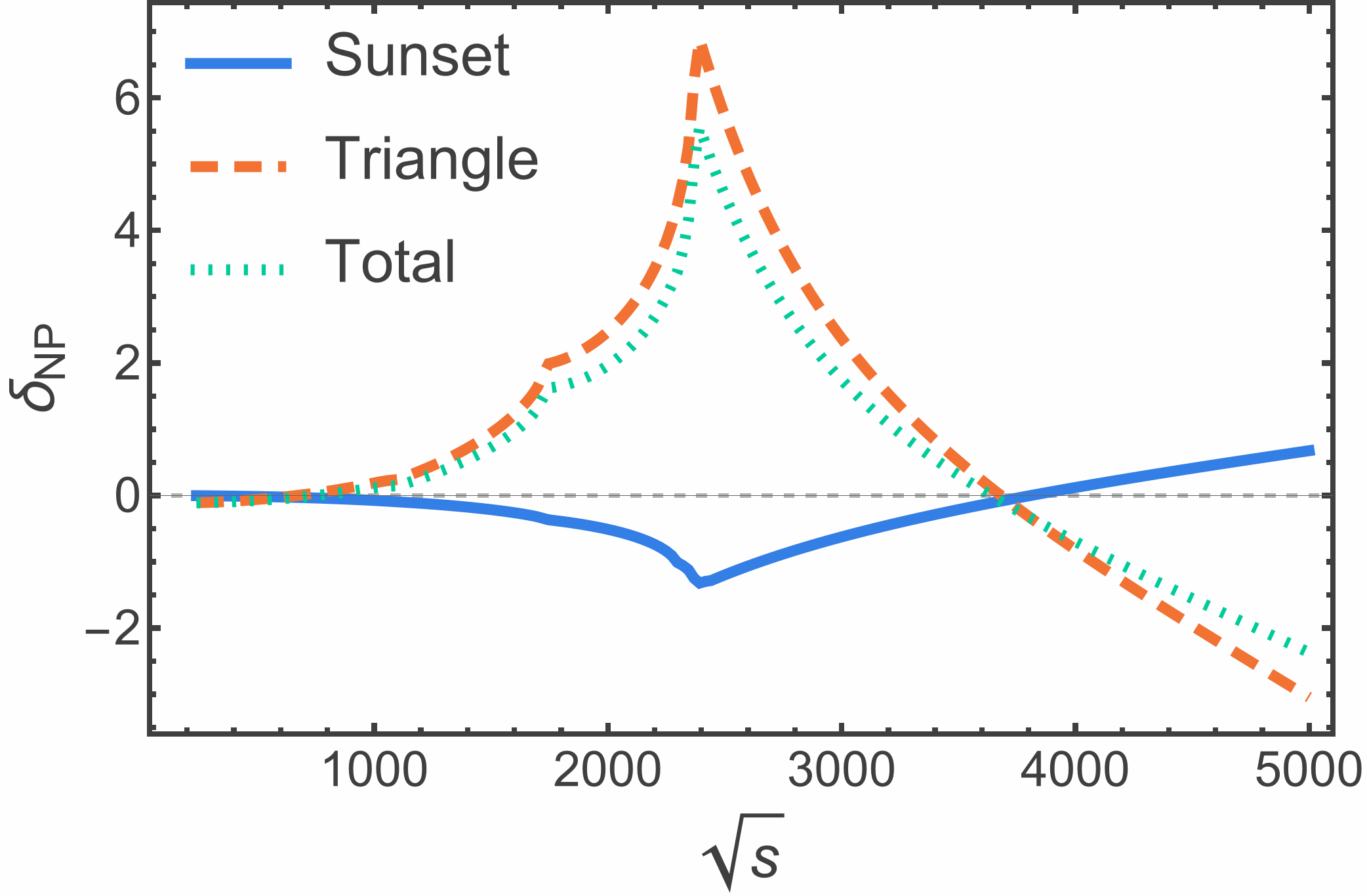}
    \label{fig:results-MII-sunset-triangle-BP2}
\end{subfigure}
\\[6pt]
\begin{subfigure}[b]{0.48\textwidth}
    \centering
    \includegraphics[width=\textwidth]{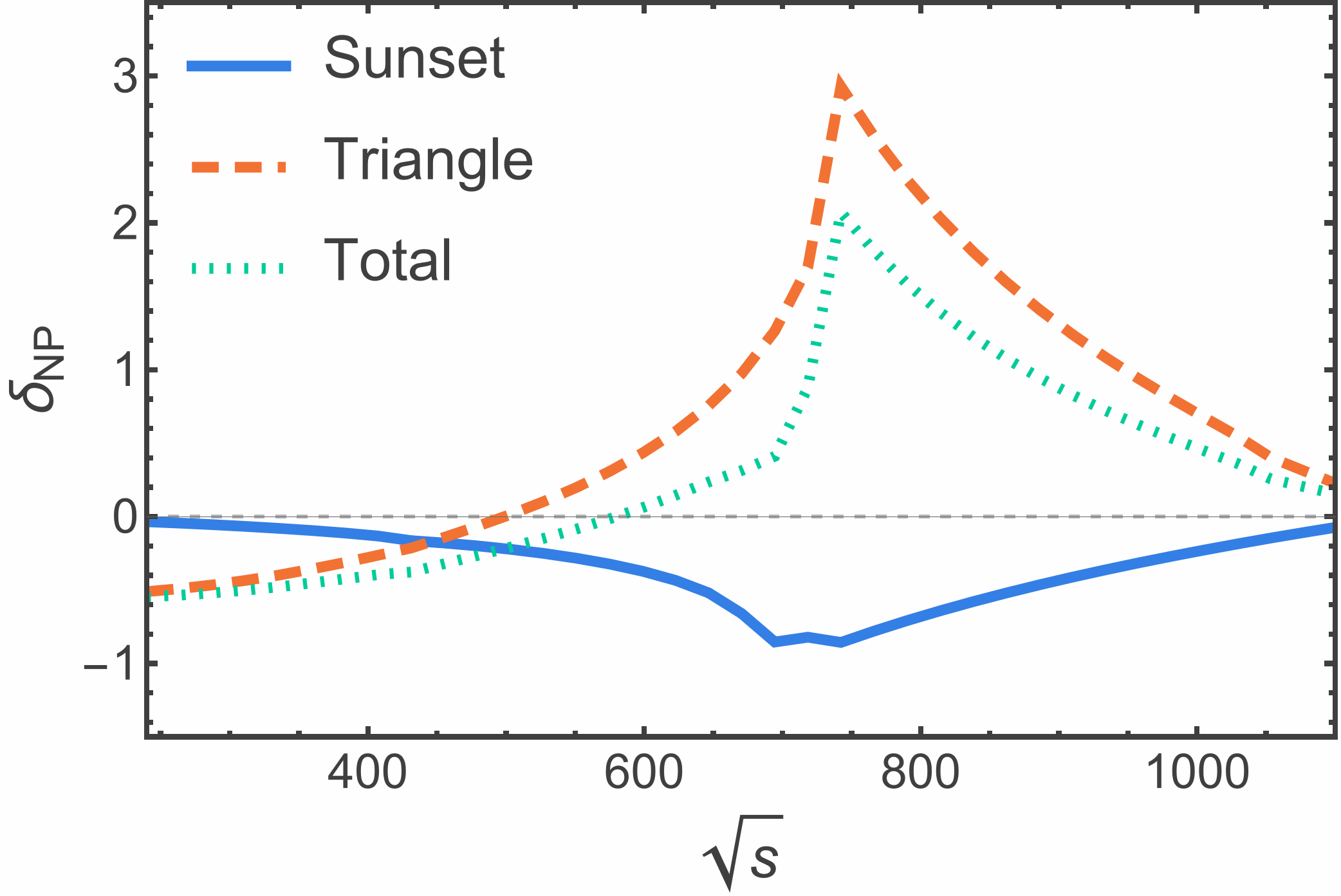}
    \label{fig:results-MII-sunset-triangle-BP3}
\end{subfigure}
\hfill
\begin{subfigure}[b]{0.48\textwidth}
    \centering
    \includegraphics[width=\textwidth]{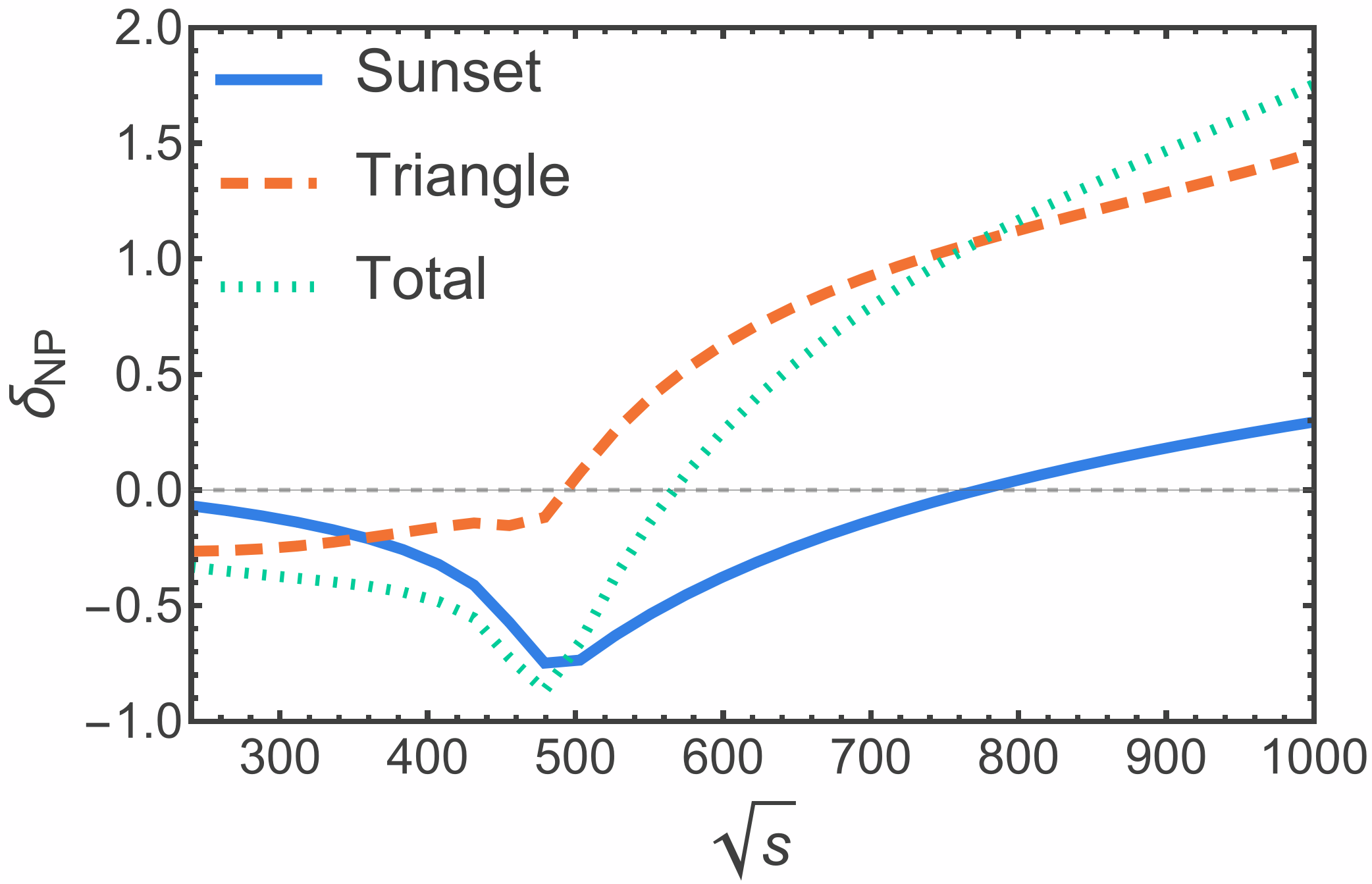}
    \label{fig:results-MII-sunset-triangle-BP4}
\end{subfigure}
\caption{As in Fig.~\ref{fig:results-MI-sunset-triangle} but for Model~\MII, sunset and triangle diagram contributions to $\delta_{\rm NP}$, and their sum, for each benchmark point detailed in Table~\ref{table:results-MII-benchmarks}. The contribution to $\delta_{\rm NP}$ from the sunset diagram is shown in solid blue, the contribution from the triangle diagram in dashed orange, and the total $\delta_{\rm NP}$ in dotted green.}
\label{fig:results-MII-sunset-triangle}
\end{figure}

At $\sqrt{s}=240$~GeV all four BPs again give $\delta_{\rm NP}<0$. As in Model~\MI, the vertex correction (Eq.~\eqref{Mod2Vertex}) dominates at BP2, BP3, and BP4, overwhelmingly so in each case, while the self-energy contribution remains the larger of the two at BP1, though by a much narrower margin (roughly a factor of three) than at Model~\MI's BP1 (roughly a factor of five). That fits with losing the $SU(3)_C$ factor: the isospin-splitting self-energy that competed with the vertex correction in the quark sector has no color enhancement here, so it is weaker relative to the vertex correction across the board. Past $\sqrt{s}=240$~GeV the four BPs split into two behaviors, mirroring what we saw in Model~\MI: BP1 and BP4 each show one modest dip, matching $2m_{E^-}$ or $2m_{N^-}$ for their respective light states. For BP2 and BP3 the triangle contribution develops a genuine resonant peak before falling away again. In the 
case of BP2, the total yield turns negative again before the range ends. For BP2 we can register a peculiar pile-up with its four heaviest pair-production thresholds within a narrow window of each other, unlike BP1's and BP4's, whose thresholds stay widely separated. BP3's peak is comparably sharp but does not reduce to the same clustering of two-body thresholds, pointing instead to the internal structure of the triangle diagram itself.

\subsection{Model~\MIII: neutral VL singlet with $\tau$-neutrino mixing}\label{subsec:results-MIII}

The two independent inputs are the physical VL neutrino mass $m_N$ and the mixing Yukawa $\lambda_\nu$, the Lagrangian mass $M_N$ following from Eq.~\eqref{eq:lagrangian_mass_model3}. We scan
\begin{equation}
    m_N \;\in\; [200,\;1500]~\text{GeV}\,,\qquad
    \lambda_\nu \;\in\; \bigl[0,\;\lambda_\nu^{\rm max}(\Lambda_i)\bigr]\,,
    \label{eq:scan-MIII}
\end{equation}
up to the perturbativity bounds of Sec.~\ref{subsec:perturbativity}. In Model~\MIII the single new parameter $\lambda_\nu$ is constrained from above by the EW non-unitarity bound $\theta_\nu\leq 0.042$, by perturbativity, and by the kinematic reality condition in Eq.~\eqref{eq:model3_lambda_bound}. Because $\tan\theta_\nu = v\lambda_\nu/(\sqrt{2}M_N)$ [Eq.~\eqref{eq:rotation_angle_mass_matrices_model_3}], large Yukawa values are kinematically accessible only for heavy $M_N$, so the allowed strip in the $(M_N,\lambda_\nu)$ plane opens up with mass. As already found in Sec.~\ref{subsec:perturbativity}, the EW bound is by far the strictest of the three, limiting $\lambda_\nu$ to values roughly an order of magnitude below both the perturbativity ceiling and the reality bound shown in the scan below. Since $\delta_{\rm NP}$ scales steeply (roughly quadratically) with the mixing angle and hence with $\lambda_\nu$, the resulting shift never comes close to the per-mille level FCC-ee could resolve: within the EW-allowed window Model~\MIII is not a viable target for this measurement. 
We nonetheless scan $\lambda_\nu$ up to the perturbativity bound in what follows, since doing so exposes the full size and shape of the loop effect. This can be justified by the possibility of relaxing the
bound in appropriate SM extensions without affecting the Higgs-strahlung process~\cite{deBlas:2025pco}. The neutral nature of $N$ means that neither the $H\to\gamma\gamma$ amplitude nor the $\gamma$-$Z$ mixing receives a one-loop contribution from this sector. The new-physics imprint on Higgs-strahlung comes from the $Z$-boson self-energy (Eq.~\eqref{model3ZZ}) and from the penguin diagrams involving the mixed $\tau$--$N$ states.

\begin{figure}[t]
\centering
\includegraphics[width=0.5\textwidth]{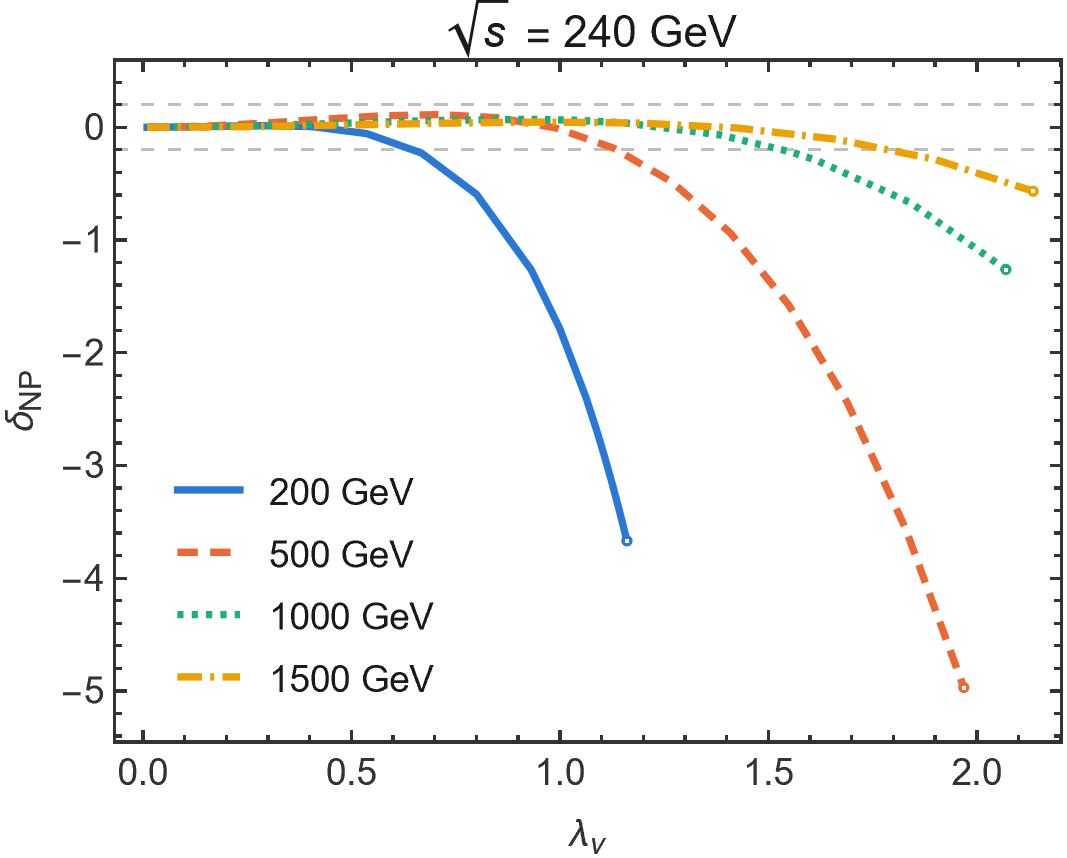}
\caption{Fractional Higgs-strahlung cross-section shift $\delta_{\rm NP}$ for Model~\MIII as a function of $\lambda_\nu$ at $\sqrt{s}=240$~GeV, for fixed values of $m_N$.}
\label{fig:results-MIII-scan-masses}
\end{figure}

\begin{figure}[h]
\centering
\begin{subfigure}[b]{0.45\textwidth}
    \centering
    \includegraphics[width=\textwidth]{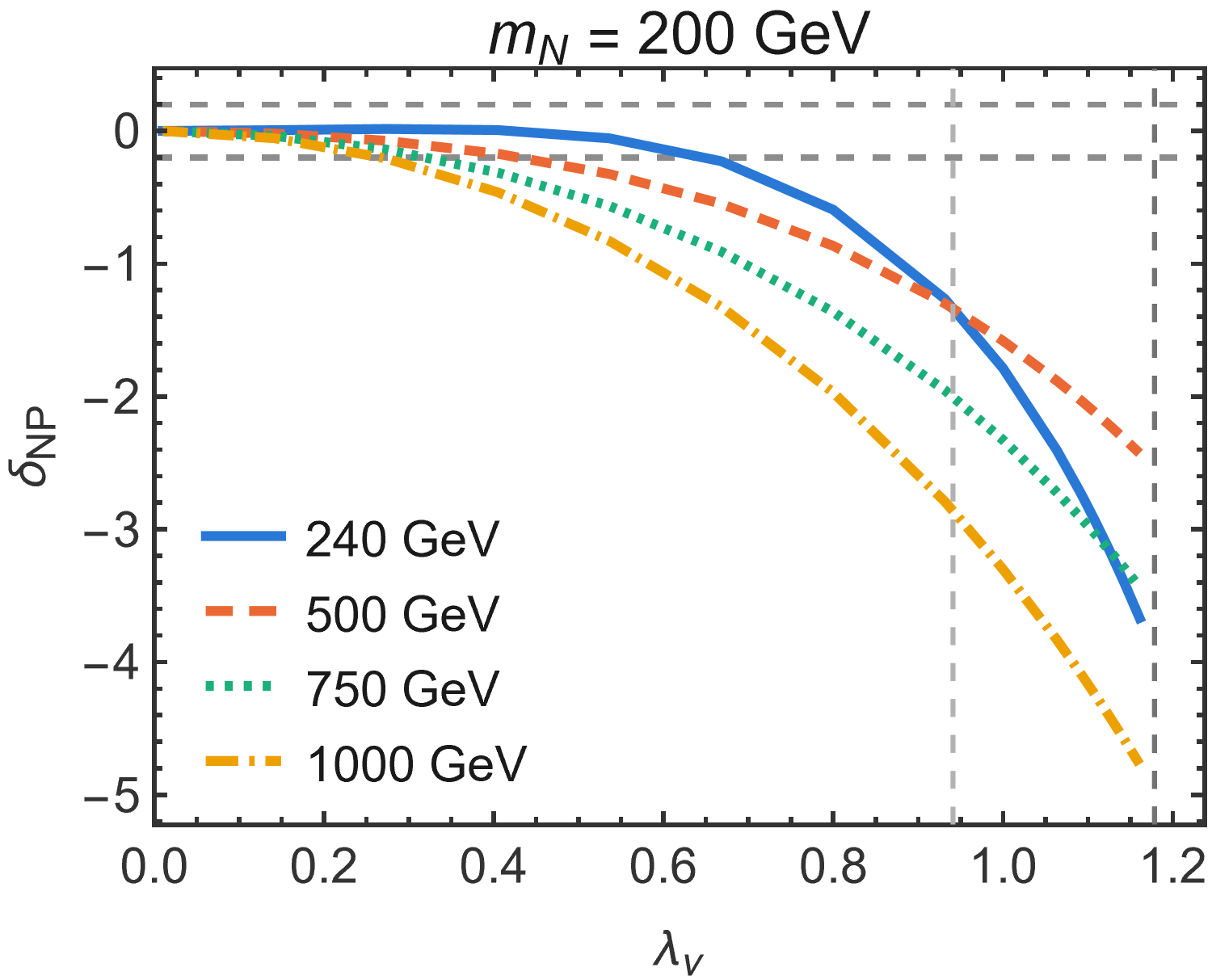}
\end{subfigure}
\hfill
\begin{subfigure}[b]{0.45\textwidth}
    \centering
    \includegraphics[width=\textwidth]{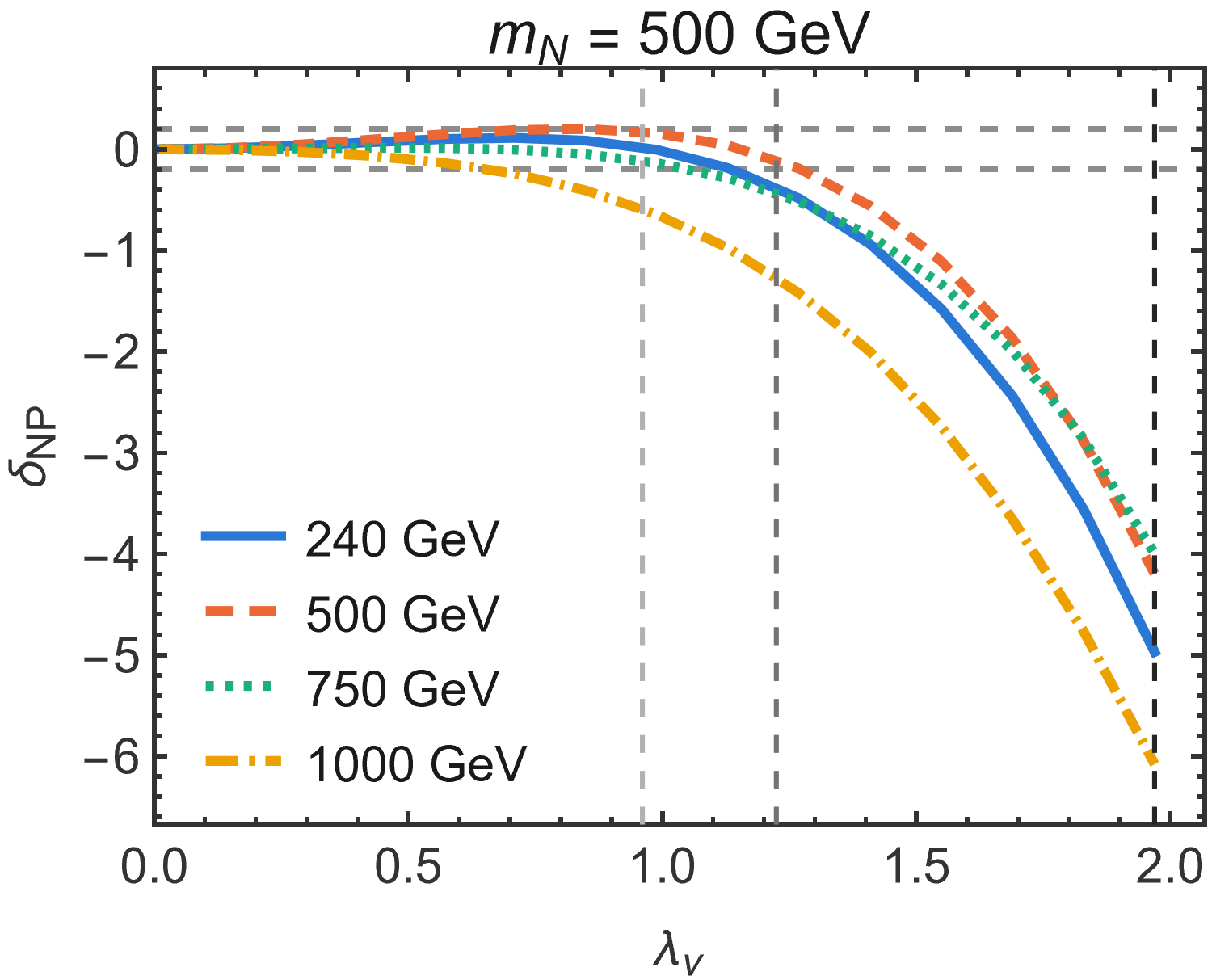}
\end{subfigure}
\\[6pt]
\begin{subfigure}[b]{0.45\textwidth}
    \centering
    \includegraphics[width=\textwidth]{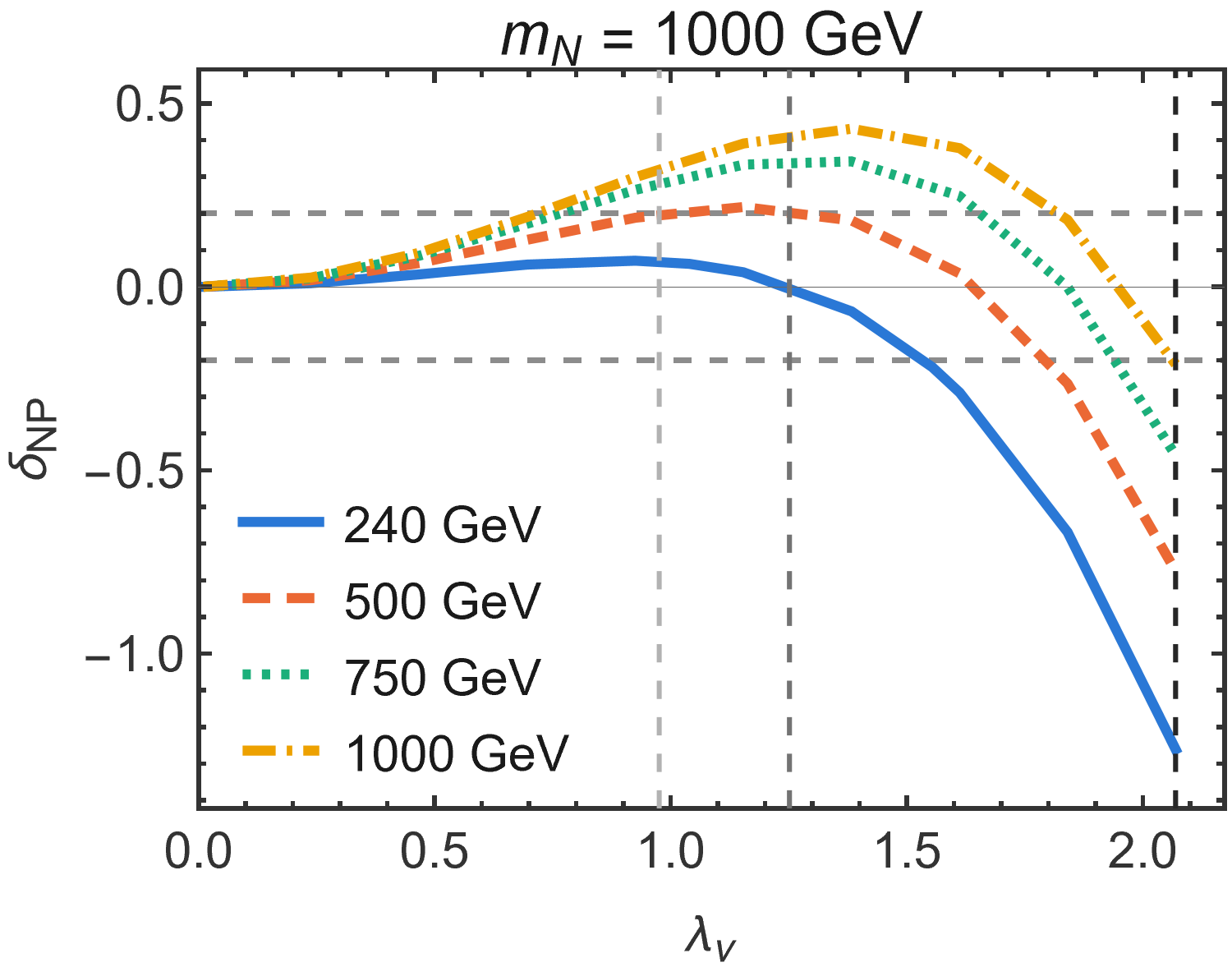}
\end{subfigure}
\hfill
\begin{subfigure}[b]{0.45\textwidth}
    \centering
    \includegraphics[width=\textwidth]{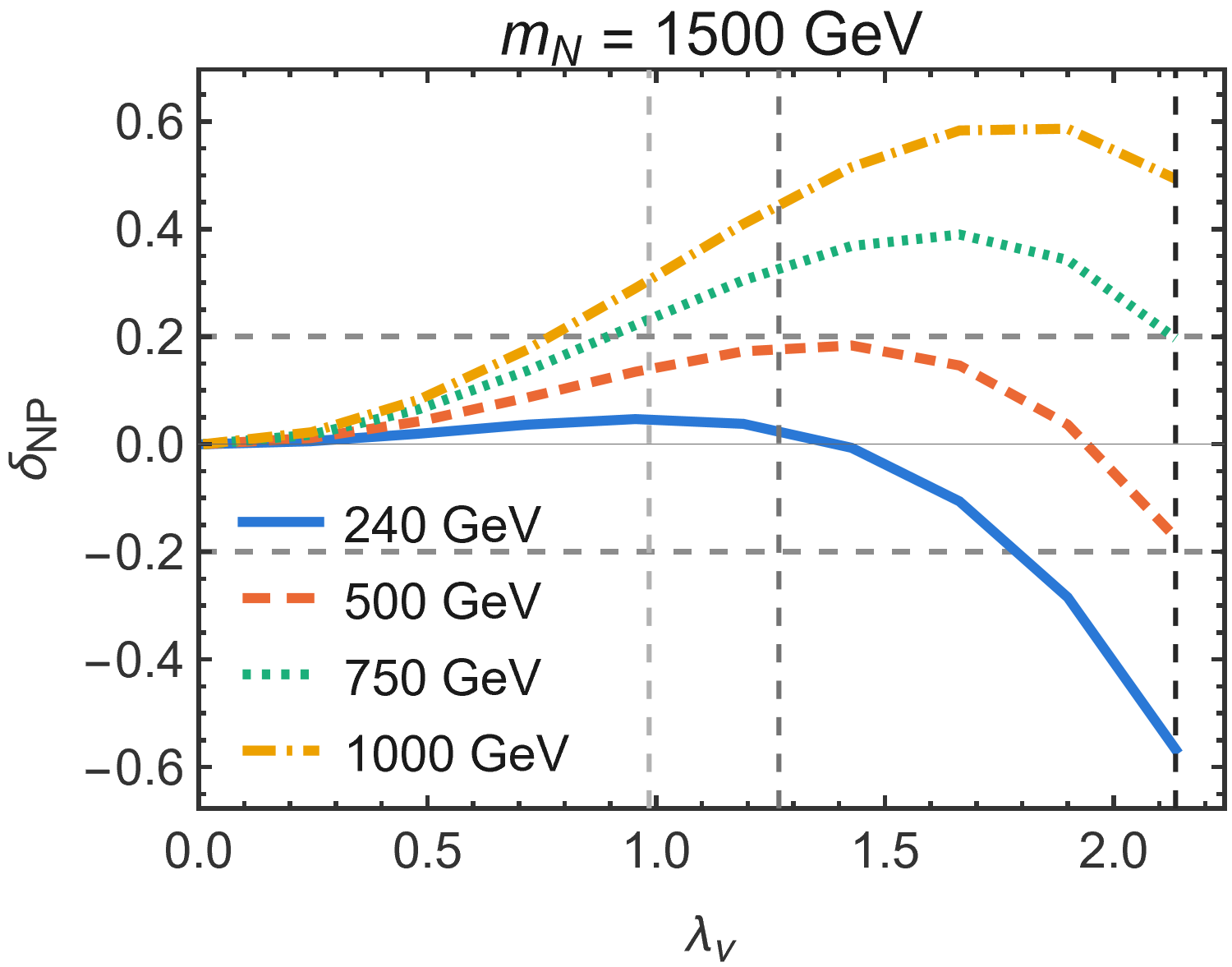}
\end{subfigure}
\caption{Fractional Higgs-strahlung cross-section shift $\delta_{\rm NP}$ for Model~\MIII as a function of $\lambda_\nu$, at fixed $m_N$ and for different $\sqrt{s}$. The vertical dashed lines mark the perturbativity ceiling $\lambda_\nu^{\rm max}$ at the benchmark cutoffs $\Lambda_1 \, , \Lambda_2 \, ,\Lambda_3$ (darkest to lightest). The horizontal dashed lines mark $\delta_{\rm NP}=\pm0.2\%$ for reference.}
\label{fig:results-MIII}
\end{figure}

Figure~\ref{fig:results-MIII-scan-masses} shows $\delta_{\rm NP}$ as a function of $\lambda_\nu$ at $\sqrt{s}=240$~GeV for the fixed values of $m_N$ presented in Table~\ref{table:perturbativity-merged}, and Fig.~\ref{fig:results-MIII} extends this to four values of $\sqrt{s}$ per mass, together with the perturbativity bounds $\lambda_\nu^{\rm max}(\Lambda_i)$ of Sec.~\ref{subsec:perturbativity} as vertical reference lines.

At $\sqrt{s}=240$~GeV, $\delta_{\rm NP}$ decreases monotonically with $\lambda_\nu$ for every $m_N$ shown in Fig.~\ref{fig:results-MIII-scan-masses}. The heavier $m_N=1000$ and $1500$~GeV curves extend to larger $\lambda_\nu$ before terminating but settle at smaller magnitudes than the lighter benchmarks, since a larger perturbativity limit does not by itself translate into a larger cross-section shift once the mass-suppression of the mixing angle is accounted for.

In every panel of Fig.~\ref{fig:results-MIII}, the value of $\delta_{\rm NP}$ in each curve rises briefly above zero at small-to-moderate $\lambda_\nu$ before turning negative, a positive excursion that grows more pronounced from the lightest to the heaviest panel. This hints to a possible small-$\lambda_\nu$ cancellation between the self-energy and penguin contributions, similar to the partial cancellation already seen for some Model~\MI benchmarks. The dependence on $m_N$ changes the shape of the curves more than their overall scale: the heavier panels reach a larger positive peak but a comparatively milder negative value at the edge of their perturbativity-allowed range, while the lightest panel shows almost no positive excursion and instead falls to the most negative value of the four. This follows from the perturbativity ceiling $\lambda_\nu^{\rm max}(\Lambda_i)$ itself growing with $m_N$ (Table~\ref{table:perturbativity-merged}): the heavier panels admit much larger $\lambda_\nu$ before the scan cuts off, so each panel effectively probes a different portion of the same underlying curve shape. As discussed above, the points reached at the upper end of these curves lie far outside the EW-allowed window on $\lambda_\nu$: they illustrate the loop sensitivity of $\delta_{\rm NP}$ rather than a realistic prediction.

\subsection{Model~\MIV: charged VL singlet with $\tau$ mixing}\label{subsec:results-MIV}

The two independent inputs are the VL charged-lepton mass $m_E$ and the mixing Yukawa $\lambda_e$; the SM $\tau$ Yukawa $Y_e$ is fixed by $m_\tau$. We scan
\begin{equation}
    m_E \;\in\; [200,\;1500]~\text{GeV}\,,\qquad
    \lambda_e \;\in\; \bigl[0,\;\lambda_e^{\rm max}(\Lambda_i)\bigr]\,,
    \label{eq:scan-MIV}
\end{equation}
up to the perturbativity ceiling of Sec.~\ref{subsec:perturbativity}, without imposing the mixing-angle bound $\theta_L\leq 0.045$ [Eq.~\eqref{eq:model4_tan2thetaL}] on $\lambda_e$, for the same reason as in Model~\MIII.

Model~\MIV is the charged-lepton counterpart of Model~\MIII. As in Model~\MIII, $\lambda_e$ is constrained from above by the left-handed mixing-angle bound $\theta_L\leq0.045$, by perturbativity, and by the kinematic reality condition [Eq.~\eqref{eq:model4_lambda_bound}], with the mixing-angle bound again the strictest of the three by roughly an order of magnitude (Sec.~\ref{subsec:perturbativity}). The same caveat therefore applies here, and we scan $\lambda_e$ up to the perturbativity ceiling to expose the size and shape of the loop effect rather than a realistic prediction. 

\begin{figure}[h]
\centering
\includegraphics[width=0.5\textwidth]{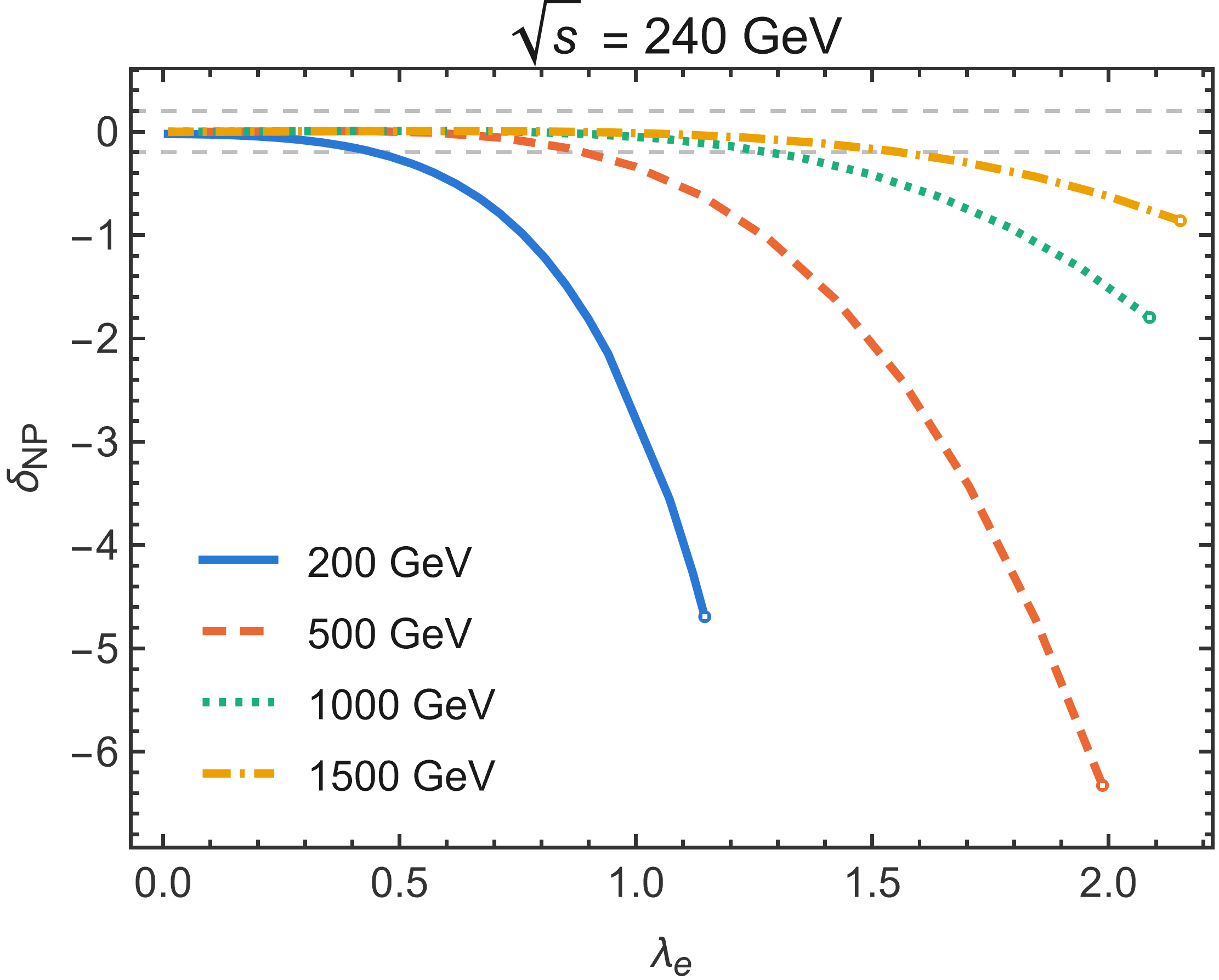}
\caption{Fractional Higgs-strahlung cross-section shift $\delta_{\rm NP}$ for Model~\MIV as a function of $\lambda_e$ at $\sqrt{s}=240$~GeV, for fixed values of $m_E$.}
\label{fig:results-MIV-scan-masses}
\end{figure}

\begin{figure}[h]
\centering
\begin{subfigure}[b]{0.48\textwidth}
    \centering
    \includegraphics[width=\textwidth]{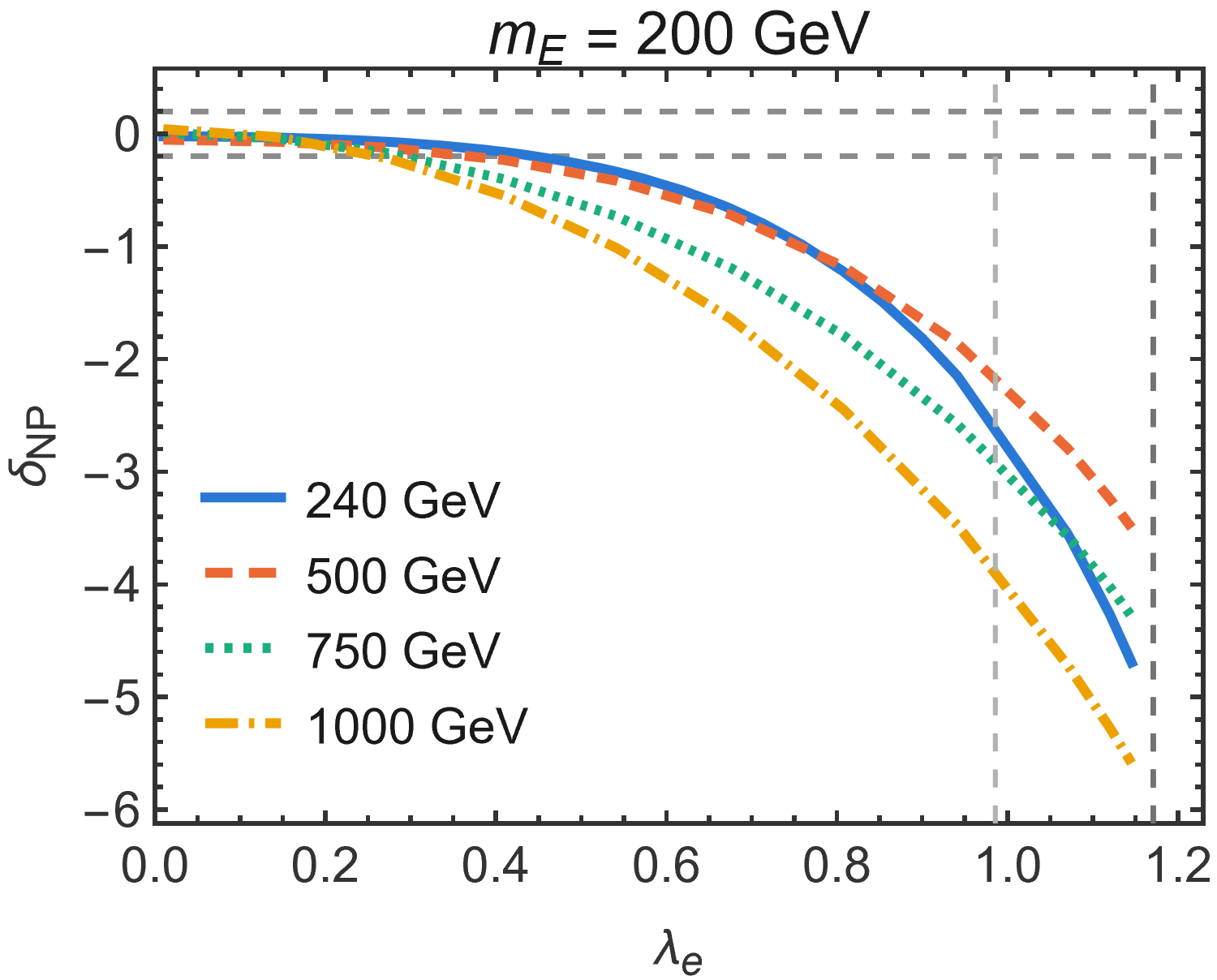}
\end{subfigure}
\hfill
\begin{subfigure}[b]{0.48\textwidth}
    \centering
    \includegraphics[width=\textwidth]{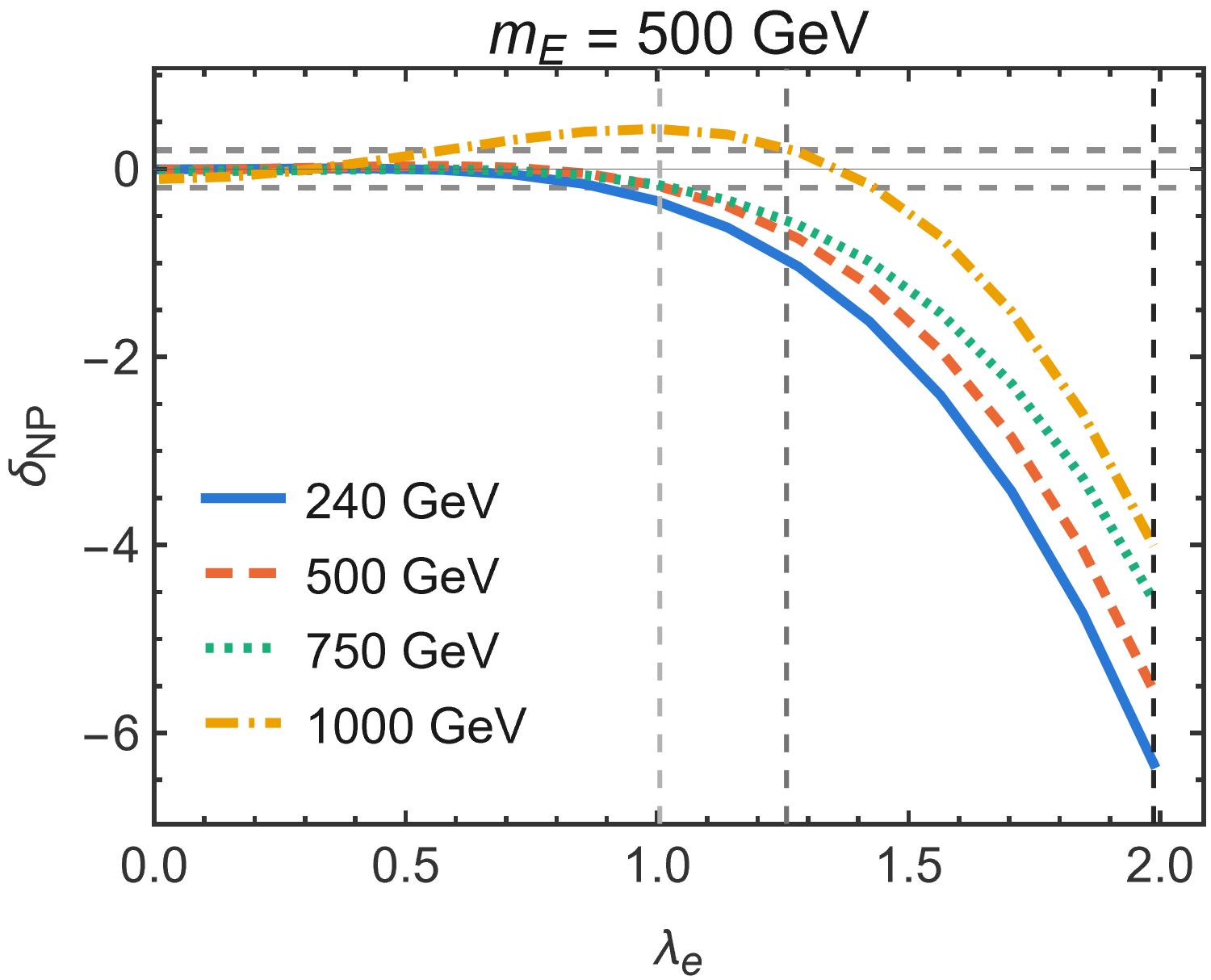}
\end{subfigure}
\\[6pt]
\begin{subfigure}[b]{0.48\textwidth}
    \centering
    \includegraphics[width=\textwidth]{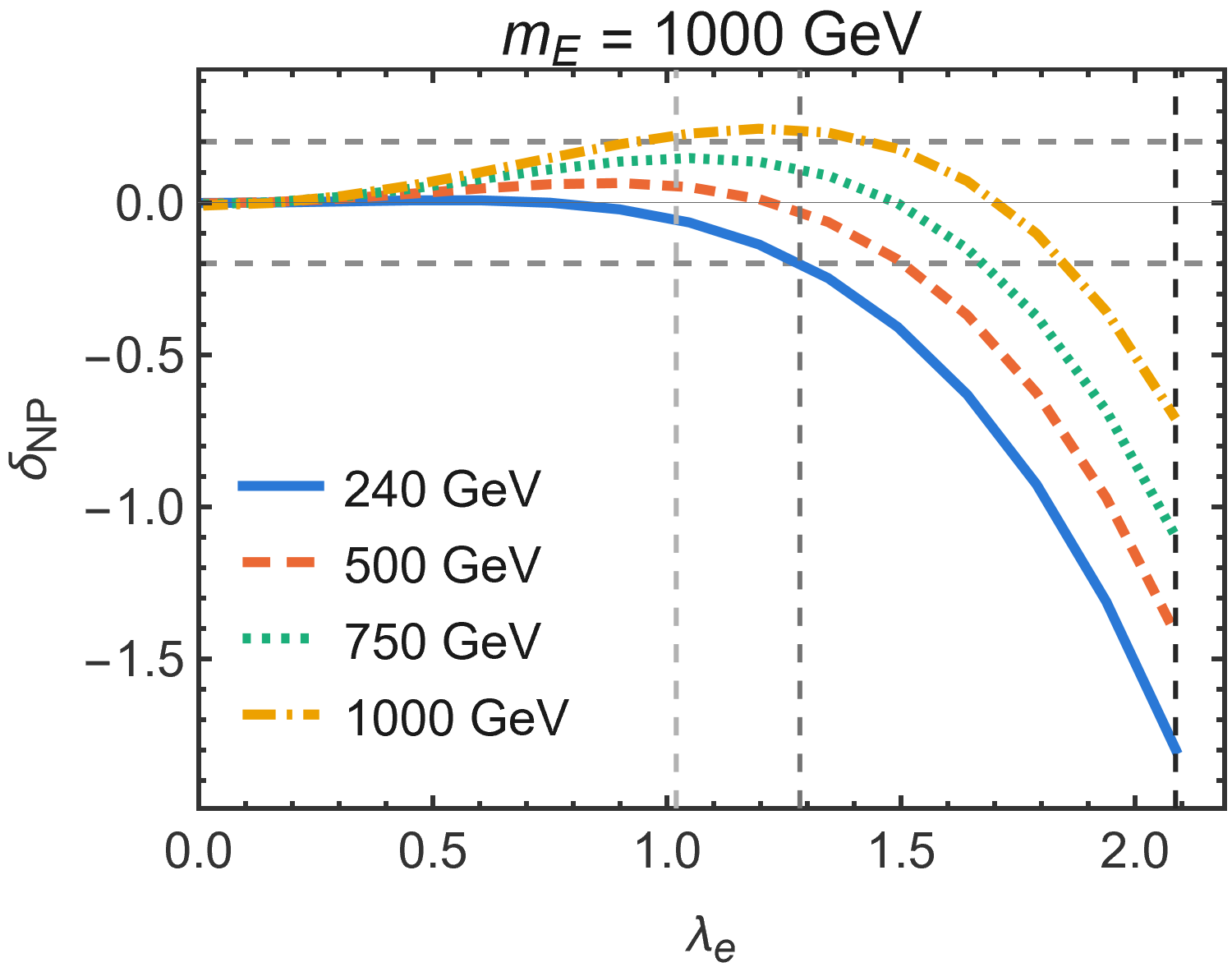}
\end{subfigure}
\hfill
\begin{subfigure}[b]{0.48\textwidth}
    \centering
    \includegraphics[width=\textwidth]{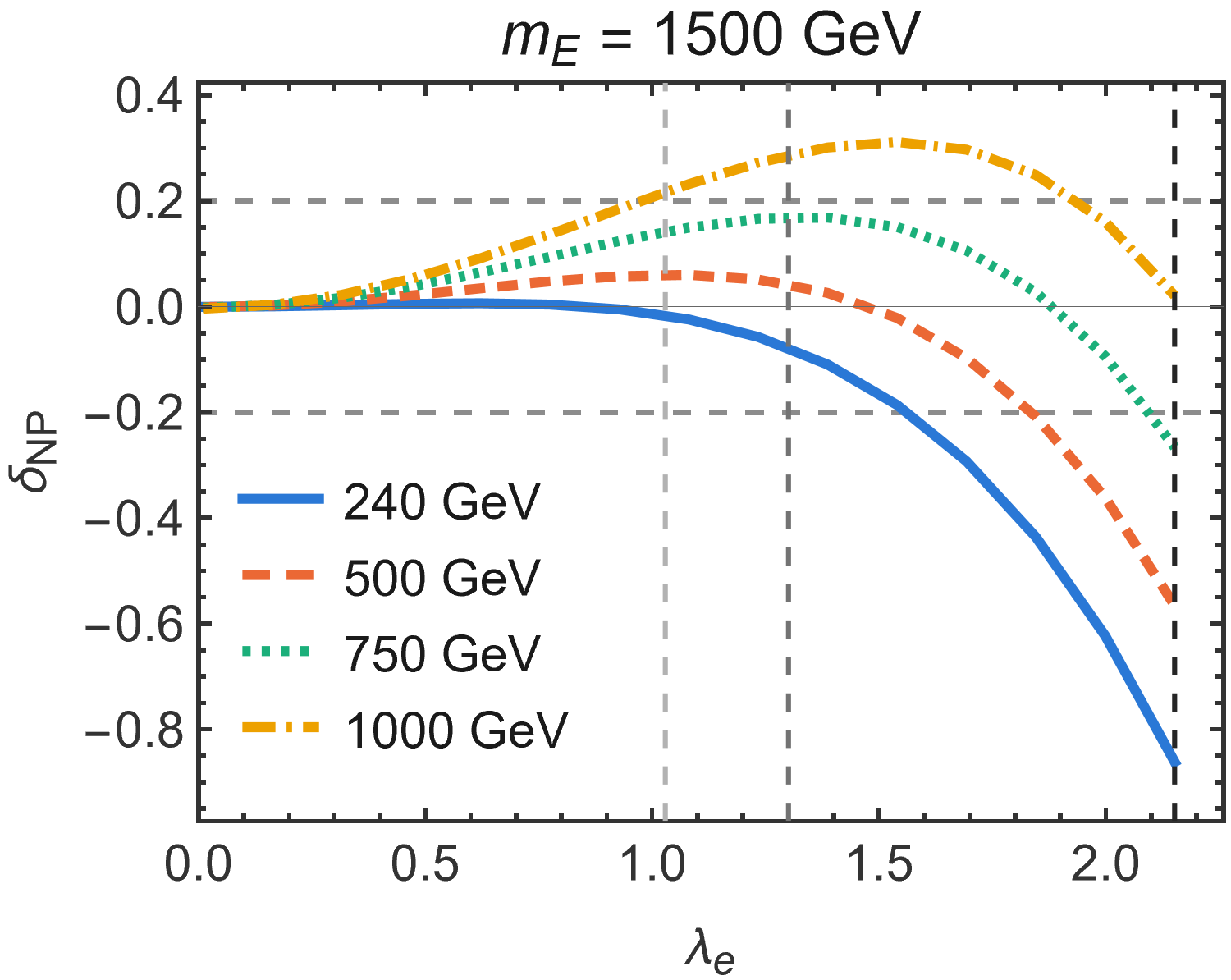}
\end{subfigure}
\caption{Fractional Higgs-strahlung cross-section shift $\delta_{\rm NP}$ for Model~\MIV as a function of $\lambda_e$, at fixed $m_E$ and for different $\sqrt{s}$. The vertical dashed lines mark the perturbativity ceiling $\lambda_e^{\rm max}$ at the benchmark cutoffs $\Lambda_1 \, , \Lambda_2 \, ,\Lambda_3$ (darkest to lightest). The horizontal dashed lines mark $\delta_{\rm NP}=\pm0.2\%$ for reference.}
\label{fig:results-MIV-scan-energy}
\end{figure}

Figure~\ref{fig:results-MIV-scan-masses} shows $\delta_{\rm NP}(\lambda_e)$ at $\sqrt{s}=240$~GeV for four fixed values of $m_E$, and Fig.~\ref{fig:results-MIV-scan-energy} extends this to four values of $\sqrt{s}$ per mass, mirroring Fig.~\ref{fig:results-MIII} for Model~\MIII.

At $\sqrt{s}=240$~GeV, $\delta_{\rm NP}$ decreases monotonically with $\lambda_e$ for every $m_E$ shown in Fig.~\ref{fig:results-MIV-scan-masses}, as in Model~\MIII (Fig.~\ref{fig:results-MIII-scan-masses}), and consistent with $\theta_L$'s bound on $\lambda_e$ the same way $\theta_\nu$ bounds $\lambda_\nu$ there [Eq.~\eqref{eq:model4_tan2thetaL}]. The magnitude reached at the edge of the perturbativity-allowed range does not track $m_E$ monotonically, however: it is largest for the $m_E=500$~GeV benchmark and smaller at both the lighter ($200$~GeV) and heavier ($1000$ and $1500$~GeV) ends of the scanned range, the same qualitative pattern already seen in Model~\MIII's single-panel figure. This reflects the same tension as there: a larger perturbativity ceiling lets the heavier benchmarks reach bigger $\lambda_e$ before the scan cuts off, but the mass suppression built into the $m_E$-dependence of the mixing angle works against a correspondingly larger cross-section shift, so the two effects only partially compensate.

Beyond $\sqrt{s}=240$~GeV, Fig.~\ref{fig:results-MIV-scan-energy} shows that Model~\MIV departs qualitatively from Model~\MIII's purely monotonic behavior [Fig.~\ref{fig:results-MIII}]. For $m_E=200$~GeV the four curves decrease monotonically with $\lambda_e$, differing from Model~\MIII mainly in that the ordering by $\sqrt{s}$ is not monotonic itself (the $\sqrt{s}=1000$~GeV curve is not always the extreme one). From $m_E=500$~GeV upward, the highest-$\sqrt{s}$ curve begins to turn \emph{positive} at intermediate $\lambda_e$ before eventually declining again, an excursion that grows larger and extends to more curves (the $750$~GeV curve as well as $1000$~GeV) as $m_E$ increases to $1000$ and especially $1500$~GeV, echoing the mass-growing positive excursion already noted for Model~\MIII.  As in Model~\MIII, the perturbativity limit $\lambda_e^{\rm max}(\Lambda_i)$ growing with $m_E$ (Table~\ref{table:perturbativity-merged}) lets the heavier panels probe further along the same underlying curve shape, and the points reached at the upper end of these curves again lie far outside the EW-allowed window on $\lambda_e$: they illustrate the loop sensitivity of $\delta_{\rm NP}$ rather than a realistic prediction.

\section{Conclusions}\label{sec:conc}
In this work we have profiled the imprint of one-loop corrections generated by VLFs on the Higgs-strahlung process. This is a necessary theoretical requirement in order to match the expected 
precision of future $e^+e^-$ Higgs factories, such as FCC-ee, and provide a complementary angle to the direct search program. VLFs are among the most economical and best-motivated extensions of the SM, subject to recent direct searches at the LHC that have constrained their masses but left the associated Yukawa sector comparatively unconstrained. 
In this paper we have presented  a possible route to pinpoint signatures of such loosely constrained parameter space of VLFs. 

We organized the general renormalizable Lagrangian for one VLQ doublet, one VLL doublet, and their singlets (Sec.~\ref{sec:the-model}) into four benchmark submodels chosen to isolate qualitatively distinct sources of NP. Models~\MI and~\MII decouple the VL sector from the SM fermions entirely and probe the sensitivity of the cross-section to the internal VL Yukawa couplings $y_U,y_D$ (quarks) and $y_E,y_N$ (leptons) alone. Models~\MIII and~\MIV instead switch off this internal coupling and turn on a single mixing coupling, $\lambda_\nu$ or $\lambda_e$, between a VL singlet and the third-generation SM lepton, so that SM and NP states propagate together in the loop. In all four cases we worked in the OS renormalization scheme (Sec.~\ref{sec:cross-section-calculation}) and imposed a perturbativity ceiling on the relevant couplings under RG running up to selected UV cutoffs $\Lambda_{1,2,3}$, computed with \texttt{RGBeta}~\cite{Thomsen:2021ncy} (Sec.~\ref{subsec:perturbativity}). Throughout, we neglected CP violation and took the VL Yukawa couplings to be real and equal for the two chiralities, following Ref.~\cite{Adhikary:2024esf}. We also restricted the mixing of Models~\MIII and~\MIV to the third generation and did not attempt to address neutrino masses, treating the SM neutrinos as massless throughout, so the mixing introduced in Model~\MIII is unrelated to any neutrino-mass mechanism. 

Beyond $\sigma(ZH)$ itself, we used the same one-loop building blocks to compute the oblique parameters $S$ and $T$ and the loop-induced $H\to\gamma\gamma$ rate, as cross-checks against, respectively, electroweak-precision data and the LHC diphoton measurements. In every benchmark model the $S$ and $T$ fit excludes only points already flagged as excluded on other grounds, and the diphoton rate remains within its current experimental range. Neither observable provides a constraint beyond those already imposed, so our discussion of the numerical results (Sec.~\ref{sec:numerical-analysis-and-results}) focused on the genuine NP shift $\delta_{\rm NP}$ of the Higgsstrahlung cross-section itself.

The results reveal a clear asymmetry between the Yukawa-driven and mixing-driven classes of models. In Models~\MI and~\MII, where the only handle is the internal VL Yukawa coupling, the mass dependence of $\delta_{\rm NP}$ dominates over the coupling dependence. A sizable fraction of the scanned parameter space, including points that satisfy the current LHC pair-production and multilepton bounds, produces shifts at or above our reference sensitivity of $\mathcal{O}(10^{-3})$. Across the benchmark points, the vertex contribution typically dominates over the self-energy contribution, in some cases by an order of magnitude or more. The two contributions are comparable in size only at the lowest mass-splitting benchmark point in each model, where the self-energy term is instead the larger of the two, and their relative sign can also flip between benchmarks rather than one diagram dominating with a fixed sign throughout.

Models~\MIII and~\MIV, by contrast, are already essentially excluded as targets for this measurement. The electroweak-precision bounds on the $\tau$-flavored mixing angles $\theta_\nu$ and $\theta_L$ restrict $\lambda_\nu$ and $\lambda_e$ to values so small that the resulting
shift never approaches the per-mille level, irrespective of the VL mass. We nonetheless scanned both couplings up to the perturbativity ceiling, beyond the region allowed by the mixing-angle bounds, to expose the underlying loop dynamics and to allow for extensions that could relax such bound without affecting our Higgs-strahlung signal~\cite{deBlas:2025pco}.

Several directions could extend this work. First, we could introduce CP phases, here neglected due to our assumption of real couplings, and consequently broaden the signatures available, enriching our radiative profiling. Secondly, a scalar extension could bridge the visible sector of the parameter space to a parallel signature of first order phase transition. We intend to explore these avenues in future studies. 

\acknowledgments
The authors are supported by the Estonian Research Council grants
TARISTU24-TK10, TARISTU24-TK3, and the CoE grant TK202 ``Foundations of the Universe''. C.~M. acknowledges support by the Estonian Research Council grant PRG1677.

%\newpage 

\appendix
	
\section{On-shell renormalization with VL states}\label{sec:1-loop-renormalization}
While a standard OS approach was adopted to reorganize the one-loop divergent quantities into finite observables, 
we have found it useful to illustrate the main steps and the conventions used for the Higgs-strahlung's determination. 
This has the dual purpose of guiding the reader through the counterterms listed in the related ancillary files as well as 
to track the implicit dependence on NP parameters that such counterterms introduce.   

By studying a process with SM external states at one-loop order, we can restrict the renormalization procedure to the EW sector and ignore the renormalization 
of NP parameters. In practice, we trade the set of Lagrangian
parameters of the gauge ($g_1, g_w$), Yukawa ($Y_u, Y_d, Y_{e}$) and scalar ($v, \lambda, \mu_H^2$) sectors for one more directly linked to the masses of EW bosons
and fermions together with their universal electromagnetic interaction (${\rm e}$) and the Higgs tree-level tadpole ($t$): 
\begin{align} \label{varSM}
&x \in \left[m^2_Z, m^2_W, m^2_H, m_{u_i}, m_{d_i}, m_{l=e,\mu,\tau}, {\rm e}, t  \right] \, . 
\end{align}
Counterterms are then generated by the splitting $x \rightarrow x_{r} + \delta x_{\rm UV}$ and demanding $\delta x_{\rm UV}$ 
to absorb the regulator dependence, while leaving the amplitude a function of the finite coupling $x_r$. Applying this splitting to the set in Eq.~(\ref{varSM})   
\begin{align} \label{CoRen}
& m^2_Z \rightarrow m^2_Z  + \delta m^2_Z, \,\, m^2_W \rightarrow m^2_W  + \delta m^2_W,  \\
& m_f \rightarrow m_f  + \delta m_f, \quad (f = u,d,l)  \\
& m^2_H \rightarrow m^2_H  + \delta m^2_H, \,\, t \rightarrow t  + \delta t \, ,  \\
& {\rm e} \rightarrow {\rm e} \left(1 + \delta {\rm e} \right) \,,
\end{align}
we can find the Feynman rules of counterterms by expanding their OS definitions
\begin{equation}
\begin{aligned}
g_1 &= {\rm e}\,\frac{m_Z}{m_W}, \quad
g_w = {\rm e}\,\sqrt{\frac{m_Z^2}{m_Z^2-m_W^2}}, \quad
\lambda = \frac{{\rm e}^2 m_H^2 m_Z^2}{2 m_W^2(m_Z^2-m_W^2)},\\
v &= \frac{2}{{\rm e}}\sqrt{\frac{m_W^2(m_Z^2-m_W^2)}{m_Z^2}}, \quad
\mu_H^2 = \frac{m_H^2}{2}+\frac{t}{v}, \quad m_e = \frac{v}{\sqrt{2}} Y_e .
\end{aligned}
\label{eq:OSparameters}
\end{equation}
Notice that the last identity, which we have conveniently restricted to the electrons involved in the Higgs-strahlung process, cannot be generalized to 
all SM fermions when mixing is present, as in the case of Models~\MIII and~\MIV. 
The arbitrariness in the renormalization scale is then removed, in the OS scheme, by identifying the set $m_Z,m_W,m_H,m_f$ with the pole of the corresponding 
propagator at all orders, and by having the QED coupling ${\rm e}$ renormalize the electron-photon scattering in the Thomson limit.
Finally, the tadpole parameter $t$ is fixed by enforcing the OS relation $m^2_H = 2 \mu_H^2$ from the more general one in Eq.(\ref{eq:OSparameters}).
On top of the OS counterterms, our ancillary files also list the counterterms $\{\delta_W, \delta_Z, \delta_A, \delta_H, \delta_{ZA}, \delta_{AZ} \}$ generated
by the field rescaling
\begin{align} \label{FiRen}
& W_{\mu} \rightarrow \left( 1 + \frac{\delta_W}{2}\right)W_{\mu}  ,  \\
& Z_{\mu} \rightarrow \left( 1 + \frac{\delta_Z}{2}\right)Z_{\mu} + \frac{1}{2}\delta_{ZA} A_{\mu},  \\ 
& A_{\mu} \rightarrow\left( 1 + \frac{\delta_A}{2}\right)A_{\mu} + \frac{1}{2}\delta_{AZ} Z_{\mu},  \\ 
& \Phi \rightarrow \left( 1 + \frac{\delta_H}{2}\right)\Phi , \quad (\Phi = H, G^0, G^{\pm})  \\
& \psi \rightarrow \left( 1 + \delta_{\psi}\right)\psi, \quad (\psi = e_L, e_R, \nu_L) \, ,
\end{align}
%%%%
and which are needed to secure the finiteness of the (off-shell) amplitude.
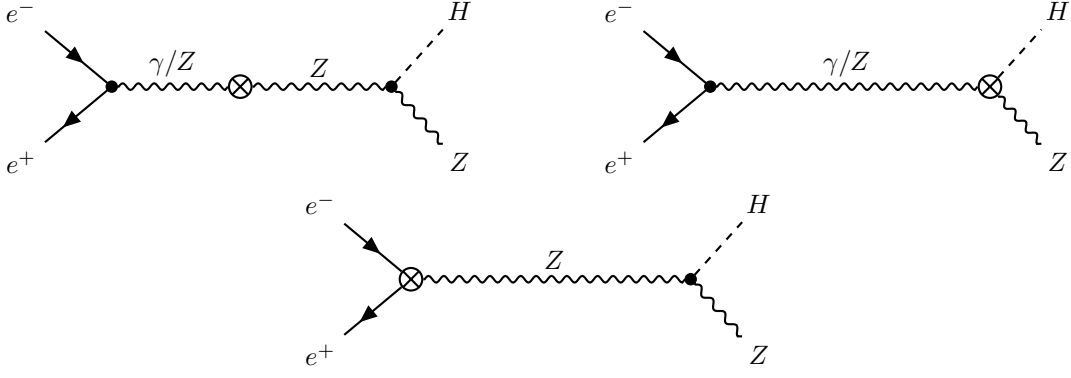
\begin{figure}[t]
\centering

\tikzset{
  counterterm/.style={
    draw, thick, circle, fill=white, minimum size=8pt, inner sep=3pt,
    path picture={
      \draw[thick]
        (path picture bounding box.north west) -- (path picture bounding box.south east)
        (path picture bounding box.north east) -- (path picture bounding box.south west);
    }
  }
}

\tikzset{
  big cross/.style={
    % 1. Create a baseline transparent circle for anchoring the lines
    circle, minimum size=4mm, inner sep=0pt,
    path picture={
      % 2. Draw a thick white circle backdrop to hide the propagator lines
      \fill[white] (path picture bounding box.center) circle (3mm);
      % 3. Draw the thick heavy cross lines on top
      \draw[line width=1.5pt, black] 
        (path picture bounding box.south west) -- (path picture bounding box.north east);
      \draw[line width=1.5pt, black] 
        (path picture bounding box.north west) -- (path picture bounding box.south east);
    }
  }
}

\begin{tikzpicture}

  \begin{feynman}[every dot={/tikz/fill=black}]
 
    \vertex (i1) at (-2.6,  1.0) {\(e^-\)};
    \vertex (i2) at (-2.6, -1.0) {\(e^+\)};
    \vertex (f1) at ( 3.2,  1.0) {\(H\)};
    \vertex (f2) at ( 3.2, -1.0) {\(Z\)};
 
    \vertex [dot] (va) at (-1.4, 0) {};   % ee Z* vertex  (was -1.8)
    \vertex [counterterm] (v1) at ( 0.3, 0) {};   % left  self-energy vertex (unchanged)
    %\vertex [dot] (v2) at ( 2.3, 0) {};   % right self-energy vertex (unchanged)
    \vertex [dot] (vb) at ( 2.3, 0) {};   % ZZH vertex     (was  4.1)
 
          \diagram* {
      %--- Incoming leptons
      (i1) -- [fermion, thick]                             (va),
      (i2) -- [anti fermion, thick]                        (va),
 
      %--- Off-shell Z from ee vertex to self-energy
      (va) -- [boson, thick, edge label=\(\gamma / Z\)]           (v1),
 
      %--- Off-shell Z from self-energy to production vertex
      (v1) -- [boson, thick, edge label=\(Z\)]           (vb),
 
      %--- Outgoing Higgs (dashed) and Z (wavy)
      (vb) -- [scalar, thick]            (f1),
      (vb) -- [boson, thick]            (f2),
    };

  \end{feynman}
\end{tikzpicture}
\qquad \qquad
\begin{tikzpicture}

  \begin{feynman}[every dot={/tikz/fill=black}]

  \vertex (i1) at (-2.6,  1.0) {\(e^-\)};
    \vertex (i2) at (-2.6, -1.0) {\(e^+\)};
    \vertex (f1) at ( 3.2,  1.0) {\(H\)};
    \vertex (f2) at ( 3.2, -1.0) {\(Z\)};
 
    \vertex [dot] (va) at (-1.4, 0) {};   % ee Z* vertex  (was -1.8)
  % \vertex [big cross] (v1) at ( 0.3, 0) {};   % left  self-energy vertex (unchanged)
    %\vertex [dot] (v2) at ( 2.3, 0) {};   % right self-energy vertex (unchanged)
    \vertex [counterterm] (vb) at ( 2.3, 0) {};   % ZZH vertex     (was  4.1)
 
    \diagram* {
      %--- Incoming leptons
      (i1) -- [fermion, thick]                             (va),
      (i2) -- [anti fermion, thick]                        (va),
 
      %--- Off-shell Z from ee vertex to self-energy
      (va) -- [boson, thick, edge label=\(\gamma / Z\)]           (vb),
 
      %--- Off-shell Z from self-energy to production vertex
 
      %--- Outgoing Higgs (dashed) and Z (wavy)
      (vb) -- [scalar, thick]            (f1),
      (vb) -- [boson, thick]            (f2),
    };

  \end{feynman}
\end{tikzpicture}
\begin{tikzpicture}

  \begin{feynman}[every dot={/tikz/fill=black}]

  \vertex (i1) at (-2.6,  1.0) {\(e^-\)};
    \vertex (i2) at (-2.6, -1.0) {\(e^+\)};
    \vertex (f1) at ( 3.2,  1.0) {\(H\)};
    \vertex (f2) at ( 3.2, -1.0) {\(Z\)};
 
    \vertex [counterterm] (va) at (-1.4, 0) {};   % ee Z* vertex  (was -1.8)
  % \vertex [big cross] (v1) at ( 0.3, 0) {};   % left  self-energy vertex (unchanged)
    %\vertex [dot] (v2) at ( 2.3, 0) {};   % right self-energy vertex (unchanged)
    \vertex [dot] (vb) at ( 2.3, 0) {};   % ZZH vertex     (was  4.1)
 
    \diagram* {
      %--- Incoming leptons
      (i1) -- [fermion, thick]                             (va),
      (i2) -- [anti fermion, thick]                        (va),
 
      %--- Off-shell Z from ee vertex to self-energy
      (va) -- [boson, thick, edge label=\(Z\)]           (vb),
 
      %--- Off-shell Z from self-energy to production vertex
 
      %--- Outgoing Higgs (dashed) and Z (wavy)
      (vb) -- [scalar, thick]            (f1),
      (vb) -- [boson, thick]            (f2),
    };

  \end{feynman}
\end{tikzpicture}
\caption{\small Counterterm diagrams contributing to the renormalization of $e^+ e^- \rightarrow ZH$.}
\label{fig:VLNLOct}
\end{figure}
The primitive vertices and self-energy counterterms enter the renormalization of the Higgs-strahlung amplitude in the form of propagator and vertex corrections, as illustrated in Fig.~\ref{fig:VLNLOct}. 
Notice how, in the OS scheme adopted, the counterterm for vertices is often, and somewhat counter-intuitively, bundled into self-energy diagrams. This is a consequence of the fact that the OS renormalization conditions are mainly imposed, with just the exceptions of electron-photon scattering and Higgs vacuum renormalization,
 on the propagator poles. This is particularly noticeable in the $ZZH$ vertex counterterm, which has the following structure

 \begin{align}
  \ctVZH \;&=\;
  \frac{i e\,g^{\mu\nu}}
       {2\,m_W^3\,\sqrt{1-m_W^2/m_Z^2}\;\bigl(m_W^2-m_Z^2\bigr)}
  \Bigl[\,\delta m_Z^2\,m_W^2\bigl(3m_W^2-2m_Z^2\bigr) \nonumber\\[0.5ex]
  &\hphantom{=\;}
  +\bigl(2\delta_e+\delta Z_{H}+2\delta_Z \bigr)m_W^2m_Z^2\bigl(m_W^2-m_Z^2\bigr)
  \nonumber\\[0.5ex]
  &\hphantom{=\;}
  +\;\delta m_W^2\,m_Z^2\bigl(m_Z^2-2m_W^2\bigr)\Bigr] \,  ,
\end{align}
signaling the implicit dependence on the renormalization of the $Z$ and $W$ boson masses and the large interdependence among physical parameters brought by EW spontaneous symmetry breaking.

As in \cite{Marzo:2022nrw,Marzo:2023uxv}, we have relied on \texttt{FeynRules} \cite{Christensen:2008py,Alloul:2013bka} supplemented by custom routines in order to automatically generate the counterterms.

\section{Renormalization group equations}\label{sec:RGE}
The one-loop $\beta$-functions quoted below were obtained with the \texttt{RGBeta} package \cite{Thomsen:2021ncy} for the renormalizable Lagrangian of Sec.~\ref{sec:the-model}. We use the standard convention $16\pi^2\,\mu\,dC/d\mu = 16\pi^2\,\beta_C$ for a generic coupling $C$, with the gauge coupling $g_1$ GUT-normalized ($g_1=\sqrt{5/3}\,g'$). In all four submodels the gauge couplings and the top Yukawa $y_t$ run alongside the new-physics coupling(s) introduced by that submodel; we list the three gauge $\beta$-functions, then $\beta_{y_t}$, then the new Yukawa coupling(s), in the same order for every model.

\subsection*{Model~\texorpdfstring{\MI}{I}}

\begin{align}
16\pi^2\beta_{g_1} &= \frac{167}{30}\,g_1^3 \, , \label{eq:beta-g1-MI} \\
16\pi^2\beta_{g_2} &= -\frac{7}{6}\,g_2^3 \, , \label{eq:beta-g2-MI} \\
16\pi^2\beta_{g_3} &= -\frac{13}{3}\,g_3^3 \, , \label{eq:beta-g3-MI} \\
16\pi^2\beta_{y_t} &= y_t\left[-\frac{17}{20}g_1^2-\frac{9}{4}g_2^2-8g_3^2+\frac{9}{2}y_t^2+6\left(y_D^2+y_U^2\right)\right] \, , \label{eq:beta-yt-MI} \\
16\pi^2\beta_{y_U} &= y_U\left[-\frac{17}{20}g_1^2-\frac{9}{4}g_2^2-8g_3^2+3y_t^2+\frac{9}{2}y_D^2+\frac{15}{2}y_U^2\right] \, , \label{eq:beta-yU-MI} \\
16\pi^2\beta_{y_D} &= y_D\left[-\frac{1}{4}g_1^2-\frac{9}{4}g_2^2-8g_3^2+3y_t^2+\frac{15}{2}y_D^2+\frac{9}{2}y_U^2\right] \, . \label{eq:beta-yD-MI}
\end{align}

\subsection*{Model~\texorpdfstring{\MII}{II}}

\begin{align}
16\pi^2\beta_{g_1} &= \frac{53}{10}\,g_1^3 \, , \label{eq:beta-g1-MII} \\
16\pi^2\beta_{g_2} &= -\frac{5}{2}\,g_2^3 \, , \label{eq:beta-g2-MII} \\
16\pi^2\beta_{g_3} &= -7\,g_3^3 \, , \label{eq:beta-g3-MII} \\
16\pi^2\beta_{y_t} &= y_t\left[-\frac{17}{20}g_1^2-\frac{9}{4}g_2^2-8g_3^2+\frac{9}{2}y_t^2+2\left(y_E^2+y_N^2\right)\right] \, , \label{eq:beta-yt-MII} \\
16\pi^2\beta_{y_E} &= y_E\left[-\frac{9}{4}g_1^2-\frac{9}{4}g_2^2+3y_t^2+\frac{7}{2}y_E^2+\frac{1}{2}y_N^2\right] \, , \label{eq:beta-yE-MII} \\
16\pi^2\beta_{y_N} &= y_N\left[-\frac{9}{20}g_1^2-\frac{9}{4}g_2^2+3y_t^2+\frac{1}{2}y_E^2+\frac{7}{2}y_N^2\right] \, . \label{eq:beta-yN-MII}
\end{align}

\subsection*{Model~\texorpdfstring{\MIII}{III}}

\begin{align}
16\pi^2\beta_{g_1} &= \frac{41}{10}\,g_1^3 \, , \label{eq:beta-g1-MIII} \\
16\pi^2\beta_{g_2} &= -\frac{19}{6}\,g_2^3 \, , \label{eq:beta-g2-MIII} \\
16\pi^2\beta_{g_3} &= -7\,g_3^3 \, , \label{eq:beta-g3-MIII} \\
16\pi^2\beta_{y_t} &= y_t\left[-\frac{17}{20}g_1^2-\frac{9}{4}g_2^2-8g_3^2+\frac{9}{2}y_t^2+\lambda_\nu^2\right] \, , \label{eq:beta-yt-MIII} \\
16\pi^2\beta_{\lambda_\nu} &= \lambda_\nu\left[-\frac{9}{20}g_1^2-\frac{9}{4}g_2^2+3y_t^2+\frac{5}{2}\lambda_\nu^2\right] \, . \label{eq:beta-lambdanu-MIII}
\end{align}

\subsection*{Model~\texorpdfstring{\MIV}{IV}}

\begin{align}
16\pi^2\beta_{g_1} &= \frac{49}{10}\,g_1^3 \, , \label{eq:beta-g1-MIV} \\
16\pi^2\beta_{g_2} &= -\frac{19}{6}\,g_2^3 \, , \label{eq:beta-g2-MIV} \\
16\pi^2\beta_{g_3} &= -7\,g_3^3 \, , \label{eq:beta-g3-MIV} \\
16\pi^2\beta_{y_t} &= y_t\left[-\frac{17}{20}g_1^2-\frac{9}{4}g_2^2-8g_3^2+\frac{9}{2}y_t^2+\lambda_e^2\right] \, , \label{eq:beta-yt-MIV} \\
16\pi^2\beta_{\lambda_e} &= \lambda_e\left[-\frac{9}{4}g_1^2-\frac{9}{4}g_2^2+3y_t^2+\frac{5}{2}\lambda_e^2\right] \, . \label{eq:beta-lambdae-MIV}
\end{align}
		
\newpage
\bibliographystyle{apsrev4-1}
		
\bibliography{main}

\end{document}